\documentclass[journal]{IEEEtran}

\usepackage{amsmath} % substack
\usepackage{array} % m{...} and b{...} in tabular
\usepackage{balance} % balance
\usepackage{comment} % todo
\usepackage{graphicx} % includegraphics, resizebox, scalebox
\usepackage{multirow} % multirow
\usepackage{pifont} % ding
\usepackage{rotating} % sidewaystable, rotatebox
\usepackage[caption=false]{subfig}
\usepackage{ulem}
\usepackage[hyphens]{url} % url

\usepackage{enumitem}

\setlist[itemize]{noitemsep, topsep=0.05cm}%{noitemsep,topsep=0pt,parsep=0pt,partopsep=0pt}%
\setlist[enumerate]{noitemsep, topsep=0.02cm}%{noitemsep,topsep=0pt,parsep=0pt,partopsep=0pt}%

\setlist[itemize]{leftmargin=*}
\setlist[enumerate]{leftmargin=*}

\numberwithin{equation}{subsubsection}

\newcommand*\rot{\rotatebox{90}}

\newcommand*\NO{\ding{55}}
\newcommand*\OKB{\ding{52}}

\newcommand{\tabred}{-4pt}

\DeclareMathOperator*{\argmax}{\arg\!\max}

\usepackage{todonotes}

\newcommand{\rikiRM}[1]{{\color{blue}\sout{}}}

\newcommand{\qqRM}[1]{{\color{purple}\sout{}}}

\newcommand{\aRM}[1]{{\color{red}\sout{}}}

\definecolor{blue}{rgb}{0.03, 0.1, .90}
\newcommand{\revOne}[1]{#1}

\begin{document}

% at, but, by, for, in, nor, of, on, or, the, to, and up, which are usually
\title{The Dark Side(-Channel) of Mobile Devices: \\A Survey on Network Traffic Analysis}
%The Dark Side(-Channel) of Smartphones and Tablets: A Survey on Mobile Network Traffic Analysis}

% Note positions of commas and nonbreaking spaces (~): LaTeX will not break
% Use \thanks to gain access to the first footnote area.
% A separate \thanks must be used for each paragraph as LaTeX2e's \thanks
% was not built to handle multiple paragraphs.
\author{Mauro~Conti, \IEEEmembership{Senior~Member,~IEEE},
        QianQian~Li,
        Alberto~Maragno,
        and~Riccardo~Spolaor*, \IEEEmembership{Member,~IEEE}.% <- this % stops a space
        
\thanks{* Corresponding author.}
\thanks{M. Conti, QQ. Li, A. Maragno, and R. Spolaor are with the SPRITZ Research Group, the Department of Mathematics, and the Human Inspired Technology Research Centre, University of Padua, Padua, Italy.}
%\thanks{R. Spolaor is also with the Department of Computer Science, University of Oxford, UK.}
\thanks{M. Conti is also Affiliated Professor with the Department of Electrical Engineering, University of Washington, USA.}
%\thanks{M. Conti, QQ. Li, and R. Spolaor are also with the Human Inspired Technology Research Centre, University of Padua, Padua, Italy.}
%
%\thanks{QQ. Li and R. Spolaor were also with the Department of General Psychology,  University of Padua, Padua, Italy.}
%
\thanks{E-mails: \texttt{\{conti,qianqian,rspolaor\}@math.unipd.it}, \texttt{\revOne{alberto.maragno@gmx.com}}.}% <- this % stops a space
%\thanks{Manuscript received month day, year; revised month day, year.}
}
\maketitle

% in the abstract.
\begin{abstract}
In recent years, mobile devices (e.g., smartphones and tablets) have met an increasing commercial success and have become a fundamental element of the everyday life for billions of people all around the world. Mobile devices are used not only for traditional communication activities (e.g., voice calls and messages) but also for more advanced tasks made possible by an enormous amount of multi-purpose applications (e.g., finance, gaming, and shopping). 
As a result, those devices generate a significant network traffic (a consistent part of the overall Internet traffic). For this reason, the research community has been investigating security and privacy issues that are related to the network traffic generated by mobile devices, which could be analyzed to obtain information useful for a variety of goals (ranging from fine-grained user profiling to device security and network optimization).

In this paper, we review the works that contributed to the state of the art of network traffic analysis targeting mobile devices. In particular, we present a systematic classification of the works in the literature according to three criteria: (i) the goal of the analysis; (ii) the point where the network traffic is captured; and (iii) the targeted mobile platforms. In this survey, we consider points of capturing such as Wi-Fi access points, software simulation, and inside real mobile devices or emulators. 
For the surveyed works, we review and compare analysis techniques, validation methods,  and achieved results. 
We also discuss possible countermeasures, challenges, and possible directions for future research on mobile traffic analysis and other emerging domains (e.g., Internet of Things).
We believe our survey will be a reference work for researchers and practitioners in this research field.
\end{abstract}


\section{Introduction}
\label{sec:introduction}

\IEEEPARstart{T}{he} last decade has been marked by the rise of mobile
devices which are nowadays widely spread among people.
The most diffused examples of such mobile devices are smartphones and tablets.
When compared with traditional cell phones, smartphones and tablets (henceforth also referred as \textit{mobile devices}) have an enormously increased computational power, more available memory, a larger display, and Internet connectivity via both Wi-Fi and cellular networks. 
Moreover, such devices run mobile operating systems which are able to experience multimedia contents, as well as to run mobile applications (also called \textit{apps}). 
Combined together, these elements enable both smartphones and tablets to have the same functionalities typically offered by laptops and desktop computers. 

According to the statistics reported in~\cite{StatistaSmartphone2016}, smartphone users were $25.3\%$ of the global population in 2015, and this percentage is expected to grow till $37\%$ in 2020. 
Similarly, the statistics about tablets reported in~\cite{StatistaTablet2016} indicate a global penetration of $13.8\%$ in 2015, expected to reach $19.2\%$ in 2020. 
The driving forces of this tremendous success are the ubiquitous Internet connectivity, thanks to the worldwide deployment of cellular and Wi-Fi networks, and a large number of apps available in the official (and unofficial) marketplaces. 
A mobile device typically hosts a lot of sensitive information about its owner, such as contacts, photos and videos, and GPS position. 
Such information must be properly protected, especially when it is transmitted to remote services.
Since an important fraction of the overall Internet traffic is due to mobile devices, it is not surprising that attackers and network traffic analysts have soon started to target them. 
For this reason, the research community investigates network traffic analysis techniques to improve both security and privacy on mobile devices.

Network traffic analysis (henceforth simply referred as \textit{traffic analysis}) is the branch of computer science that studies inferential methods which take the network traces of a group of devices (from a few to many thousands) as input, and give information about those devices, their users, their apps, or the traffic itself as output. 
Network traces can be captured at different layers (e.g., data-link layer, application layer), different points (e.g., within a Wi-Fi network, within the devices), and their content is often encrypted (making analysis even more challenging). 
Typically, researchers follow two different approaches to analyze mobile network traffic: (i) taking pre-existent methods designed for traditional Internet traffic, and adapting them to the mobile scenario; or (ii) developing new methods tailored to mobile Internet traffic properties.
It is worth to underline that this survey focuses on Internet traffic only. 
We do not consider other types of mobile traffic (e.g., Call Detail Records) or data transmission technologies (e.g., Bluetooth, infrared).

\textit{Contributions} --
In this paper, we survey the state of the art of network traffic analysis on mobile devices, giving the following contributions: 
\begin{itemize}
\item We categorize each work according to three criteria: 
\begin{enumerate}
\item the goal of the analysis;
\item the point where the network traffic is captured (henceforth simply referred to as \textit{point of capturing}); and 
\item the targeted mobile platforms. 
\end{enumerate}
Moreover, we provide further insights on the models and methods that can be used to perform traffic analysis targeting mobile devices. 

\item The objective of this survey is three-fold. 
On the one hand, we provide a systematic classification of state-of-the-art techniques for the analysis of the network traffic of mobile devices. 
On the other hand, we provide an overview of methodologies adopted for the analyses and information about the datasets used for validating the obtained results.
We also discuss possible countermeasures to thwart mobile traffic analysis and provide meaningful insights about challenges and pitfalls related to the topics that have been investigated, as well as identify possible future research directions. 
We believe that our work will both help new researchers in this field and foster future research trends.

\item To the best of our knowledge, we are the first to survey the works that analyze datasets of mobile traffic that are either: 
(i) logged on one or more mobile devices; 
(ii) extracted from wired network traces; 
(iii) sniffed at one or more access points of a Wi-Fi network; 
(iv) eavesdropped by one or more Wi-Fi monitors; 
(v) produced by one or more mobile device emulators; or 
(vi) generated via a software simulation.
The work by Naboulsi et al.~\cite{Naboulsi2016} is the only published survey that reviews the works in which the analyzed datasets are collected within the network infrastructure of one or more cellular providers (e.g., 3G and HSDPA).
In fact, our survey is complementary to the one in~\cite{Naboulsi2016}, and together they provide a complete treatment of the research field of traffic analysis targeting mobile devices.
 
\end{itemize}

\begin{figure}
\centering
\includegraphics[scale=0.65]{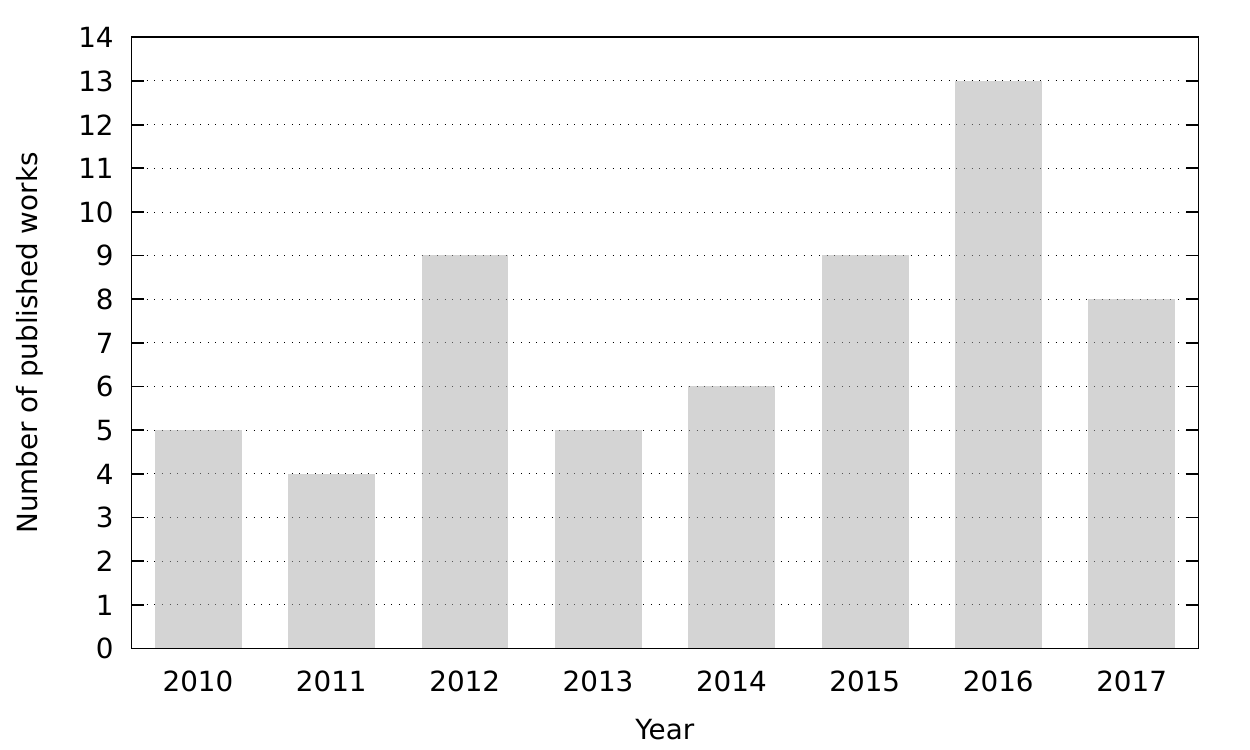}
\caption{Number of published works that we identified and considered relevant to this survey, sorted by year of publication.}
\label{fig:papers_by_year}
\end{figure}

Overall, we survey \revOne{$59$} works, published between 2010 and 2017. 
Figure \ref{fig:papers_by_year} shows that the number of publications in the considered research field has significantly increased in the last years. 
We believe that this amount of work will grow in the future as the global spreading of mobile devices is increasing and their contribution to the worldwide Internet traffic is becoming more significant.

\textit{Organization} --
The rest of the document is organized as follows. 
\revOne{Section~\ref{sec:categorization} provides a road-map of the survey that helps the reader with understanding the classifications adopted to report the surveyed works.} 
In the following sections, we survey the works according to three criteria: 
in Section~\ref{sec:goals}, the goal of the analysis performed on the mobile traffic; 
in Section~\ref{sec:points_of_capturing}, the point of capturing \revOne{used to collect the mobile traffic}; 
and, in Section~\ref{sec:mobile_platforms}, the targeted mobile platforms. 
In Section~\ref{sec:models_methods}, we review the models and methods applied in the surveyed works to perform traffic analysis targeting mobile devices. 
\revOne{In Section~\ref{sec:validation}, we describe the validation datasets used in the evaluation and discuss the obtained results. 
We discuss the effect of network traffic encryption and other countermeasures in Section~\ref{sec:countermeasures}.
In Section~\ref{sec:present_future}, we outline the current situation in the field of traffic analysis targeting mobile devices, as well as the trends that are likely to drive research in the next future.} 
Finally, we conclude the paper in Section~\ref{sec:conclusions}.

\section{Categorization of Work}
\label{sec:categorization}

In this section, we present an overview of the classification criteria we follow to categorize the works considered in our survey: the goal of the analysis performed on the mobile traffic (Section~\ref{sec:classification_by_goal}); the point of capturing of the mobile traffic (Section~\ref{sec:classification_by_PoC}); and the targeted mobile platform (Section~\ref{sec:classification_by_platform}). 
In Table \ref{tab:all_papers}, we report the surveyed works according to these criteria. 
For each work, we also indicate whether the proposed analyses are still applicable in case of traffic encryption via either SSL/TLS or IPsec (see Section~\ref{sec:points_of_capturing} for more details about how traffic encryption affects the analyses presented in the surveyed works). 
It is worth to notice that a few works (i.e., Wei et al.~\cite{Wei2012ProfileDroid}, Le et al. \cite{Le2015}, Alan et al. \cite{Alan2016}, Tadrous and Sabharwal~\cite{Tadrous2016}, and Wang et al. \cite{Wang2016}) propose multiple traffic analysis techniques, each affected by traffic encryption in a different way.

\begin{table*}
\caption{All surveyed works.}
\label{tab:all_papers}
\centering
\begin{tabular}{
|>{\hspace{\tabred}}c<{\hspace{\tabred}}
|>{\hspace{\tabred}}l<{\hspace{\tabred}}
|>{\hspace{\tabred}}c<{\hspace{\tabred}}
|>{\hspace{\tabred}}c<{\hspace{\tabred}}
|>{\hspace{\tabred}}c<{\hspace{\tabred}}
|>{\hspace{\tabred}}c<{\hspace{\tabred}}
|>{\hspace{\tabred}}c<{\hspace{\tabred}}|}
\hline
\textbf{Year} & \multicolumn{1}{c|}{\textbf{Paper}} & \textbf{Goal of the Analysis} & \textbf{Point of Capturing} & \textbf{Targeted Mobile Platform} & \textbf{SSL/TLS} & \textbf{IPsec} \\ \hline
% 2010
\multirow{5}{*}{2010} & Afanasyev et al. \cite{Afanasyev2010} & Characterization, Usage & APs, Wired & Platform-independent & \NO & \NO \\ \cline{2-7}
& Falaki et al. \cite{Falaki2010} & Characterization, Usage & Devices & Android, Windows Mobile & \OKB & \NO \\ \cline{2-7}
& Husted et al. \cite{Husted2010} & \revOne{Position} & Simulator & Platform-independent & \OKB & \OKB \\ \cline{2-7}
& Maier et al. \cite{Maier2010} & Characterization, Usage & Wired & Platform-independent & \NO & \NO \\ \cline{2-7}
& Shepard et al. \cite{Shepard2010} & Characterization & Devices & iOS & \OKB & \NO\\ \hline
% 2011
\multirow{5}{*}{2011} & Finamore et al. \cite{Finamore2011} & Characterization, Usage & Wired & Platform-independent & \NO & \NO\\ \cline{2-7}
& Gember et al. \cite{Gember2011} & Characterization, Usage & APs & Platform-independent & \NO & \NO\\ \cline{2-7}
& \multirow{2}{*}{Lee et al. \cite{Lee2011}} & App & \multirow{2}{*}{Wired} & Android, iOS & \NO & \NO \\ \cline{3-3} \cline{5-7}
& & Characterization, Usage & & Platform-independent & \NO & \NO \\ \cline{2-7}
& {Rao et al. \cite{Rao2011}} & {Characterization} & {Wired} & {Android, iOS} & \NO & \NO \\ \hline
% 2012
\multirow{9}{*}{2012} & Baghel et al. \cite{Baghel2012} & Characterization & Wired & Android & \OKB & \NO \\ \cline{2-7}
& Chen et al. \cite{Chen2012} & Characterization & Wired & Platform-independent & \NO & \NO \\ \cline{2-7}
& Ham et al. \cite{Ham2012} & Usage & Devices & Android & \OKB & \OKB \\ \cline{2-7}
& Musa et al. \cite{Musa2012} & \revOne{Position} & Monitors & Platform-independent & \OKB & \OKB \\ \cline{2-7}
& {Shabtai et al. \cite{Shabtai2012}} & {Malware} & {Devices} & {Android} & {\OKB} & {\OKB} \\ \cline{2-7}
& {Stevens et al. \cite{Stevens2012}} & {PII Leakage} & {APs} & {Android} & \NO & \NO \\ \cline{2-7}
& Su et al. \cite{Su2012} & Malware & Devices & Android & \OKB & \NO\\ \cline{2-7}
& Wei et al. \cite{Wei2012} & Malware & Wired & Android & \NO & \NO\\ \cline{2-7}
& {Wei et al. \cite{Wei2012ProfileDroid}} & {Characterization} & {Devices} & {Android} & \OKB & \OKB/\NO\\ \hline
% 2013
\multirow{5}{*}{2013} & Barbera et al. \cite{Barbera2013} & Sociological & Monitors & Platform-independent & \OKB & \OKB\\ \cline{2-7}
& Kuzuno et al. \cite{Kuzuno2013} & PII Leakage & Devices & Android & \NO & \NO\\ \cline{2-7}
& Qazi et al. \cite{Qazi2013} & App & APs, Devices & Android & \OKB & \NO\\ \cline{2-7}
& {Rao et al. \cite{Rao2013}} & {App, PII Leakage} & {Wired} & {Android, iOS} & \OKB & \NO\\ \cline{2-7}
& {Watkins et al. \cite{Watkins2013}} & {User Actions} & {APs} & {Android} & {\OKB} & {\OKB} \\ \hline
% 2014
\multirow{7}{*}{2014} & \multirow{2}{*}{Chen et al. \cite{Chen2014}} & OS & APs & Android, iOS & \OKB & \NO\\ \cline{3-7}
& & Tethering & Monitors, Wired & Platform-independent & \OKB & \NO\\ \cline{2-7}
& Coull et al. \cite{Coull2014} & User Actions, OS & Devices & iOS & \OKB & \OKB\\ \cline{2-7}
& Crussell et al. \cite{Crussell2014} & Ad Fraud & Emulators & Android & \NO & \NO\\ \cline{2-7}
& Lindorfer et al. \cite{Lindorfer2014} & Characterization & Emulators & Android & \NO & \NO \\ \cline{2-7}
& Shabtai et al. \cite{Shabtai2014} & Malware & Devices & Android & \OKB & \OKB\\ \cline{2-7}
& Verde et al. \cite{Verde2014} & User Fingerprinting & Wired & Platform-independent & \OKB & \OKB\\ \hline
% 2015
\multirow{9}{*}{2015} & Chen et al. \cite{Chen2015} & Characterization & Wired & Android & \NO & \NO \\ \cline{2-7}
& Fukuda et al. \cite{Fukuda2015} & Characterization, Usage & Devices & Android, iOS & \OKB & \OKB \\ \cline{2-7}
& Le et al. \cite{Le2015} & App, PII Leakage & Devices & Android & \OKB/\NO & \NO\\ \cline{2-7}
& Park et al. \cite{Park2015} & User Actions & Wired & Android & \OKB & \NO \\ \cline{2-7} %check
& Soikkeli et al. \cite{Soikkeli2015} & Usage & Devices & Platform-independent & \OKB & \OKB \\ \cline{2-7} %%%%MANCA DESCRIZIONE su cosa faccia
& Song et al. \cite{Song2015} & PII Leakage & Devices & Android & \OKB & \NO\\ \cline{2-7}
& Wang et al. \cite{Wang2015} & App & Monitors & iOS & \OKB & \OKB\\ \cline{2-7}
& Yao et al. \cite{Yao2015} & App & APs, Emulators & Android, iOS, Symbian & \NO & \NO\\ \cline{2-7}
& Zaman et al. \cite{Zaman2015} & Malware & Devices & Android & \NO & \NO\\ \hline
% 2016
\multirow{13}{*}{2016} & Alan et al. \cite{Alan2016} & App & APs & Android & \OKB & \OKB/\NO\\ \cline{2-7}
& Conti et al. \cite{Conti2016} & User Actions & Wired & Android & \OKB & \NO\\ \cline{2-7}
& \revOne{Fu et al. \cite{Fu2016}} & \revOne{User Actions} & \revOne{APs} & \revOne{Android} & \revOne{\OKB} & \revOne{\OKB} \\ \cline{2-7}
& {Mongkolluksamee et al. \cite{Mongkolluksamee2016}} & {App} & {Devices} & {Android} & \OKB & \NO\\ \cline{2-7}
& Narudin et al. \cite{Narudin2016} & Malware & Devices, Emulators & Android & \NO & \NO\\ \cline{2-7}
& {Nayam et al. \cite{Nayam2016}} & {Characterization} & {Wired} & {Android, iOS} & \NO & \NO \\ \cline{2-7}
& Ren et al. \cite{Ren2016} & PII Leakage & Wired & Android, iOS, Windows Phone & \OKB & \NO\\ \cline{2-7}
& Ruffing et al. \cite{Ruffing2016} & OS & Monitors & Android, iOS, Windows Phone, Symbian & \OKB & \OKB \\ \cline{2-7}
& Saltaformaggio et al. \cite{Saltaformaggio2016} & User Actions & APs & Android, iOS & \OKB & \OKB\\ \cline{2-7}
& Spreitzer et al. \cite{Spreitzer2016} & Website Fingerprinting & Devices & Android & \OKB & \OKB\\ \cline{2-7}
& Tadrous et al. \cite{Tadrous2016} & Characterization & APs & Android, iOS & \OKB & \OKB/\NO \\ \cline{2-7}
& {Vanrykel et al. \cite{Vanrykel2016}} & {PII Leakage, User Fingerprinting} & {Wired} & {Android} & \NO & \NO\\ \cline{2-7}
& {Wang et al. \cite{Wang2016}} & {Malware} & {Wired} & {Android} & \OKB/\NO & \NO\\ \hline
% 2017
\multirow{8}{*}{2017} & {Arora et al. \cite{Arora2017}} & {Malware} & {Devices} & {Android} & \OKB & \OKB\\ \cline{2-7}
& \revOne{Chen et al. \cite{Chen2017}} & \revOne{App} & \revOne{Emulators} & \revOne{Android} & \revOne{\NO} & \revOne{\NO} \\ \cline{2-7}
& \revOne{Cheng et al. \cite{Cheng2017}} & \revOne{PII Leakage} & \revOne{Wired} & \revOne{Android} & \revOne{\OKB} & \revOne{\NO} \\ \cline{2-7}
& {Continella et al. \cite{Continella2017}} & {PII Leakage} & {Wired} & {Android} & \OKB & \NO\\ \cline{2-7}
& {Espada et al. \cite{Espada2017}} & {Characterization} & {Devices} & {Android} & \OKB & \NO \\ \cline{2-7}
& {Malik et al. \cite{Malik2017}} & {OS} & {APs} & {Android, iOS, Windows Phone} & {\OKB} & {\OKB} \\ \cline{2-7}
& Taylor et al. \cite{Taylor2017} & App & Wired & Android & \OKB & \NO\\ \cline{2-7}
& {Wei et al. \cite{Wei2017}} & {Characterization, Usage} & {Wired} & {Platform-independent} & {\NO} & {\NO} \\ \hline
\end{tabular}
\end{table*}

\subsection{Classification by Goal of the Analysis}
\label{sec:classification_by_goal}

\begin{figure*}[t]
\centering
\includegraphics[width=0.9\textwidth]{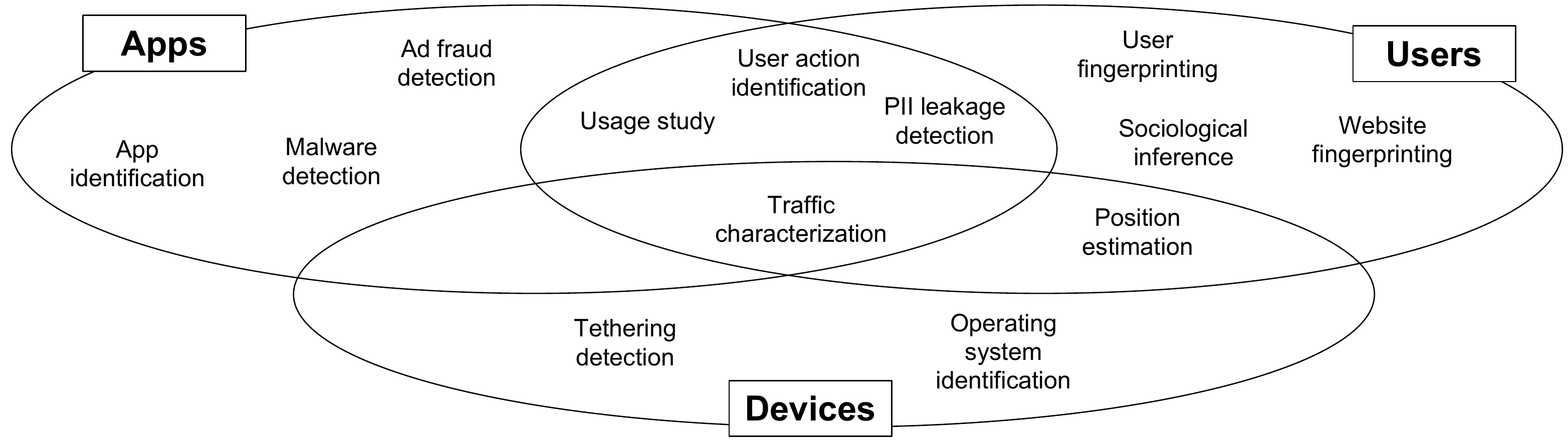}
\caption{The goals of traffic analysis targeting mobile devices, and their pertinence to apps, mobile users, and mobile devices.}
\label{fig:goals}
\end{figure*}

The first classification takes into account the goal of the analysis performed on the captured mobile traffic. For each surveyed work, Table \ref{tab:all_papers} provides this information in the \textit{Goal of the Analysis} column. We survey more in detail the works according to this classification in Section~\ref{sec:goals}.

Overall, we are able to identify \revOne{thirteen} goals. In Figure \ref{fig:goals}, we depict such goals by their field of pertinence: apps, mobile users, and mobile devices. In what follows, we list and briefly describe each goal, sorted by the number of works:
\begin{itemize}
\item \textbf{Traffic characterization} (\textit{Characterization} in Table \ref{tab:all_papers}): to infer the network properties of mobile traffic. The knowledge of such properties is crucial to effectively deploy and configure resources in cellular networks, as well as in Wi-Fi networks serving mobile devices. More details in Section~\ref{sec:traffic_characterization}.
\item \textbf{App identification} (\textit{App} in Table \ref{tab:all_papers}): to identify the network traffic belonging to a specific mobile app. This type of analysis can help network administrators in resource planning and management, as well as in app-specific policy enforcement (e.g., forbidding a social network app within an enterprise network). Moreover, app identification can be employed to uncover the presence of sensitive apps (e.g., dating, health, religion) in the mobile device of a target user. See Section~\ref{sec:app_identification}.
\item \textbf{Usage study} (\textit{Usage} in Table \ref{tab:all_papers}): to infer the usage habits of mobile users (e.g., which are the most frequently used apps). As an example, the knowledge of the places where mobile devices are mostly used can drive the deployment of cellular stations and Wi-Fi hotspots. See Section~\ref{sec:usage_study}.
\item \textbf{PII leakage detection} (\textit{PII Leakage} in Table \ref{tab:all_papers}): to detect and/or prevent the leakage of a mobile user's Personal Identifiable Information (PII). This type of analysis can be employed to assess the behavior of a mobile app from a privacy point of view, by checking which PII it actually discloses to remote hosts. Detecting PII leakage is also the first step to prevent such problem, since it is then possible to block network transmissions carrying PII, or replace sensitive information with bogus data. See Section~\ref{sec:PII_leakage_detection}.

\item \textbf{Malware detection} (\textit{Malware} in Table \ref{tab:all_papers}): to detect whether a mobile app behaves maliciously (e.g., downloading and installing malicious code from the network). This type of analysis can be used to assess the security of an app submitted by a developer to a mobile marketplace. In such case, the result of the security tests decides whether the app can be released to the public. Moreover, malware detection algorithms can be embedded into anti-virus apps that mobile users can use to check whether an installed app is malicious. See Section~\ref{sec:malware_detection}.

\item \textbf{User action identification} (\textit{User Actions} in Table \ref{tab:all_papers}): to identify a specific action that a mobile user performed on her mobile device (e.g., uploading a photo on Instagram), or to infer some information about that specific action (e.g., the length of a mobile user's message sent through an instant messaging app). Researchers can employ such analysis to discover the identity behind an anonymous social network profile. This can be accomplished by verifying whether there is a match between the events reported on that profile's page, and the actions a suspect performed while using the mobile app of that social network. Alternatively, it is possible to build behavioral profiles of mobile users, which are useful for user reconnaissance within networks and, in aggregated form, for marketing studies. See Section~\ref{sec:user_action_identification}.

\item \textbf{Operating system identification} (\textit{OS} in Table \ref{tab:all_papers}): to discover the operating system of a mobile device. This type of analysis is usually a preliminary phase for more advanced attacks against mobile devices: the adversary tries to infer the operating system of the target mobile device in order to subsequently exploit an ad hoc vulnerability for that specific OS. Moreover, operating system identification carried out on a large mobile user population can be a starting point for other types of analysis not directly related to computer science (e.g., sociological studies). See Section~\ref{sec:OS_identification}.

\item \textbf{Position estimation} (\textit{Position} in Table \ref{tab:all_papers}): to estimate the position and/or the trajectory (i.e., the movements) of a mobile device within a geographical area. This type of analysis helps infer social status, interests, and habits of the owner of a mobile device. As a further step, the profiles of several mobile users can be aggregated for marketing, as well as sociological studies. Besides, position estimation can aid road traffic prediction along urban streets, by leveraging the most frequent trajectories followed by the citizens that move along the city. See Section~\ref{sec:position_estimation}.

\item \textbf{User fingerprinting} (\textit{User Fingerprinting} in Table \ref{tab:all_papers}): to detect the traffic belonging to a specific mobile user. This type of analysis can be employed to trace a mobile user, by approximating her position with the location of the Wi-Fi hotspot or cellular station to which her mobile device is connected. From this information, it is then possible to build a behavioral profile of that mobile user. 
Alternatively, it is possible to examine a mobile traffic dataset in order to extract and group together the network traces generated by a specific mobile user. 
Such data can be subsequently used for other types of traffic analysis targeting that user. 
See Section~\ref{sec:user_fingerprinting}.

\item \textbf{Ad fraud detection} (\textit{Ad Fraud} in Table \ref{tab:all_papers}): to detect ad fraud by a mobile app, i.e., to recognize whether a mobile app is trying to trick the advertising business model (e.g., fabricating false user clicks on ads). This type of analysis is valuable to ad providers, which can rely on it to protect themselves from dishonest app developers. See Section~\ref{sec:ad_fraud_detection}.

\item \textbf{Sociological inference} (\textit{Sociological} in Table \ref{tab:all_papers}): to infer some kind of sociological information about mobile users (e.g., language, religion, health condition, sexual preference, wealth), from one or more properties related to their mobile devices (e.g., list of installed apps, associated Wi-Fi networks). See Section~\ref{sec:sociological_inference}.

\item \textbf{Tethering detection} (\textit{Tethering} in Table \ref{tab:all_papers}): to detect whether a mobile device is tethering, i.e., it is sharing its Internet connectivity with other devices, for which it acts as an access point. Tethering constitutes a problem for cellular network providers, since it significantly increases the volume of network traffic generated by a single client. Such providers are therefore interested in tethering detection techniques that can be used to prevent their customers from sharing their Internet connectivity, or simply require them to pay an extra fee to do that. See Section~\ref{sec:tethering_detection}.

\item \textbf{Website fingerprinting} (\textit{Website Fingerprinting} in Table \ref{tab:all_papers}): to infer which websites and/or webpages are visited by a mobile user while navigating via the web browser of her mobile device. Similarly to sociological inference, this type of analysis can reveal interests, social habits, religious belief, as well as sexual and political orientations of a mobile user. See Section~\ref{sec:website_fingerprinting}.
\end{itemize}

\subsection{Classification by Point of Capturing}
\label{sec:classification_by_PoC}

The second classification considers where and how the mobile traffic is captured. For each surveyed work, Table \ref{tab:all_papers} provides this information in the \textit{Point of Capturing} column. It is worth to notice that: (i) we focus on the (hardware and/or software) equipment that captures the traffic; and (ii) we report the point of capturing only for those datasets for which the authors give enough details about the collection process. We survey more in detail the works according to this classification in Section~\ref{sec:points_of_capturing}.

Overall, we identify six different points of capturing. 
In what follows, we list and briefly describe each of them, sorted by the number of works in which a point of capturing is employed:
\begin{itemize}
\item At one or more wired network equipments (\textit{Wired} in Table~\ref{tab:all_papers}). The size of the population of monitored mobile devices varies according to the type of considered network equipments: thousands of mobile users in the case of edge routers (i.e., routers connecting customers to the ISP's backbone) and Internet gateways; from tens to a few hundreds in the case of VPN servers and forwarding servers (i.e., traditional desktop computers \revOne{set up to log all traffic traversing a wired link that connects a Wi-Fi hotspot serving mobile devices to the Internet}). More details in Section~\ref{sec:wired_network_equipments}.
\item Within one or more mobile devices, i.e., client-side (\textit{Devices} in Table \ref{tab:all_papers}). This type of point of capturing is particularly useful if we want to target a specific mobile app (e.g., Facebook), or a particular network interface (e.g., cellular). We specify that this category also includes the case of a network traffic logger installed within either: (i) a mobile device emulator; and (ii) a machine to which the mobile traffic is mirrored using a remote virtual network interface. See Section~\ref{sec:mobile_devices}.
\item At one or more access points of a Wi-Fi network (\textit{APs} in Table \ref{tab:all_papers}). This type of point of capturing allows the number of monitored mobile devices to vary from tens to a few thousands, and it is suitable to capture the traffic of mobile devices while their users are performing network-intensive tasks (e.g., watching streaming videos, updating apps). See Section~\ref{sec:Wi-Fi_access_points}.
\item At one or more machines running virtual mobile devices, i.e., emulators (\textit{Emulators} in Table \ref{tab:all_papers}). 
Running multiple virtual instances of mobile devices and controlling them via automated tools make possible to collect network traffic on a large-scale. On real mobile devices, same data collection would be far more expensive. 
It is important to highlight that the traffic logging is performed by the host machines or their virtualization managers. 
We do not consider the case in which the traffic logging takes place within the emulated mobile devices (such case is covered by the \textit{Devices} category). See Section~\ref{sec:emulators}.
\item At one or more Wi-Fi monitors (\textit{Monitors} in Table \ref{tab:all_papers}). Researchers usually employ this type of capturing devices to focus the network traffic collection process on a specific geographical area of interest (e.g., a train station). Such approach is often the only viable solution whether it is not possible to directly access a target mobile device, or the network to which it is connected. See Section~\ref{sec:Wi-Fi_monitors}.
\item At one or more virtual capturing points within a simulated environment generated and managed by a software program (\textit{Simulator} in Table \ref{tab:all_papers}). This point of capturing can help study particular deployments of mobile devices that are not observable in a real-world scenario because of technical, economical, or legal constraints. See Section~\ref{sec:simulator}.
\end{itemize}

\subsection{Classification by Targeted Mobile Platform}
\label{sec:classification_by_platform}

The third classification considers the mobile platforms that are targeted by the traffic analysis. For each surveyed work, Table \ref{tab:all_papers} provides this information in the \textit{Targeted Mobile Platform} column. It is worth to specify that we classify a work as \textit{platform-independent} if its authors do not provide information about the targeted mobile platforms, or such information is not relevant to the analysis they perform on the mobile traffic \revOne{(we discuss this type of works in Section~\ref{sec:platform-independent_works})}. We survey more in detail the works according to this classification in Section~\ref{sec:mobile_platforms}.

Overall, we find four distinct mobile platforms \revOne{(in what follows, listed by their popularity in the surveyed works)}: Android, Google's open-source mobile operating system (we discuss it in Section~\ref{sec:Android}); iOS, the operating system of Apple's mobile devices (Section~\ref{sec:iOS}); Windows Mobile/Phone, the mobile counterpart of Microsoft's desktop operating system (Section~\ref{sec:Windows_Mobile/Phone}); and Symbian, the first released modern mobile operating system (Section~\ref{sec:Symbian}).

\section{Goals of Traffic Analysis Targeting Mobile Devices}
\label{sec:goals}

\begin{figure}
\centering
\includegraphics[scale=0.70]{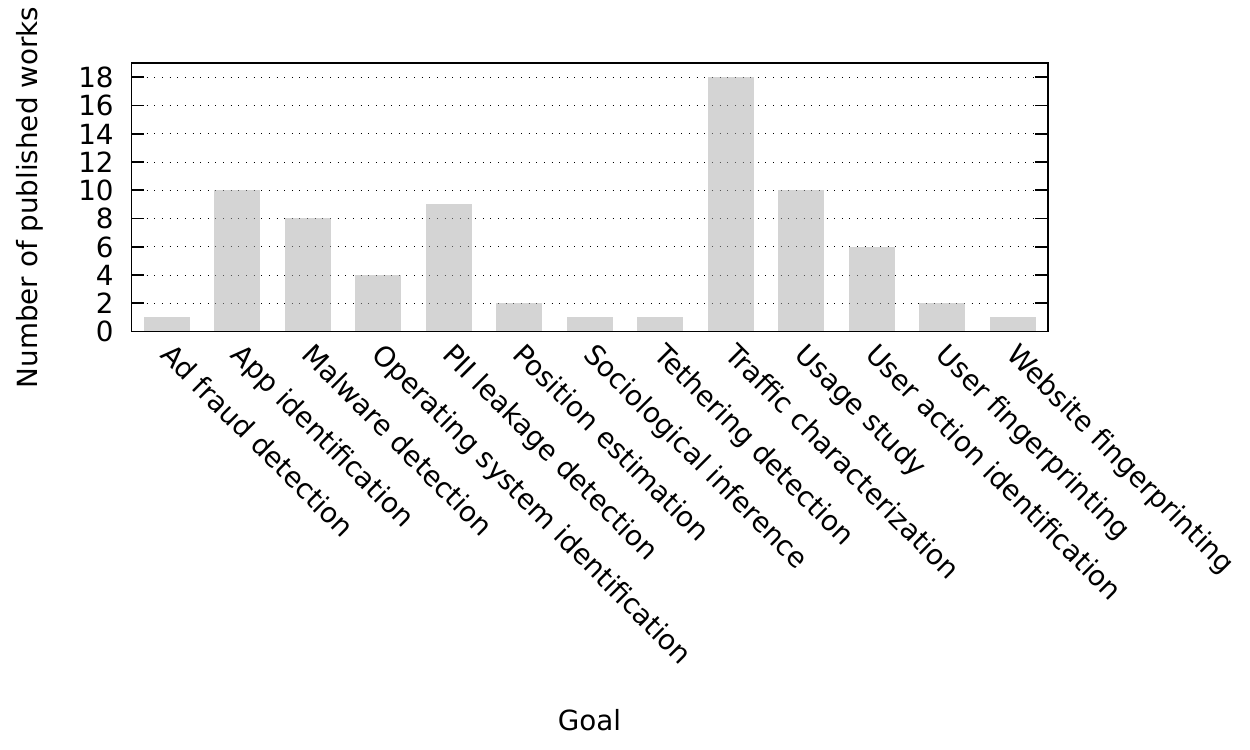}
\caption{Number of published works contributing traffic analysis methods targeting mobile devices, sorted by goal of the analysis.}
\label{fig:papers_by_goal}
\end{figure}

In this section, we survey the works according to the goal of the analysis that is performed on the mobile traffic. 
Table~\ref{tab:goals} summarizes the goals of the surveyed works. 
As shown in Figure \ref{fig:papers_by_goal}, the most frequently pursued goal is traffic characterization (eighteen works), followed by app identification and usage study (ten works each), \revOne{PII leakage detection (nine works), malware detection (eight works)}, user action identification (\revOne{six} works), operating system identification (four works), and \revOne{position estimation and} user fingerprinting (two works \revOne{each}). Each of the following goals counts one work only: ad fraud detection, sociological inference, tethering detection, and website fingerprinting. As shown in Table \ref{tab:goals}, twelve works pursue two goals, and one work even three. 

In the following sections, we present the goal(s) and achieved results for each surveyed work. 
\revOne{For each goal, we report relevant aspects and findings in the state of the art, and we discuss whether the proposed analyses work} on encrypted network traffic.
For the sake of simplicity, encryption methods that make TCP headers and IP headers unavailable to the analysis are referred as to SSL/TLS and IPsec, respectively. 
\revOne{We enter in more detail about encryption as a countermeasure against mobile traffic analysis in Section~\ref{sub:countermeasures_encryption}. 
The treatment of each goal takes place in its own section, and all sections are ordered by goal popularity in the surveyed works.}

\begin{table}
\caption{The surveyed works by goal of the analysis performed on the mobile traffic.}
\label{tab:goals}
\centering
\scalebox{0.85}{
\begin{tabular}{|>{\hspace{\tabred}}c<{\hspace{\tabred}}|>{\hspace{\tabred}}l<{\hspace{\tabred}}|
>{\hspace{\tabred}}c<{\hspace{\tabred}}|
>{\hspace{\tabred}}c<{\hspace{\tabred}}|>{\hspace{\tabred}}c<{\hspace{\tabred}}|
>{\hspace{\tabred}}c<{\hspace{\tabred}}|>{\hspace{\tabred}}c<{\hspace{\tabred}}|
>{\hspace{\tabred}}c<{\hspace{\tabred}}|>{\hspace{\tabred}}c<{\hspace{\tabred}}|
>{\hspace{\tabred}}c<{\hspace{\tabred}}|>{\hspace{\tabred}}c<{\hspace{\tabred}}|
>{\hspace{\tabred}}c<{\hspace{\tabred}}|>{\hspace{\tabred}}c<{\hspace{\tabred}}|
>{\hspace{\tabred}}c<{\hspace{\tabred}}|>{\hspace{\tabred}}c<{\hspace{\tabred}}|}
\hline
\textbf{Year} & \multicolumn{1}{c|}{\textbf{Paper}} & \rot{\textbf{Ad Fraud Detection}} & \rot{\textbf{App Identification}} & \rot{\textbf{Malware Detection}} & \rot{\textbf{Operating System Identification }} & \rot{\textbf{PII Leakage Detection}} & \rot{\revOne{\textbf{Position Estimation}}} & \rot{\textbf{Sociological Inference}} & \rot{\textbf{Tethering Detection}} & \rot{\textbf{Traffic Characterization}} & \rot{\textbf{Usage Study}} & \rot{\textbf{User Action Identification}} & \rot{\textbf{User Fingerprinting}} & \rot{\textbf{Website Fingerprinting}} \\ \hline
% 2010
\multirow{5}{*}{2010} & Afanasyev et al. \cite{Afanasyev2010} & & & & & & & & & \OKB & \OKB & & & \\ \cline{2-15}
& Falaki et al. \cite{Falaki2010} & & & & & & & & & \OKB & \OKB & & & \\ \cline{2-15}
& Husted et al. \cite{Husted2010} & & & & & & \revOne{\OKB} & & & & & & & \\ \cline{2-15}
& Maier et al. \cite{Maier2010} & & & & & & & & & \OKB & \OKB & & & \\ \cline{2-15}
& Shepard et al. \cite{Shepard2010} & & & & & & & & & \OKB & & & & \\ \hline
% 2011
\multirow{4}{*}{2011} & Finamore et al. \cite{Finamore2011} & & & & & & & & & \OKB & \OKB & & & \\ \cline{2-15}
& Gember et al. \cite{Gember2011} & & & & & & & & & \OKB & \OKB & & & \\ \cline{2-15}
& Lee et al. \cite{Lee2011} & & \OKB & & & & & & & \OKB & \OKB & & & \\ \cline{2-15}
& {Rao et al. \cite{Rao2011}} & & & & & & & & & {\OKB} & & & & \\ \hline
% 2012
\multirow{9}{*}{2012} & Baghel et al. \cite{Baghel2012} & & & & & & & & & \OKB & & & & \\ \cline{2-15}
& Chen et al. \cite{Chen2012} & & & & & & & & & \OKB & & & & \\ \cline{2-15}
& Ham et al. \cite{Ham2012} & & & & & & & & & & \OKB & & & \\ \cline{2-15}
& Musa et al. \cite{Musa2012} & & & & & & \revOne{\OKB} & & & & & & & \\ \cline{2-15}
& {Shabtai et al. \cite{Shabtai2012}} & & & {\OKB} & & & & & & & & & & \\ \cline{2-15}
& {Stevens et al. \cite{Stevens2012}} & & & & & {\OKB} & & & & & & & & \\ \cline{2-15}
& Su et al. \cite{Su2012} & & & \OKB & & & & & & & & & & \\ \cline{2-15}
& Wei et al. \cite{Wei2012} & & & \OKB & & & & & & & & & & \\ \cline{2-15}
& {Wei et al. \cite{Wei2012ProfileDroid}} & & & & & & & & & {\OKB} & & & & \\ \hline
% 2013
\multirow{5}{*}{2013} & Barbera et al. \cite{Barbera2013} & & & & & & & \OKB & & & & & & \\ \cline{2-15}
& Kuzuno et al. \cite{Kuzuno2013} & & & & & \OKB & & & & & & & & \\ \cline{2-15}
& Qazi et al. \cite{Qazi2013} & & \OKB & & & & & & & & & & & \\ \cline{2-15}
& {Rao et al. \cite{Rao2013}} & & {\OKB} & & & {\OKB} & & & & & & & & \\ \cline{2-15}
& {Watkins et al. \cite{Watkins2013}} & & & & & & & & & & & {\OKB} & & \\ \hline
% 2014
\multirow{6}{*}{2014} & Chen et al. \cite{Chen2014} & & & & \OKB & & & & \OKB & & & & & \\ \cline{2-15}
& Coull et al. \cite{Coull2014} & & & & \OKB & & & & & & & \OKB & & \\ \cline{2-15}
& Crussell et al. \cite{Crussell2014} & \OKB & & & & & & & & & & & & \\ \cline{2-15}
& Lindorfer et al. \cite{Lindorfer2014} & & & & & & & & & \OKB & & & & \\ \cline{2-15}
& Shabtai et al. \cite{Shabtai2014} & & & \OKB & & & & & & & & & & \\ \cline{2-15}
& Verde et al. \cite{Verde2014} & & & & & & & & & & & & \OKB & \\ \hline
% 2015
\multirow{9}{*}{2015} & Chen et al. \cite{Chen2015} & & & & & & & & & \OKB & & & & \\ \cline{2-15}
& Fukuda et al. \cite{Fukuda2015} & & & & & & & & & \OKB & \OKB & & & \\ \cline{2-15}
& Le et al. \cite{Le2015} & & \OKB & & & \OKB & & & & & & & & \\ \cline{2-15}
& Park et al. \cite{Park2015} & & & & & & & & & & & \OKB & & \\ \cline{2-15}
& Soikkeli et al. \cite{Soikkeli2015} & & & & & & & & & & \OKB & & & \\ \cline{2-15}
& Song et al. \cite{Song2015} & & & & & \OKB & & & & & & & & \\ \cline{2-15}
& Wang et al. \cite{Wang2015} & & \OKB & & & & & & & & & & & \\ \cline{2-15}
& Yao et al. \cite{Yao2015} & & \OKB & & & & & & & & & & & \\ \cline{2-15}
& Zaman et al. \cite{Zaman2015} & & & \OKB & & & & & & & & & & \\ \hline
% 2016
\multirow{13}{*}{2016} & Alan et al. \cite{Alan2016} & & \OKB & & & & & & & & & & & \\ \cline{2-15}
& Conti et al. \cite{Conti2016} & & & & & & & & & & & \OKB & & \\ \cline{2-15}
& \revOne{Fu et al. \cite{Fu2016}} & & & & & & & & & & & \revOne{\OKB} & & \\ \cline{2-15}
& {Mongkolluksamee et al. \cite{Mongkolluksamee2016}} & & {\OKB} & & & & & & & & & & & \\ \cline{2-15}
& Narudin et al. \cite{Narudin2016} & & & \OKB & & & & & & & & & & \\ \cline{2-15}
& {Nayam et al. \cite{Nayam2016}} & & & & & & & & & {\OKB} & & & & \\ \cline{2-15}
& Ren et al. \cite{Ren2016} & & & & & \OKB & & & & & & & & \\ \cline{2-15}
& Ruffing et al. \cite{Ruffing2016} & & & & \OKB & & & & & & & & & \\ \cline{2-15}
& Saltaformaggio et al. \cite{Saltaformaggio2016} & & & & & & & & & & & \OKB & & \\ \cline{2-15}
& Spreitzer et al. \cite{Spreitzer2016} & & & & & & & & & & & & & \OKB \\ \cline{2-15}
& Tadrous et al. \cite{Tadrous2016} & & & & & & & & & \OKB & & & & \\ \cline{2-15}
& {Vanrykel et al. \cite{Vanrykel2016}} & & & & & {\OKB} & & & & & & & {\OKB} & \\ \cline{2-15}
& {Wang et al. \cite{Wang2016}} & & & {\OKB} & & & & & & & & & & \\ \hline
% 2017
\multirow{8}{*}{2017} & {Arora et al. \cite{Arora2017}} & & & {\OKB} & & & & & & & & & & \\ \cline{2-15}
& \revOne{Chen et al. \cite{Chen2017}} & & \revOne{\OKB} & & & & & & & & & & & \\ \cline{2-15}
& \revOne{Cheng et al. \cite{Cheng2017}} & & & & & \revOne{\OKB} & & & & & & & & \\ \cline{2-15}
& {Continella et al. \cite{Continella2017}} & & & & & {\OKB} & & & & & & & & \\ \cline{2-15}
& {Espada et al. \cite{Espada2017}} & & & & & & & & & {\OKB} & & & & \\ \cline{2-15}
& {Malik et al. \cite{Malik2017}} & & & & {\OKB} & & & & & & & & & \\ \cline{2-15}
& Taylor et al. \cite{Taylor2017} & & \OKB & & & & & & & & & & & \\ \cline{2-15}
& {Wei et al. \cite{Wei2017}} & & & & & & & & & {\OKB} & {\OKB} & & & \\ \hline
\end{tabular}
}
\end{table}

\subsection{Traffic Characterization}
\label{sec:traffic_characterization}

Network management can benefit from knowing the properties of the Internet traffic that traverses the network. 
Such information can be used to efficiently deploy the hardware equipments, as well as to setup them in order to provide the best Quality of Service (QoS) to the users. 
This statement particularly holds for networks serving mobile devices, since such devices generate traffic with peculiar properties. 
In light of the rapid evolution of mobile devices, the characterization of their Internet traffic is crucial to provide network administrators the information they need for resource planning, deployment, and management.

We define as \textit{traffic characterization} the analysis of the network traffic generated by mobile devices in order to infer its properties. 
In Table \ref{tab:traffic_characterization}, we group the works that study the mobile traffic according to the scope of the analysis:
\begin{itemize}
\item The works that study the network traffic of specific apps and/or mobile services. 
We survey {nine} works belonging to this category. 
Rao et al. in~\cite{Rao2011} study the Android and iOS native apps of two video streaming services, namely Netflix and YouTube.
YouTube is targeted also by Finamore et al. in~\cite{Finamore2011}. 
The Android apps of Facebook and Skype are considered in~\cite{Baghel2012}. 
In~\cite{Wei2012ProfileDroid}, Wei et al. focus on $27$ Android apps ($19$ free and $8$ paid).
\revOne{The analysis by Lindorfer et al.} in~\cite{Lindorfer2014} covers over $1{,}000{,}000$ unique Android apps. 
In~\cite{Chen2015}, Chen et al. analyze $5560$ malicious Android apps (from $177$ malware families). 
In~\cite{Nayam2016}, Nayam et al. study $63$ Android and $35$ iOS free apps, all belonging to the ``Health \& Fitness'' category. 
The work presented by Tadrous et al. in~\cite{Tadrous2016} focuses on five interactive apps for both Android and iOS. 
In~\cite{Espada2017}, Espada et al. present a framework for traffic characterization of Android apps, and choose Spotify as case study.
\item The works that study the network traffic generated by a population of mobile devices. 
We can further divide such works into two subsets:
\begin{itemize}
\item The works that compare mobile traffic with non-mobile one. 
We survey four works belonging to this subcategory. 
The work in~\cite{Maier2010} focuses on the network traffic of mobile devices when they are connected to home Wi-Fi networks, while the works in~\cite{Afanasyev2010,Gember2011,Lee2011} carry out the same analysis for campus Wi-Fi networks.
\item The works that only consider mobile traffic. 
We survey \revOne{five} works belonging to this subcategory. 
The works in~\cite{Chen2012,Wei2017} target campus Wi-Fi networks, while the works in~\cite{Falaki2010,Shepard2010,Fukuda2015} leverage client-side measurements collected through logging apps.
\end{itemize}
\end{itemize}

\begin{table}
\caption{The surveyed works that deal with traffic characterization.}
\label{tab:traffic_characterization}
\centering
\begin{tabular}{|>{\hspace{\tabred}}c<{\hspace{\tabred}}|
>{\hspace{\tabred}}l<{\hspace{\tabred}}|
>{\hspace{\tabred}}c<{\hspace{\tabred}}|
>{\hspace{\tabred}}c<{\hspace{\tabred}}|
>{\hspace{\tabred}}c<{\hspace{\tabred}}|}
\hline
\multirow{3}{*}{\textbf{Year}} & \multicolumn{1}{c|}{\multirow{3}{*}{\textbf{Paper}}} &  & \multicolumn{2}{c|}{\textbf{Mobile Devices}} \\ \cline{4-5}
& & \textbf{Apps/}& \textbf{Comparison } & \textbf{Only} \\ 
& & \textbf{Services} & \textbf{with non-mobile} & \textbf{Mobile} \\ \hline
% 2010
\multirow{4}{*}{2010} & Afanasyev et al. \cite{Afanasyev2010} & & \OKB & \\ \cline{2-5}
& Falaki et al. \cite{Falaki2010} & & & \OKB \\ \cline{2-5}
& Maier et al. \cite{Maier2010} & & \OKB & \\ \cline{2-5}
& Shepard et al. \cite{Shepard2010} & & & \OKB \\ \hline
% 2011
\multirow{4}{*}{2011} & Finamore et al. \cite{Finamore2011} & \OKB & & \\ \cline{2-5}
& Gember et al. \cite{Gember2011} & & \OKB & \\ \cline{2-5}
& Lee et al. \cite{Lee2011} & & \OKB & \\ \cline{2-5}
& {Rao et al. \cite{Rao2011}} & {\OKB} & & \\ \hline
% 2012
\multirow{3}{*}{2012} & Baghel et al. \cite{Baghel2012} & \OKB & & \\ \cline{2-5}
& Chen et al. \cite{Chen2012} & & & \OKB \\ \cline{2-5}
& {Wei et al. \cite{Wei2012ProfileDroid}} & {\OKB} & & \\ \hline
% 2014
2014 & Lindorfer et al. \cite{Lindorfer2014} & \OKB & & \\ \hline
% 2015
\multirow{2}{*}{2015} & Chen et al. \cite{Chen2015} & \OKB & & \\ \cline{2-5}
& Fukuda et al. \cite{Fukuda2015} & & & \OKB \\ \hline
% 2016
\multirow{2}{*}{2016} & {Nayam et al. \cite{Nayam2016}} & {\OKB} & & \\ \cline{2-5}
& Tadrous et al. \cite{Tadrous2016} & \OKB & & \\ \hline
% 2017
\multirow{2}{*}{2017} & {Espada et al. \cite{Espada2017}} & {\OKB} & & \\ \cline{2-5}
& {Wei et al. \cite{Wei2017}} & & & {\OKB} \\ \hline
\end{tabular}
\end{table}

%In Table \ref{tab:traffic_characterization}, we provide a view of the classification described above. 
%In what follows, we summarize the main properties of mobile traffic that stem from the works we survey:
\revOne{The aforementioned works provide interesting results and observations about mobile network traffic characteristics. 
In what follows, we highlight the main properties that emerged from the works we survey:}

\begin{itemize}
\item Compared to residential broadband traffic, the daily volume of traffic per mobile user is roughly one order of magnitude smaller~\cite{Falaki2010}.
\item Mobile devices generate \revOne{more downlink than uplink traffic, 
%showing a clear 
clearly following a client-server behavior}~\cite{Falaki2010,Wei2012ProfileDroid,Fukuda2015,Tadrous2016}.
\item At the network layer, IP flows of mobile devices have \revOne{a shorter duration, a much higher number of packets, and a much smaller packets,} compared to IP flows of non-mobile devices~\cite{Lee2011}.
\item Most of the transport-layer traffic is carried over TCP~\cite{Falaki2010,Gember2011,Chen2012}, and more \revOne{than half of the overall TCP traffic} is encrypted~\cite{Falaki2010}. 
Transfers within TCP connections are small in size~\cite{Falaki2010,Gember2011}, causing a high overhead for lower-layer protocols, particularly when transport-layer encryption is in place~\cite{Falaki2010}.
\item Most of the application-layer traffic is carried over HTTP or HTTPS~\cite{Falaki2010,Gember2011,Lee2011,Chen2012,Lindorfer2014,Chen2015,Wei2017}. 
Moreover, the analysis carried out by Chen et al. in~\cite{Chen2012} highlights that: (i) the adoption of HTTPS is increasing (a trend confirmed by Nayam et al. in~\cite{Nayam2016} and by Wei et al. in~\cite{Wei2017}); and (ii) Akamai and Google servers serve nearly $40\%$ of \revOne{the global mobile traffic.}
\item Mobile devices {contact} a less diverse set of hosts compared to non-mobile devices~\cite{Gember2011,Chen2012}.
\item Mobile devices experience a low loss rate on Wi-Fi networks~\cite{Chen2012}. 
Instead, \revOne{mobile traffic on cellular networks suffers high delays and losses,} as well as low throughputs~\cite{Falaki2010}.
\item An important fraction of mobile traffic is due to video streaming~\cite{Maier2010,Gember2011}, mainly on the YouTube platform~\cite{Finamore2011}.
\item Android apps typically do not encrypt their network traffic (i.e., they simply rely on HTTP), they connect to several different hosts, and \revOne{part of their network traffic is toward Google's services~\cite{Wei2012ProfileDroid}.}
\item A significant part of the network traffic generated by Android and iOS free apps is due to advertisement and tracking services~\cite{Nayam2016}.
\item \revOne{Netflix and YouTube apps for Android tend to} periodically buffer large portions of the video to be played, while their counterparts for iOS tend to initially buffer a large amount of data, then periodically buffer small portion of the video to keep playback ongoing (although the YouTube app employs large-block buffering under favorable network conditions)~\cite{Rao2011}. 
\revOne{Moreover, Netflix and Youtube apps for iOS create a large number of TCP flows to provide a single video to cope with TCP timeouts caused by the delays of cellular networks. 
This causes an overhead} that is not necessary when mobile devices are connected to Wi-Fi networks~\cite{Rao2011,Chen2012}.
\end{itemize}

\subsection{App Identification}
\label{sec:app_identification}

The Internet connectivity and multi-purpose apps are two key aspects of the success and widespread adoption of mobile devices. 
Most of the apps can send and receive data through the network interfaces of mobile devices (i.e., Wi-Fi and cellular), and often this capability is mandatory for apps to work properly. 

The network traffic patterns related to an app (or type of app) constitute a behavioral network fingerprint which can be recognized in unseen network traces. 
We refer to \revOne{this analysis} as \textit{app identification}. 
It is worth to notice that this approach also takes into consideration the network traffic generated by an app that is not directly related to any user actions (e.g., the data exchanged because of background activities).

App identification brings several benefits to network management, but it also has privacy implications:
\begin{itemize}
\item The knowledge of the apps used by the clients of a network can help the administrators to tune the network equipments and parameters in order to deliver the best achievable Quality of Service (QoS).
\item In an enterprise network where some particular apps is not allowed to be used (e.g., Facebook, Twitter), app detection can help the administrators enforce such policy by blocking traffic belonging to the forbidden apps.
\item It is possible to target a high profile user and discover whether she uses privacy-sensitive (e.g., health, dating) apps. 
\item \revOne{Knowing the set of apps installed on a mobile device can reveal sensitive information about the user such as relationship status, spoken languages, country, and religion~\cite{Seneviratne2014}.}
\end{itemize}

In Table \ref{tab:app_identification_platforms}, we report the surveyed works that deal with app identification \cite{Lee2011,Qazi2013,Rao2013,Le2015,Wang2015,Yao2015,Alan2016,Mongkolluksamee2016,Chen2017,Taylor2017}. 
The number of apps selected for profiling and fingerprinting varies from {less than ten} to many thousands.

\begin{table}
\caption{Targeted mobile platforms and number of considered apps in the surveyed works that deal with app identification.}
\label{tab:app_identification_platforms}
\centering
\scalebox{0.95}{
\begin{tabular}{|c|l|c|c|c|}
\hline
\multicolumn{1}{|c|}{\multirow{2}{*}{\textbf{Year}}} & \multicolumn{1}{c|}{\multirow{2}{*}{\textbf{Paper}}} & \multicolumn{3}{c|}{\textbf{Number of Targeted Apps}} \\ \cline{3-5}
& & \textbf{Android} & \textbf{iOS} & \textbf{Symbian} \\ \hline
% 2011
2011 & Lee et al. \cite{Lee2011} & $50$ & $50$ & None \\ \hline
% 2013
\multirow{2}{*}{2013} & Qazi et al. \cite{Qazi2013} & $40$ & None & None \\ \cline{2-5}
& {Rao et al. \cite{Rao2013}} & {$832$} & {$209$} & {None} \\ \hline
% 2015
\multirow{3}{*}{2015} & Le et al. \cite{Le2015} & $70$ & None & None \\ \cline{2-5}
& Wang et al. \cite{Wang2015} & None & $13$ & None \\ \cline{2-5}
& Yao et al. \cite{Yao2015} & $651{,}000$ & $68{,}000$ & $10{,}000$ \\ \hline
% 2016
\multirow{2}{*}{2016} & Alan et al. \cite{Alan2016} & $1{,}595$ & None & None \\ \cline{2-5}
& {Mongkolluksamee et al. \cite{Mongkolluksamee2016}} & {$5$} & {None} & {None} \\ \hline
% 2017
\multirow{2}{*}{2017} & \revOne{Chen et al. \cite{Chen2017}} & \revOne{$5{,}000$} & \revOne{None} & \revOne{None} \\ \cline{2-5}
& Taylor et al. \cite{Taylor2017} & $110$ & None & None \\ \hline
\end{tabular}
}
\end{table}

Despite the core topic of the work by Lee et al. in~\cite{Lee2011} is a comparison between smartphone traffic and traditional Internet traffic, the authors also perform app identification targeting \revOne{the} Android and iOS platforms. 
Indeed, they select the top $50$ apps of both Apple App Store and Google Play Store, generate their payload signatures, and use such signatures to recognize the traffic generated by such apps in network traces.
%The classifier matches only $15.37\%$ of flows with one of the considered apps, testifying that smartphone users install a variety of apps on their devices. 
Unfortunately, this approach is based on payload signatures, thus it cannot deal with the apps that encrypt their network traffic.

Qazi et al. in~\cite{Qazi2013} present the Atlas framework, which incorporates application identification into Software-Defined Networking (SDN). 
\revOne{Prototyped on HP Labs wireless network, the identification performance of Atlas is tested on the top $40$ popular Android apps from Google Play Store.}
%and collect over $100{,}000$ network flows which are used to train a machine learning classifier that reaches $94\%$ F-measure on average. 
Since it requires to inspect transport-layer information, Atlas cannot process network traffic protected by IPsec.

{Rao et al. in~\cite{Rao2013} present Meddle, a cross-platform system for collecting and analyzing the network traffic of mobile devices. 
The idea is to leverage VPN tunnels (which are natively supported by modern mobile \revOne{OSes}) to redirect the network traffic of the target mobile devices to the Meddle proxy server, where software middleboxes are responsible for traffic processing and analysis. 
Thanks to this man-in-the-middle approach, Meddle can inspect the network traffic protected by SSL/TLS, but it cannot work with data transmissions encrypted via IPsec. 
The authors employ Meddle for app identification (and also PII leakage detection, see Section \ref{sec:PII_leakage_detection} for details) \revOne{based on fields of HTTP messages (i.e., Host and User-Agent).}
%On a dataset for which ground truth is available, the identification rate \revOne{(i.e., the number of identified apps over the total number of considered apps)} is $64\%$ for Android apps, and $89\%$ for iOS ones. 
%On a real-world dataset with no ground truth and no SSL decryption (because of privacy reasons), over $92\%$ of the flows, as well as 40\% of the SSL ones, are mapped to the considered Android and iOS apps.}

Le et al. in~\cite{Le2015} propose AntMonitor, a system for collecting and analyzing network traffic from Android devices. 
Among other types of analysis, AntMonitor can perform app identification. 
\revOne{The authors select $70$ Android apps to evaluate the performance of their solution.}
%and leverage a machine learning classifier which takes into account $84$ network-level features. 
%The system achieves $70.1\%$ F-measure. 
Among the considered features, there are the flags of TCP segments, which are hidden if IPsec is employed, thus the proposed framework does not work if network-layer encryption is in place.

The app identification framework proposed by Wang et al. in~\cite{Wang2015} is based on extracting side-channel information from Wi-Fi traffic belonging to a target mobile device. 
The authors depict a passive adversary as follows: 
(i) she is able to sniff the traffic on the same WLAN channel as the access point to which the target device is connected; 
(ii) leveraging the MAC address of the target device, she can elicit its traffic from the collected network traces; and 
(iii) she cannot break the encryption scheme of the sniffed traffic (i.e., the app identification can also target secure WLANs). 
To evaluate their solution, the authors choose the iOS platform, considering thirteen popular apps from a wide range of different app categories. 
%The reached accuracy is more than $90\%$.

Yao et al. in~\cite{Yao2015} present SAMPLES (Self Adaptive Mining of Persistent LExical Snippets), an app identification framework that leverages the occurrences of app identifiers within HTTP headers (thus it cannot handle network traffic protected by IPsec or SSL/TLS). 
SAMPLES models such occurrences into generalized conjunctive rules, {which are used} to identify the app that generated a given network flow.
%SAMPLES models such occurrences into generalized conjunctive rules, and uses the resulting set of such rules to identify the app that generated a given network flow.
To evaluate their system, the authors consider over $700{,}000$ apps from Google Play Store, Apple App Store, and Nokia OVI Store (details are given in Table \ref{tab:app_identification_platforms}). 
%The proposed solution identifies over $90\%$ of these apps with an average accuracy of $99\%$. 
%Surprisingly, these results are obtained by training SAMPLES with less than $2\%$ of the apps for each of the three considered marketplaces. 
%Moreover, SAMPLES outperforms the solutions proposed by Tongaonkar et al. in~\cite{Tongaonkar2013} and Xu et al. in~\cite{Xu2011}.

Alan and Kaur in~\cite{Alan2016} investigate the feasibility of identifying Android apps from their launch-time network traffic by only leveraging the information available in TCP/IP headers. 
The authors collect the launch-time traffic of $1{,}595$ apps.
%and use it to train and evaluate three machine learning classifiers, achieving a maximum accuracy of $87\%$ (more details on the classifiers are given in Section~\ref{sec:machine_learning_app_identification}). 
\revOne{This work proposes three different methods that can handle network traffic encrypted via SSL/TLS, but only two of them can also deal with IPsec encryption.}

{Mongkolluksamee et al. in~\cite{Mongkolluksamee2016} (which is an extended version of a previous work by the same authors~\cite{Mongkolluksamee2015}) \revOne{apply} machine learning to build an app identification system for Android apps. 
The authors leverage graphlet- and histogram-based features, and employ a random forest classifier (more details in Section~\ref{sec:machine_learning_app_identification}). 
%The proposed system achieves $0.96$ F-measure but it is evaluated on five Android apps only. 
However, \revOne{this analysis requires} to access TCP and UDP headers, which is infeasible for apps that employ IPsec to encrypt their network traffic. 
It is worth to notice that despite this work focuses on 3G traffic, 
%we survey this work despite the analysis carried out by the authors considers 3G traffic only. 
%The reason is that 
the actual capturing of the network traffic is performed within a mobile device via tcpdump (more details in Section~\ref{sec:mobile_devices}).}

\revOne{
Chen et al. in~\cite{Chen2017} present an innovative method to identify the invariant tokens (e.g., URLs, key-value pairs, developer IDs) that are present in the network traffic of a specific app; such tokens can then be exploited to carry out app identification via deep packet inspection (which is applicable to unencrypted traffic only). 
The described framework requires to perform an advanced static analysis on the targeted app in order to find the parts of code that trigger network activities; compared to dynamic analysis and UI fuzzing, this method permits to cover almost all ($98.54\%$) of the app's network activities in a very short period of time (less than twenty seconds). 
The authors focus on the Android platform, and evaluate their solution on $2{,}500$ apps from Google Play Store and other third-party Android marketplaces, and $2{,}500$ malicious apps from VirusTotal.
}

Taylor et al. in~\cite{Taylor2017} (which is an extension of a previous work by the same authors~\cite{Taylor2016}) propose AppScanner, an app identification system based on machine learning (more details in Section~\ref{sec:machine_learning_app_identification}). 
The authors profile $110$ popular Android apps crawled from Google Play Store \revOne{and re-identify them in real-time.} % with up to $96\%$ accuracy. 
Moreover, the authors study how the classification performance is affected by varying the duration of the network traffic capturing, the mobile device that generates the collected data, and the version of the fingerprinted apps. 
AppScanner leverages the information within IP and TCP headers, thus being able to process the network traffic encrypted via SSL/TLS, but not the one protected by IPsec.

\subsection{Usage Study}
\label{sec:usage_study}

The habits of mobile users have significantly changed with the evolution of cellphones toward smartphones and tablets. 
First of all, the adoption of the touchscreen display has revolutionized the human-device interaction. 
Moreover, the development of mobile operating systems supporting multitasking and third-party apps has enhanced the capabilities of mobile devices well beyond the requirements for communication activities. 
In this scenario, many researchers investigate how mobile users interact with their mobile devices. This to improve the usability of mobile OSes and apps, as well as to properly set up networks serving mobile devices. 
For instance, the knowledge of places where mobile devices are mostly used can drive the deployment of free Wi-Fi hotspots to reduce the traffic load on cellular networks.

We define as \textit{usage study} the analysis of the network traffic of mobile devices that aims at inferring the usage habits of mobile users. 
The works we review in this section leverage network-side measurements~\cite{Afanasyev2010,Maier2010,Finamore2011,Gember2011,Lee2011,Wei2017}, as well as data collected within mobile devices~\cite{Falaki2010,Ham2012,Fukuda2015,Soikkeli2015}. 
In Table~\ref{tab:usage_study}, we show the three analysis perspectives adopted by the surveyed works that investigate the usage habits of mobile users:
\begin{itemize}
%on time the users are active during the day, duration of activity, amount of traffic generated, most frequently used network interfaces
\item \textit{The network}. As an example, observing the time in which users are active during the day (i.e., sending and receiving data), duration of activity, amount of traffic generated, and most frequently used network interfaces (i.e., Wi-Fi or cellular).
\item \textit{The apps and/or mobile services}. For instance, analyzing which are the most frequently used apps/services, and which is the traffic volume of a specific app.
\item \textit{The geographical positions and mobility patterns}. As an example, \revOne{studying the locations} where mobile devices are most frequently used, and where they generate most of their traffic.
\end{itemize}

\begin{table}
\caption{The surveyed works that deal with usage study.}
\label{tab:usage_study}
\centering
\begin{tabular}{|>{\hspace{\tabred}}c<{\hspace{\tabred}}|
>{\hspace{\tabred}}l<{\hspace{\tabred}}|
>{\hspace{\tabred}}c<{\hspace{\tabred}}|
>{\hspace{\tabred}}c<{\hspace{\tabred}}|
>{\hspace{\tabred}}c<{\hspace{\tabred}}|}
\hline
& & & \textbf{Apps/} & \textbf{Geography/} \\ 
\textbf{Year} & \multicolumn{1}{c|}{\textbf{Paper}} & \textbf{Network} & \textbf{Services} & \textbf{Mobility} \\ \hline

% 2010
\multirow{3}{*}{2010} & Afanasyev et al. \cite{Afanasyev2010} & \OKB & \OKB & \OKB \\ \cline{2-5}
& Falaki et al. \cite{Falaki2010} & & \OKB & \\ \cline{2-5}
& Maier et al. \cite{Maier2010} & & \OKB & \\ \hline
% 2011
\multirow{3}{*}{2011} & Finamore et al. \cite{Finamore2011} & & \OKB & \\ \cline{2-5}
& Gember et al. \cite{Gember2011} & & \OKB & \\ \cline{2-5}
& Lee et al. \cite{Lee2011} & \OKB & \OKB & \\ \hline
% 2012
2012 & Ham et al. \cite{Ham2012} & \OKB & \OKB & \\ \hline
% 2015
\multirow{2}{*}{2015} & Fukuda et al. \cite{Fukuda2015} & \OKB & & \\ \cline{2-5}
& Soikkeli et al. \cite{Soikkeli2015} & \OKB & & \OKB \\ \hline
% 2017
{2017} & {Wei et al. \cite{Wei2017}} & {\OKB} & & \\ \hline
\end{tabular}
\end{table}

%In Table~\ref{tab:usage_study}, we show the types of usage study carried out in the surveyed works that deal with this kind of traffic analysis.
In what follows, we summarize their findings reported by the works in usage study:
\begin{itemize}
\item The most frequently used apps are the ones related to multimedia content (e.g., YouTube, Spotify) and web browsers~\cite{Falaki2010,Maier2010,Gember2011,Lee2011,Ham2012,Fukuda2015}. 
\revOne{Nonetheless, social} network and instant messaging apps are also popular~\cite{Fukuda2015}.
\item The predominance of cellular over Wi-Fi network traffic observed for mobile devices by Ham et al. in~\cite{Ham2012} in 2012 is gradually disappearing. 
As reported in~\cite{Fukuda2015}, in 2015 more than half of mobile traffic is carried over Wi-Fi. 
In particular, mobile users prefer switching to Wi-Fi connectivity whenever a Wi-Fi access point is available~\cite{Fukuda2015}.
\item The usage of mobile devices is low at nighttime and high in daytime~\cite{Afanasyev2010,Lee2011,Ham2012,Soikkeli2015}.
\item Cellular traffic peaks \revOne{during commute times}, while Wi-Fi traffic peaks in the evening~\cite{Fukuda2015}. 
Cellular traffic is lighter on weekends than weekdays, while Wi-Fi traffic follows the opposite trend~\cite{Fukuda2015}.
\item Mobile users tend to generate more network traffic when they are out of home, and when their devices have high battery level~\cite{Soikkeli2015}.
\item The volume of traffic generated by the mobile users of a Wi-Fi network varies greatly, from less than $100$ MB to several GBs, according to users' habits and needs~\cite{Wei2017}.
\item According to Finamore et al. in~\cite{Finamore2011}, YouTube mobile users: (i) similarly to non-mobile users, they prefer short videos ($40\%$ of the watched videos are shorter than three minutes, and only $5\%$ are longer than ten minutes); 
(ii) similarly to non-mobile users, they rarely change video resolution and, whenever they do that, it is to switch to a higher resolution (although full screen mode is not frequently used); 
and (iii) more frequently than non-mobile users, they early stop watching the video (within the first fifth of its duration for $60\%$ of the videos).
\end{itemize}

\subsection{PII Leakage Detection}
\label{sec:PII_leakage_detection}

As we introduced in Section~\ref{sec:introduction}, a mobile device is a source of sensitive information about its owner (e.g., phone number, contacts, photos, videos, GPS position). 
\revOne{In addition to that}, apps often require to access such information to deliver their services. 
As an example, an instant messaging app (e.g., WhatsApp, Telegram, WeChat) requires to access the contacts saved in the device's address book. 
As another example, a social network app (e.g., Facebook, Instagram) requires to inspect the device's memory to find photos.

To disclose sensitive information to a remote host, an app must be authorized to: 
(i) access some kind of sensitive information (e.g., the GPS position); 
and (ii) connect to the Internet. 
\revOne{The disclosure of such information can be either allowed or illicit, depending on three factors: 
(i) the level of sensitivity of the disclosed information; 
(ii) the reason why the app transmits such information to a remote host;
and (ii) whether the user is aware of this transmission of sensitive data.}

In this section, we focus on \textit{Personal Identifiable Information} (\textit{PII}), which is information that can be used to identify, locate, or contact an individual. 
In the domain of mobile devices, \revOne{we can identify four types of PII:}
\begin{itemize}
\item Information related to mobile devices, such as IMEI (International Mobile Equipment Identity, a unique identifier associated to each mobile device), Android Device ID (an identifier randomly generated on the first boot of any Android device), and MAC address (a unique identifier assigned to each network interface).
\item Information related to SIM cards, such as IMSI (International Mobile Subscriber Identity, a unique identifier assigned to each subscriber of a cellular service), and SIM Serial ID (the identifier assigned to each SIM card).
\item Information related to users, such as name, gender, date of birth, address, phone number, and email.
\item Information about user's location, such as GPS position and ZIP code.
\end{itemize}

We define as \textit{PII leakage detection} the analysis of the network traffic of a mobile device in order to detect the leakage of a user's PII. 
Once a PII leakage is detected, it is possible to apply suitable countermeasures, such as blocking the network flows carrying the PII, or substituting the sensitive information with bogus data. 
The latter approach is a good solution for mobile users who want to protect their privacy while being able to enjoy the functionalities of the apps.

In Table \ref{tab:PII_leakage_detection}, we present the surveyed works that deal with PII leakage detection~\cite{Stevens2012,Kuzuno2013,Rao2013,Le2015,Song2015,Ren2016,Vanrykel2016,Cheng2017,Continella2017}. 
For each work, we summarize the targeted mobile platforms, and whether the PII leaks are simply detected or also prevented.

\begin{table*}
\caption{Works that deal with PII leakage detection.}
\label{tab:PII_leakage_detection}
\centering
\begin{tabular}{|c|l|c|c|c|c|c|}
\hline
\multicolumn{1}{|c|}{\multirow{2}{*}{\textbf{Year}}} & \multicolumn{1}{c|}{\multirow{2}{*}{\textbf{Paper}}} & \multicolumn{3}{c|}{\textbf{Targeted Mobile Platform}} & \multicolumn{2}{c|}{\textbf{Action on PII {Leaks}}} \\ \cline{3-7}
& & \textbf{Android} & \textbf{iOS} & \textbf{Windows Phone} & \textbf{Detection} & \textbf{Prevention} \\ \hline
% 2012
{2012} & {Stevens et al. \cite{Stevens2012}} & {\OKB} & & & {\OKB} & \\ \hline
% 2013
\multirow{2}{*}{2013} & Kuzuno et al. \cite{Kuzuno2013} & \OKB & & & \OKB & \\ \cline{2-7}
& {Rao et al. \cite{Rao2013}} & {\OKB} & {\OKB} & & {\OKB} & {\OKB} \\ \hline
% 2015
\multirow{2}{*}{2015} & Le et al. \cite{Le2015} & \OKB & & & \OKB & \\ \cline{2-7}
& Song et al. \cite{Song2015} & \OKB & & & \OKB & \OKB \\ \hline
% 2016
\multirow{2}{*}{2016} & Ren et al. \cite{Ren2016} & \OKB & \OKB & \OKB & \OKB & \OKB \\ \cline{2-7}
& {Vanrykel et al. \cite{Vanrykel2016}} & {\OKB} & & & {\OKB} & \\ \hline
% 2017
\multirow{2}{*}{2017} & \revOne{Cheng et al. \cite{Cheng2017}} & \revOne{\OKB} & & & \revOne{\OKB} & \\ \cline{2-7}
& {Continella et al. \cite{Continella2017}} & {\OKB} & & & {\OKB} & \\ \hline
\end{tabular}
\end{table*}

Stevens at al. in~\cite{Stevens2012} present a comprehensive study on thirteen popular ad providers for Android. 
In particular, part of this study focuses on the analysis of ad traffic in order to detect the transmission of the user's PII. 
The authors observe that at the time of writing only one of the considered ad providers leverage encryption to protect its network traffic. 
For this reason, they choose to perform a deep packet inspection to identify the leakage of the user's private information. 
The results show that several types of PII (e.g., age, gender, GPS position) are leaked in clear by ad libraries. 
Moreover, the authors highlight that although none of the considered ad providers is able \revOne{to build a complete user profile, 
the UDIDs in ad-related} traffic can be exploited by an external adversary to correlate sensitive information from different ad providers and build a complete user profile.

Kuzuno and Tonami in~\cite{Kuzuno2013} investigate the leakage of sensitive information by the advertisement libraries embedded into free Android apps. 
They focus on both original and hashed identifiers unique to mobile devices (i.e., IMEI and Android ID) and SIM cards (i.e., IMSI and SIM Serial ID), as well as on the name of the cellular operator (CARRIER). 
\revOne{To carry out their analysis,} the authors develop two components: 
(i) a server application; 
and (ii) a mobile app that can be installed on an Android device. 
The server application takes as input the network traffic of a set of apps that leak sensitive information, and applies a clustering method (see Section~\ref{sec:machine_learning_PII_leakage_detection} for details) to generate traffic signatures. 
The mobile app leverages such signatures to identify the sensitive information leaked by the other apps installed on the device. 
To evaluate their solution, the authors employ the network traffic of $1{,}188$ free Android apps and achieve the following results: 
$94\%$ of HTTP messages containing sensitive information are correctly detected, 
with $5\%$ false negatives (i.e., undetected HTTP messages carrying sensitive information), and less than $3\%$ false positives (i.e., HTTP messages without sensitive information, incorrectly identified as sensitive). 
Since the signature generation phase requires to inspect HTTP messages looking for sensitive information, the system cannot work on \revOne{encrypted traffic (i.e., neither SSL/TLS nor IPsec).}

Rao et al. in~\cite{Rao2013} and Ren et al. in~\cite{Ren2016} present ReCon, a cross-platform system that allows mobile users to control the PII leaked in the network traffic of their devices. 
ReCon is based on Meddle (we described it in Section \ref{sec:app_identification}), therefore it can inspect mobile traffic even if it is encrypted at transport layer, but cannot cope with data transmissions protected by IPsec. 
Moreover, ReCon offers a web interface through which the user can visualize in real time which PII is leaked, and optionally modify such PII or block the connection carrying it. 
In \cite{Rao2013}, Rao et al. target Android and iOS OSes, and the PII leakage detection mechanism is based on a domain blacklist. 
In \cite{Ren2016}, Ren et al. also include Windows Phone among the considered OSes, and PII leaks are detected using properly trained machine learning classifiers (more details in Section~\ref{sec:machine_learning_PII_leakage_detection}). 
The works in \cite{Rao2013,Ren2016} expose an extensive leakage of sensitive information belonging to all the types of PII we listed above, as well as the transmission of usernames and passwords in both plain-text (HTTP) and encrypted (HTTPS) traffic.

Le et al. in~\cite{Le2015} present AntMonitor, a system for collecting and analyzing network traffic from Android devices (we already mentioned it in Section~\ref{sec:app_identification}). 
Among other types of analysis, AntMonitor can perform PII leakage detection. 
The authors capture the network traffic of nine Android users for a period of five weeks, then inspect the collected dataset searching the following PII: IMEI, Android Device ID, phone number, email address, and device location. 
Overall, $44\%$ and $66\%$ of the analyzed apps leak IMEI and Android Device ID, respectively, while PII related to the user is rarely disclosed to remote hosts. 
It is worth to notice that the proposed analysis requires to inspect application-layer data, which is infeasible in case of traffic encryption, \revOne{neither at network (IPsec) nor transport layer (SSL/TLS).}

Song and Hengartner in~\cite{Song2015} develop PrivacyGuard, an open-source Android app that leverages the \texttt{VPNService} class of the Android API for eavesdropping the network traffic of the apps installed on the device. 
The authors employ PrivacyGuard to investigate the leakage of PII related to mobile users (e.g., phone number) and devices (e.g., IMEI) by Android apps. 
In an evaluation conducted using $53$ Android apps, PrivacyGuard detects more PII leaks than TaintDroid \cite{Enck2010}. The proposed app can optionally replace the leaked information with bogus data. 
Moreover, it can inspect transmission protected by SSL/TLS (through a man-in-the-middle approach), but cannot deal with traffic encrypted via IPsec.

Vanrykel et al. in~\cite{Vanrykel2016} investigate the leakage of sensitive identifiers in the unencrypted network traffic of Android apps. 
The authors develop a framework that automatically executes apps, collects their network traffic, inspects the HTTP data, and detects the identifiers that are transmitted in clear. 
The analysis of $1{,}260$ Android apps (from $42$ app categories) shows that: 
(i) the Android ID and Google Advertising ID are the most frequently leaked identifiers, while the SIM serial number, the IMSI, the device serial number, and the email of the registered Google account are less common in apps' network traffic; 
(ii) there is an extensive leakage of app-specific identifiers; and 
(iii) certain apps leak the user's phone number, email address, or position.

Cheng et al. in~\cite{Cheng2017} present a framework for the detection of PII leaks by Android apps. 
Such framework leverages the information available in IP and TCP headers; for this reason, it works even if the apps employ SSL/TLS to encrypt their network transmissions, and it can be blocked only using IPsec. 
Overall, the idea is to model the network traffic related to a user-app interaction into a sequence of packet sizes, then convert such sequences into feature vectors to be used for training and evaluating a machine learning classifier. 
The authors consider seven Android apps (i.e., BaidoYun, Evernote, QQ, QQMail, TouTiao, WeChat, and Weibo), plus a self-developed Android malware, called Moledroid, that implements several techniques employed by malicious apps to leak PII.

Continella et al. in~\cite{Continella2017} develop Agrigento, an open-source framework for the analysis of Android apps in order to detect PII leakage. 
Agrigento is based on differential analysis, and its workflow consists of two phases. 
In the first phase, the app under scrutiny is executed several times on a physical device to collect: 
(i) its network traffic; 
and (ii) additional system- and app-level information that is contextual to the execution (e.g., randomly-generated identifiers, timestamps). 
Subsequently, the collected information is aggregated to model the network behavior of the app. 
In the second phase, a specific PII within the operating system of the mobile device is set to a different value. 
The app is then executed once again to collect its network traffic and the contextual information. 
Finally, a PII leak is reported if the collected data does not conform to the model learned before. 
Evaluated on $1{,}004$ Android apps, Agrigento detects more privacy leaks than currently available state-of-the-art solutions (e.g., ReCon~\cite{Ren2016}), while limiting the number of false positives. 
The proposed framework requires to inspect HTTP messages and leverages a man-in-the-middle approach to deal with HTTPS traffic. 
However, Agrigento does not work on network traffic encrypted via IPsec.

\subsection{Malware Detection}
\label{sec:malware_detection}

As happened for personal computers, the success and widespread adoption of mobile devices have attracted the interest of malware developers. 
Mobile devices, and particularly smartphones, are an ideal target for attackers since: (i) they are ubiquitous, i.e., the population of potential targets is large; 
(ii) they host sensitive information about their owners (e.g., identity, contacts, GPS position); 
and (iii) they have networking capabilities and they are usually connected to the Internet.

We define as \textit{malware detection} the attempt to understand \revOne{whether} a mobile app is malicious through the analysis of the network traffic it generates. 
In this section, we present the state of the art techniques for such kind of traffic analysis. 
We point out that we do not report the works that study the properties of the network traffic generated by malicious apps, because in such case the analysis is more related to traffic characterization (see Section~\ref{sec:traffic_characterization}). 
From the surveyed works, we elicit three kinds of actor \revOne{that actually perform malware detection task:} %can be interested in malware detection: 
(i) an app marketplace~\cite{Su2012}; 
(ii) a security company~\cite{Wei2012,Zaman2015,Narudin2016,Wang2016,Arora2017}; 
or (iii) a mobile user~\cite{Shabtai2012,Shabtai2014}.

Shabtai et al. in~\cite{Shabtai2012} present an anomaly-based malware detection app for Android devices. 
%The detection model is based on machine learning. 
The proposed app monitors several aspects of the device (e.g, memory, network, power) and extracts different features, some of which are related to network traffic (e.g., the number of received packets). 
A properly trained machine-learning-based classifier is then employed to check whether an installed app is malicious. 
The proposed solution is evaluated using $40$ benign and $4$ malicious Android apps. 
The authors consider different classifiers (e.g., decision tree, Bayesian networks), as well as different metrics for feature selection (e.g., Fisher score, information gain). 
Moreover, the authors investigate how the detection accuracy is affected when: 
(i) the testing apps are not used in the training phase; 
and (ii) training and testing are performed on different devices. 
%Overall, the proposed app achieves high accuracy, despite causing a limited performance overhead on the device.

Su et al. in~\cite{Su2012} propose a framework that allows an Android marketplace to detect whether an \revOne{app submitted by a mobile developer is malicious or benign.} % is malicious. 
The system consists of two components: (i) servers, where developers can upload their new apps for verification; 
and (ii) physical Android devices, where apps are actually executed while monitoring their system calls and network traffic. 
\revOne{The gathered information is sent to a central server, which classifies each app as safe or malicious according to the response of two classifiers.
In particular, a classifier bases its decision on system call statistics, while the other considers network traffic features.}
The network traffic classifier is trained with data from $49$ malicious apps (from $22$ malware families) and $60$ benign apps, and tested with data from $50$ malicious apps (from $22$ malware families) and $70$ benign apps (from eleven app categories). 
%The best performing implementation of the classifier (i.e., random forest) reaches $96.7\%$ accuracy. 
It is worth to notice that such classifier cannot process network traffic encrypted via IPsec, since one of the features it leverages is the average TCP session duration, which is not computable without accessing TCP headers.

Wei et al. in~\cite{Wei2012} present a framework for Android malware detection. 
Using network traffic generated by malicious Android apps, \revOne{the system learns the network behavior} of Android malware with regard to the resolution of domain names. 
Then, the system is employed to automatically analyze the DNS traffic produced by a given app and state whether that app is safe or malicious. 
The authors evaluate their solution using malicious apps from a public dataset of Android malware and benign apps from the official Android marketplace. 
%The proposed classifier reaches nearly $1.0$ accuracy, precision, and recall. 
A weakness of this framework is that it requires the access to the DNS traffic of apps, which can be hidden by IPsec or SSL/TLS encryption.

Shabtai et al. in~\cite{Shabtai2014} design an anomaly-based malware detection app for Android devices. 
Such app analyzes the network behavior of the apps installed on the device in order to identify self-updating malware (i.e., benign apps that after being installed on the device, they download a malicious payload from the Internet) and popular apps republished with additional malicious code. 
The idea is to model the normal network behavior of each installed app as a set of traffic patterns, and subsequently detect any deviation from those patterns. 
The system is evaluated on several benign apps, ten self-updating malicious apps developed by the authors, and the infected version of five of the chosen benign apps. 
%The authors claim that their method achieves a \revOne{recall} over $83\%$, and a \revOne{false-alarm rate (i.e., the number of benign apps classified as malware over the total number of benign apps)} below $12\%$. 
\revOne{The system works} even with apps that encrypt their network traffic, since it needs to know only their amount of transmitted/received bytes and its percent out of the total device traffic.

The work by Zaman et al.~\cite{Zaman2015} stems from the observation that malicious apps usually send the user's sensitive information \revOne{to accomplice remote hosts.} 
The idea is to log all communications with remote hosts for each app installed on the mobile device. 
Leveraging a list of known malicious domains, it is possible to label the apps that contacted them as malware. 
This approach requires to inspect the URLs within HTTP messages, therefore \revOne{it does not work on} encrypted network traffic. 
\revOne{The authors evaluate their solution on \textit{DroidKungFu} and \textit{AnserverBot} samples (i.e., two Android malware) being able to detect only the former one.}

Narudin et al. in~\cite{Narudin2016} investigate whether an anomaly-based IDS \revOne{(Intrusion Detection System)} can successfully detect malicious Android apps by relying on traffic analysis. 
To build a comprehensive dataset of network traces, the authors run benign apps on a physical Android device and malicious apps on dynamic analysis platforms available online. 
The collected network traffic is then sent to a central server, where several machine learning classifiers (e.g., random forest, multi-layer perceptron) are trained and evaluated. 
%All such classifiers reach over $90\%$ \revOne{recall} on the network traces generated by the benign apps and several malicious apps provided by the Android Malware Genome Project, and over $73\%$ on the network traces generated by the benign apps and $30$ newly (in 2013) appeared malicious apps. 
\revOne{Unfortunately, such classifiers} cannot process encrypted traffic, since they need to inspect HTTP messages.

Wang et al. in \cite{Wang2016} present TrafficAV, an Android malware detection system based on machine learning. 
The proposed framework offers two distinct detection models, which rely on TCP- and HTTP-related network features, respectively. 
{We give more details about considered features and classifiers in Section \ref{sec:machine_learning_malware_detection}.
\revOne{The authors evaluate their models on} the network traffic of $8{,}312$ benign apps and $5{,}560$ malware samples.
\revOne{While the HTTP feature-based model cannot work on traffic encrypted with SSL/TLS since it requires to perform Deep Packet Inspection (DPI), 
both the proposed models cannot cope with apps that employ IPsec.} 
%, TrafficAV achieves a \revOne{recall} of $98.16\%$ for the TCP-based model, and $99.65\%$ for the HTTP-based model.

Arora and Peddoju in~\cite{Arora2017} (which is an extension and refinement of a previous work by the same authors~\cite{Arora2014}) {also} apply machine learning to detect Android malware. 
They collect the network traffic of malware samples from eleven families, extract $22$ network-layer features (e.g., average time interval between received packets, per-flow sent bytes), and train a naive Bayes classifier. 
\revOne{They evaluate their proposal on the network traffic} of malware samples from six families (different from the ones used for training).
%, the classifier reaches a \revOne{F-measure} of $87.25\%$. 
Moreover, the authors present a feature selection algorithm that reduces the number of features to be used, while limiting the drop in detection accuracy. 
%The initial set of features is reduced from $22$ to $9$ elements, and the classifier achieves a \revOne{F-measure} of $83.3\%$. 
The proposed framework is encryption-agnostic, \revOne{although the same authors admit that} encryption may be a possible solution to evade detection. 
%\rikiTODO{clarify the previous sentence}
We provide more details about the features and the algorithm to select them in Section~\ref{sec:machine_learning_malware_detection}.

\subsection{User Action Identification}
\label{sec:user_action_identification}

As we stated in Section \ref{sec:app_identification}, most of the apps can leverage the Wi-Fi and cellular network interfaces of mobile devices to send and receive data. 
Since users perform several actions while interacting with apps, it is likely that most of such actions generate data transmissions. 
The network traffic trace of a given action typically follows a pattern that depends on the nature of the user-app interaction of that action. 
As a practical example, browsing a user's profile on Facebook will likely produce a different traffic pattern compared to posting a message on Twitter. 
These patterns can be used to recognize specific user actions related to a particular app of interest in generic network traces. 
Moreover, it is often possible to infer specific information {about} 
% (Alberto) Don't use "from" (it changes the meaning of the sentence).
a given user action (e.g., the length of the message sent via an instant messaging app). 
\revOne{We refer to these types of traffic analysis as \textit{user action identification}.}

The possibility to identify actions of mobile users can be useful in several scenarios:
\begin{itemize}
\item It is possible to profile the habits of a mobile user (e.g., checking emails in the morning, watching YouTube videos in the evening). The user's behavioral profile can be used to later recognize the presence of that user in a network. Moreover, profiles of thousands of mobile users can be aggregated in order to infer some information for marketing or intelligence purposes.
\item It is possible to perform \textit{user de-anonymization}. Suppose a national agency is trying to discover the identity of a dissident spreading anti-government propaganda on a social network. 
It is possible to monitor a suspect and detect when she posts messages via the social network mobile app. 
The inferred posting timestamps can be matched with the time of the messages on the dissident social profile in order to understand whether the suspect is actually the dissident.
\end{itemize}

In Table~\ref{tab:user_action_identification_app_categories}, we show the app categories covered by the surveyed works \revOne{that perform user action identification}~\cite{Watkins2013,Coull2014,Park2015,Conti2016,Fu2016,Saltaformaggio2016}. 
As we can notice from Table~\ref{tab:user_action_identification_app_categories}, {almost all} the works target communication apps, which belong to the most privacy-sensitive app category. 
This category \revOne{includes both instant messaging apps (e.g., iMessage, KakaoTalk, WhatsApp) and} email clients (e.g., Gmail, Yahoo Mail). Another sensitive category is social (e.g., Facebook, Twitter, Tumblr), which is targeted in~\cite{Conti2016,Saltaformaggio2016}. 
{Apps related to multimedia contents (e.g., YouTube) are considered in~\cite{Watkins2013,Saltaformaggio2016}.} 
Moreover, Saltaformaggio et al. in~\cite{Saltaformaggio2016} also focus on other categories of apps: dating (e.g., Tinder), health (e.g., HIV Atlas), maps (e.g., Yelp), news (e.g., CNN News), and shopping (e.g., Amazon). 
The works in~\cite{Watkins2013,Conti2016} cover productivity apps {(e.g., Dropbox), and Watkins et al. in~\cite{Watkins2013} also consider mobile games (e.g., Temple Run 2) and utility apps (e.g., ZArchiver)}.

\begin{table}
\caption{App categories covered in the surveyed works that deal with user action identification.}
\label{tab:user_action_identification_app_categories}
\centering
\begin{tabular}{|
>{\hspace{\tabred}}c<{\hspace{\tabred}}|
>{\hspace{\tabred}}l<{\hspace{\tabred}}|
>{\hspace{\tabred}}c<{\hspace{\tabred}}|
>{\hspace{\tabred}}c<{\hspace{\tabred}}|
>{\hspace{\tabred}}c<{\hspace{\tabred}}|
>{\hspace{\tabred}}c<{\hspace{\tabred}}|
>{\hspace{\tabred}}c<{\hspace{\tabred}}|
>{\hspace{\tabred}}c<{\hspace{\tabred}}|
>{\hspace{\tabred}}c<{\hspace{\tabred}}|
>{\hspace{\tabred}}c<{\hspace{\tabred}}|
>{\hspace{\tabred}}c<{\hspace{\tabred}}|
>{\hspace{\tabred}}c<{\hspace{\tabred}}|
>{\hspace{\tabred}}c<{\hspace{\tabred}}|}
\hline
& & \multicolumn{11}{c|}{\textbf{Covered App Categories}} \\ \cline{3-13}
\textbf{Year} & \multicolumn{1}{c|}{\textbf{Paper}} & \rot{\textbf{Communication }} & \rot{\textbf{Dating}} & \rot{\textbf{Gaming}} & \rot{\textbf{Health}} & \rot{\textbf{Maps}} & \rot{\textbf{Media}} & \rot{\textbf{News}} & \rot{\textbf{Productivity}} & \rot{\textbf{Shopping}} & \rot{\textbf{Social}} & \rot{\textbf{Utility}} \\ \hline
% 2014
{2013} & {Watkins et al. \cite{Watkins2013}} & & & {\OKB} & & & {\OKB} & & {\OKB} & & & {\OKB} \\ \hline
% 2014
2014 & Coull et al. \cite{Coull2014} & \OKB & & & & & & & & & & \\ \hline
% 2015
2015 & Park et al. \cite{Park2015} & \OKB & & & & & & & & & & \\ \hline
% 2016
\multirow{3}{*}{2016} & Conti et al. \cite{Conti2016} & \OKB & & & & & & & \OKB & & \OKB & \\ \cline{2-13}
& \revOne{Fu et al. \cite{Fu2016}} & \revOne{\OKB} & & & & & & & & & & \\ \cline{2-13}
& Saltaformaggio et al. \cite{Saltaformaggio2016} & \OKB & \OKB & & \OKB & \OKB & \OKB & \OKB & & \OKB & \OKB & \\ \hline
\end{tabular}
\end{table}

\begin{comment}
\begin{table*}
\caption{App categories covered in the surveyed works that deal with user action identification.}
\label{tab:user_action_identification_app_categories}
\centering
\begin{tabular}{|c|l|c|c|c|c|c|c|c|c|c|c|c|}
\hline
\multicolumn{1}{|c|}{\multirow{2}{*}{\textbf{Year}}} & \multicolumn{1}{c|}{\multirow{2}{*}{\textbf{Paper}}} & \multicolumn{11}{c|}{\textbf{Covered App Categories}} \\ \cline{3-13}
& & \rot{\textbf{Communication }} & \rot{\textbf{Dating}} & \rot{\textbf{Gaming}} & \rot{\textbf{Health}} & \rot{\textbf{Maps}} & \rot{\textbf{Media}} & \rot{\textbf{News}} & \rot{\textbf{Productivity}} & \rot{\textbf{Shopping}} & \rot{\textbf{Social}} & \rot{\textbf{Utility}} \\ \hline
% 2014
{2013} & {Watkins et al. \cite{Watkins2013}} & & & {\OKB} & & & {\OKB} & & {\OKB} & & & {\OKB} \\ \hline
% 2014
2014 & Coull et al. \cite{Coull2014} & \OKB & & & & & & & & & & \\ \hline
% 2015
2015 & Park et al. \cite{Park2015} & \OKB & & & & & & & & & & \\ \hline
% 2016
\multirow{2}{*}{2016} & Conti et al. \cite{Conti2016} & \OKB & & & & & & & \OKB & & \OKB & \\ \cline{2-13}
& Saltaformaggio et al. \cite{Saltaformaggio2016} & \OKB & \OKB & & \OKB & \OKB & \OKB & \OKB & & \OKB & \OKB & \\ \hline
\end{tabular}
\end{table*}

\end{comment}

Watkins et al. in~\cite{Watkins2013} develop a framework that exploits the inter-packet time of responses to ICMP packets (i.e., pings) to infer the type of action that the target user is performing on her mobile device. 
In particular, the authors focus on three types of user action: 
(i) CPU intensive; 
(ii) I/O intensive; 
and (iii) non-CPU intensive. 
First of all, the authors check the feasibility of their approach for the Android and iOS platforms, showing that unfortunately \revOne{their solution does not work on iOS since such OS does not use CPU throttling.} 
Subsequently, \revOne{they evaluate their framework using six Android apps.}
%, achieving a minimum $93\%$ accuracy. 
Since the proposed solution exploits the timing of packets, it is not affected by traffic encryption.

Coull and Dyer in~\cite{Coull2014} target iMessage, Apple's instant messaging service, which is available as an app for iOS or a computer application for OS X. 
The proposed analysis leverages the sizes of the packets exchanged between the target user and Apple's servers, thus it works despite all iMessage communications are encrypted. 
\revOne{The authors focus on five user actions: ``start typing'', ``stop typing'', ``send text'', ``send attachment'', and ``read receipt''.
The authors also aim to infer the language (among six languages: Chinese, English, French, German, Russian, and Spanish) and length of the exchanged messages. 
The authors make two assumption necessary by their methods to work correctly: (i) for user actions identification, they assume to have correctly inferred that the target mobile device is running iOS; and (ii) for language and message length inference, they also assume to have correctly identified an iMessage action.} 
%all user actions can be classified with over $99\%$ accuracy, except for the ``read receipt'' action that is often confused with the ``start typing'' action. 
%Assuming to have correctly identified an iMessage action on a mobile device running iOS, the language classification achieves more than $80\%$ accuracy by considering the first $50$ packets. 
%Besides, the length classification achieves an average error of $6.27$ characters for text messages, and an absolute error of less than $10$ bytes for attachment transfers.

Park and Kim in~\cite{Park2015} target KakaoTalk, an instant messaging service widely used in Korea. 
They consider eleven actions that a user can perform on the Android app (e.g., join a chat room, send a message, add a friend). 
For each action, the proposed framework learns its traffic pattern as a sequence of packets. 
Such sequence is then used to recognize that specific action in unseen network traces. 
The proposed solution works %reaches $99.7\%$ accuracy 
despite KakaoTalk traffic is encrypted using SSL/TLS, but it does not work in presence of IPsec encryption.

Conti et al. in~\cite{Conti2016} present an identification framework which leverages the information available in IP and TCP headers (e.g., source and destination IP addresses) and therefore it works even if the network traffic is encrypted via SSL/TLS. 
However, the proposed approach does not work on an IPsec scenario, since it relies on (IP address, TCP port) pairs to separate traffic flows. 
\revOne{The authors target seven popular Android apps (namely Dropbox, Evernote, Facebook, Gmail, Google+, Tumblr, and Twitter). %, reaching over $95\%$ accuracy and precision for most of the actions, and 
The authors also compare their proposal with websites fingerprinting algorithms by Liberatore and Levine~\cite{Liberatore2006} and Herrmann et al. in~\cite{Herrmann2009}, outperforming them.}

Fu et al. in~\cite{Fu2016} propose CUMMA, a framework for user action identification that targets messaging apps. 
The authors focus their analysis on the Android platform, and consider the WeChat and WhatsApp apps. 
For each targeted app, several user actions are chosen for identification, such as sending a text message or sharing the GPS position. 
Since the network traffic of messaging apps is usually encrypted to protect the privacy of their users, CUMMA is designed to overcome such limitation by exploiting only the size and timing of the packets exchanged between the app and the servers of the service provider. 
The proposed framework leverages machine learning by employing a classifier that is trained and evaluated on the feature vectors extracted from the captured information. 
The authors also develop a clustering method based on hidden Markov model to deal with the fact that the same network flow can likely contain the network data related to multiple user actions (see Section~\ref{sec:machine_learning_user_action_identification}).

Saltaformaggio et al. in~\cite{Saltaformaggio2016} present NetScope, a user action identification system that can be deployed at Wi-Fi access points or other network equipments. 
Since it leverages IP headers/metadata, NetScope can be employed even if the network traffic is protected by IPsec. 
The authors evaluate their solution by considering $35$ user actions from $22$ apps across two platforms (i.e., Android and iOS) and eight app categories. 
%The identification accuracy reaches average precision and recall of $78.04\%$ and $76.04\%$ respectively, performing better for Android devices rather than iOS ones.

\subsection{Operating System Identification}
\label{sec:OS_identification}

We define as \textit{operating system identification} the attempt to discover the operating system of a mobile device by analyzing its network traffic. This type of analysis has several applications:
\begin{itemize}
\item An adversary can identify the operating system of a target mobile device, and tailor her subsequent attack to that OS (e.g., by choosing a proper security exploit). 
In such case, the operating system identification is a preparatory task for more advanced and focused attacks. 
Moreover, the overall attack strategy can be more effective if the adversary is able to infer not only the operating system of the target mobile device, but also the version of that OS.
\item It is possible to expose the adoption of mobile operating systems among a crowd of people. 
This can be a starting point for marketing, as well as sociological studies (we consider the latter in Section~\ref{sec:sociological_inference}).
\end{itemize}

In this section, we survey \revOne{four} works \cite{Chen2014,Coull2014,Ruffing2016,Malik2017}. 
Table \ref{tab:OS_identification} reports the mobile operating systems they consider.

\begin{table}
\caption{Targeted mobile OSes in the surveyed works that deal with operating system identification.}
\label{tab:OS_identification}
\centering
\begin{tabular}{|
>{\hspace{\tabred}}c<{\hspace{\tabred}}|
>{\hspace{\tabred}}l<{\hspace{\tabred}}|
>{\hspace{\tabred}}c<{\hspace{\tabred}}|
>{\hspace{\tabred}}c<{\hspace{\tabred}}|
>{\hspace{\tabred}}c<{\hspace{\tabred}}|>{\hspace{\tabred}}c<{\hspace{\tabred}}|}
\hline
& & & & \textbf{Windows} & \\ 
\textbf{Year} & \multicolumn{1}{c|}{\textbf{Paper}} & \textbf{Android} & \textbf{iOS} & \textbf{Phone} & \textbf{Symbian} \\ \hline
% 2014
\multirow{2}{*}{2014} & Chen et al. \cite{Chen2014} & \OKB & \OKB & & \\ \cline{2-6}
& Coull et al. \cite{Coull2014} & & \OKB & & \\ \hline
% 2016
2016 & Ruffing et al. \cite{Ruffing2016} & \OKB & \OKB & \OKB & \OKB \\ \hline
% 2017
{2017} & {Malik et al. \cite{Malik2017}} & {\OKB} & {\OKB} & {\OKB} & \\ \hline
\end{tabular}
\end{table}

Chen et al. in~\cite{Chen2014} develop a probabilistic classifier that leverages the information available in IP and TCP headers. 
\revOne{Therefore, their method works unless IPsec is employed} to hide the content of IP packets. 
Such classifier is evaluated using network traces captured at a Wi-Fi access point to which Android and iOS mobile devices, as well as Windows laptops are connected. 
%The results show that the classifier reaches $1.0$ precision and recall for iOS identification, and $1.0$ precision and $0.8$ recall for Android identification.

Coull and Dyer in~\cite{Coull2014} target iMessage, Apple's instant messaging service. 
They leverage the sizes of the encrypted packets exchanged between the target user and Apple's servers, in order to determine whether she is using iMessage on iOS or OS X. 
The proposed classifier needs to observe only five packets \revOne{to successfully identify the OS.} %with $100\%$ accuracy.

The work by Ruffing et al.~\cite{Ruffing2016} stems from the observation that the timing of the network traffic generated by a mobile device depends on its operating system. 
The idea is to analyze the frequency spectrum of packet timing in order to identify the frequency components that are related to OS features, and filter out the ones that bring noise. 
Since this approach does not require to inspect the content of packets, it can be successfully applied even if encryption is in place. 
The authors evaluate their solution using network traffic captured from smartphones running the following operating systems: Android, iOS, Windows Phone, and Symbian. 
%On $30$-seconds-long traces, the proposed framework achieves around $70\%$ success rate (i.e., the number of correct OS identifications over the total number of attempts). On traces lasting five minutes or more, the success rate is around $90\%$. 
%Moreover, in case of heavy multitasking, only $30$ seconds of traffic are sufficient to achieve $100\%$ success rate. 
\revOne{The authors also evaluate whether their approach is suitable to discriminate different versions of the same OS, and they choose Android and iOS for such analysis.} 
%On fifteen-minutes-long traces from the Skype app, the success rate is $98\%$ and the misclassification rate is below $10\%$. 
%On traces of the same length but generated by the YouTube app, the success rate is around $50\%$.

Malik et al. in~\cite{Malik2017} present a framework that exploits the inter-packet time of packets coming from a target mobile device in order to infer its operating system. 
In particular, the authors focus on two types of packet: (i) the response to an ICMP packet sent to the target mobile device (active measurement); and (ii) an IP packet related to a video stream involving the target mobile device (passive measurement). 
In both cases, the proposed solution effectively discriminates among three mobile operating systems, namely Android, iOS, and Windows Phone. 
Moreover, such approach is not hindered by traffic encryption, since it exploits the timing of packets. 
However, we must point out that the authors' testbed includes only three devices, one for each of the considered mobile operating systems. 
\revOne{Therefore, it is not clear whether the mobile device model and the OS version may affect the accuracy in identifying the OS.}

\revOne{
\subsection{Position estimation}
\label{sec:position_estimation}

The set of places frequently visited by a person tells a lot about her social status, interests, and habits. 
Such information can be exploited for commercial purposes (e.g., targeted advertisement), as well as intelligence activities (e.g., police investigations). 
Since most of people own a mobile device and keep it with them all day long, locating the smartphone/tablet of a target user becomes a simple yet effective way to know her position. 
Multiple position detections can be then aggregated to build a profile of the subject or reconstruct its movements. 
The movements of several mobile users can be aggregated as well, for example to aid traffic prediction along urban streets.

We define as \textit{position estimation} the inference of the position and/or trajectory (i.e., movements) of a mobile device in a geographical area, by analyzing the network traffic that device generates. 
In this section, we survey the works that propose this type of traffic analysis. 
We point out that we do not consider the works in which: 
(i) the mobile traffic is analyzed to detect the leakage of GPS coordinates (we consider this kind of works in Section~\ref{sec:PII_leakage_detection}); and (ii) the analysis performed on the network traffic is device-agnostic, i.e., it does not take into account the fact that the target devices are smartphones/tablets (this kind of works is excluded since it is too generic).

Husted and Myers in~\cite{Husted2010} investigate whether a malnet (i.e., a colluding network of malicious Wi-Fi devices) can successfully determine the location of a mobile device. 
Each malicious node looks for probe requests carrying the MAC address of the target mobile device, and uploads its findings to a central server where the data coming from all nodes is used for trilateration. 
Through a software simulation of a metropolitan population of users equipped with 802.11g mobile devices, the authors show that $10\%$ of tracking population is sufficient to track the position of the remaining users. 
Besides, the tracking benefits from extending the broadcasting range of the mobile devices. 
This suggests that the adoption of newer 802.11 standards can make it feasible to build a geolocating malnet.

Musa and Eriksson in~\cite{Musa2012} present a system for passively tracking mobile devices by leveraging the Wi-Fi probe requests they periodically transmit. 
The idea is to employ a number of Wi-Fi monitors, which look for probe requests from mobile devices and report each detection to a central server, where the detections of the same mobile device are turned into a spatio-temporal trajectory. 
To evaluate their system, the authors set up three deployments and leverage GPS ground truth to measure the accuracy of the inferred trajectories. The mean error is under $70$ meters when the distance among the monitors is over $400$ meters.
}

\subsection{User Fingerprinting}
\label{sec:user_fingerprinting}

Mobile users interact actively with their devices, leveraging the nearly ubiquitous Internet connectivity and the capabilities of the apps available in the marketplaces. 
To each mobile user, it is possible to associate a set of preferred (i.e., most frequently used) apps and, for each of these apps, \revOne{a set of preferred actions (i.e., most frequently executed).} 
Since most of the mobile apps are able to connect to the Internet, and many user actions within them trigger data transmission through the network, it becomes clear that the network traffic generated by a user is likely to present a fairly constant pattern across different devices, as well as across different networks. 
We define as \textit{user fingerprinting} the attempt to exploit such pattern in order to recognize the network traffic belonging to a specific mobile user. 
This type of analysis can be applied to:
\begin{itemize}
\item Recognize the presence of a specific mobile user within a network. Once the network is identified, it is then possible to approximate the geographical position of that user with the location of the Wi-Fi hotspot or cellular station to which her mobile device is connected.
\item Partition the network traces of a mobile traffic dataset by user. Once the transmissions related to a specific mobile user have been separated and grouped together, it is then possible to apply other types of traffic analysis targeting that user.
\end{itemize}
In this section, we review the works that deal with mobile user fingerprinting.

Verde et al. in~\cite{Verde2014} present a system being able to accurately infer when a target user is connected to a given network and her IP address, even though she is hidden behind NAT among thousands of other users. 
To achieve this objective, a machine learning classifier is trained with the NetFlow records of the target user's traffic, and then employed to analyze the NetFlow records of a given network in order to detect the presence of the target user within it (more details on the classifier are provided in Section~\ref{sec:machine_learning_user_fingerprinting}). 
\revOne{The system is evaluated as follows:
\begin{itemize}
\item Cross-validation is applied to the NetFlow records of the traffic generated by $26$ different mobile users connecting to the Internet through the same Wi-Fi access point. 
%The best performing implementation of the classifier (which is based on random forest) reaches $95\%$ true-positive rate, $7\%$ false-positive rate, $0.95$ precision, $0.93$ recall, and $0.94$ F-measure.
% (Alberto) Here we must keep "implementation of the classifier" (QianQian suggested to remove "implementation of"), since random forest, SVM, etc. is only a part of the entire classifier.
\item The authors enroll five mobile users and ask them to connect their devices to a Wi-Fi access point %under the control of the authors, 
to then try to detect the presence of such users within a real-world large metropolitan Wi-Fi network.
%The implementation of the classifier based on random forest (the best performing) reaches over $90\%$ true-positive rate, less than $8\%$ false-positive rate, and over $0.9$ precision and recall.
\end{itemize}
}%%%%%%%%%
It is worth to notice that the proposed solution is encryption-agnostic, since NetFlow records can be extracted even from encrypted traffic.

{Vanrykel et al. in~\cite{Vanrykel2016} investigate how mobile unencrypted traffic can be exploited for user surveillance. 
The authors develop a framework to automatically execute apps, collect their network traffic, inspect the HTTP data, and identify the sensitive identifiers that are transmitted in clear. 
Moreover, the authors present a graph building technique that exploits such identifiers to extract the network traces generated by a specific mobile user from a traffic dataset (more details in Section~\ref{sec:graph_analysis}). 
\revOne{The authors analyze $1{,}260$ Android apps (from $42$ app categories) showing that their proposed solution can link $57\%$} of a mobile user's unencrypted network traffic. 
In addition, the authors observe the limited effectiveness of ad-blocking apps in preventing the leakage of sensitive identifiers.}

\subsection{Ad Fraud Detection}
\label{sec:ad_fraud_detection}

Mobile apps can be partitioned into two macro categories: paid and free apps. To cover the cost of developing and maintaining a free app, developers usually rely on advertisement, which applies the following business model:
\begin{itemize}
\item The ad provider yields a library that can be embedded in the app. Such library fetches ad contents and displays them on the app's user interface.
\item The ad provider pays according to the amount of times the ads are displayed to the user (\textit{impressions}) and/or clicked by the user (\textit{clicks}).
\end{itemize}

We define as \textit{ad fraud detection} the analysis of network traffic in order to uncover apps that trick the business model described above and let their developers illicitly earn money.
\revOne{Unfortunately, despite the economic impact that such research brings in the market of mobile advertisements, only one work has been published on this topic.}

Crussell et al. in~\cite{Crussell2014} focus on the Android platform. They identify two fraudulent app behaviors:
\begin{itemize}
\item To request ads while the app is running in background. 
This generates impressions without actually displaying ads to the user.
\item To click on ads without any user interaction, which is achievable in the following ways: 
(i) the app can trick the ad library by simulating a user click on the ad with a touch event; 
and (ii) the app extracts the click URL from the ad request (i.e., the web page that will be opened when the user clicks on the ad), then makes an HTTP request to the click URL to simulate a user click.
\end{itemize}
The authors propose MAdFraud, a tool being able to automatically run Android apps in emulators and analyze their application-layer traffic in order to expose ad fraud. 
The system is employed to analyze $130{,}339$ Android apps crawled from nineteen different marketplaces, and $35{,}087$ Android apps that probably contain malware (provided by an unspecified security company). 
The authors report that $30\%$ of apps generate fake impressions (i.e., they request to display an ad while running in the background), while $27$ apps generate fake clicks (i.e., they contact a click URL without any user interaction). 
Unfortunately, MAdFraud cannot process encrypted traffic, since it relies on the HTTP and DNS data generated by apps. 
However, the authors' analysis covers most of the available ad libraries. 
This means that such libraries do not usually employ any form of encryption for \revOne{their data transfers (i.e., they simply rely on plain HTTP).}

\subsection{Sociological Inference}
\label{sec:sociological_inference}

A property of a mobile device (e.g., the list of installed apps, the Wi-Fi networks to which the device associated) characterizes its owner. 
Sociologists can leverage this kind of information to study a population of mobile users. 
In this section, we review the works that deal with \textit{sociological inference}, which we define as the analysis of the network traffic generated by mobile devices in order to infer some kind of sociological information about their users.

Barbera et al. in~\cite{Barbera2013} investigate whether sociological information about a large crowd can be inferred by inspecting the Wi-Fi probe requests generated by the mobile devices of those people. 
First of all, the authors: 
(i) devise a methodology to convert a dataset of Wi-Fi probe requests into a social graph representing the owners of the monitored mobile devices; and 
(ii) develop an automatic procedure to infer the language of a given SSID. 
Subsequently, they target gatherings of people at urban, national, and international scale, as well as a mall, a train station, and a campus. 
\revOne{As a result of their analysis, the authors report some findings:} 
(i) the social graph of all the targeted events has social-network properties; 
(ii) the distributions of languages and mobile device vendors match the nature of the monitored crowds; 
and (iii) socially interconnected people tend to adopt mobile devices of the same vendor, and appear in the same time slot.

\subsection{Tethering Detection}
\label{sec:tethering_detection}

The ability to connect to cellular networks lets mobile devices have nearly ubiquitous Internet connectivity. Moreover, mobile devices are able to share such connectivity with other devices that cannot leverage cellular networks (e.g., laptops). 
This practice is commonly referred to as \textit{tethering}, and can be carried out in many ways, such as via a USB cable, via Bluetooth, or establishing a WLAN (hotspot) for which the mobile device acts as a router.

In this section, we report the works that deal with \textit{tethering detection}, which we define as the analysis of the network traffic generated by a mobile device in order to discover whether it is sharing its Internet connection with other devices. 
This type of analysis can be valuable for a cellular network provider, since tethering can significantly increase the amount of traffic its network infrastructure has to sustain. 
An effective detection method would let the cellular ISP prevent users from sharing their mobile Internet connection, or require them to pay an extra fee.

Chen et al. in~\cite{Chen2014} develop a probabilistic classifier being able to detect tethering by leveraging several network features (e.g., the number of distinct TTLs in the packets coming from the same IP address). 
To evaluate their solution, the authors use publicly available Wi-Fi traces collected at two conferences, as well as a dataset of Wi-Fi traffic from a campus network. 
The authors simulate tethering by randomly mix packets from different IP addresses, then modifying source IP addresses accordingly. 
%On the public traces, the classifier reaches $0.68$-$0.85$ recall with target precision fixed at $0.95$, and $0.78$-̃$0.89$ recall with target precision fixed at $0.8$, outperforming alternative methods based on decision trees and linear regression. 
%On the campus traces, the classifier reaches $0.86$ precision, $0.74$ recall, and $0.8$ F-measure. 
Since the proposed solution requires to inspect the content of TCP headers, it cannot work whether the captured network traffic is protected by IPsec.

\subsection{Website Fingerprinting}
\label{sec:website_fingerprinting}

The Internet has a central role in people's everyday life, and surfing the Web has become a common task that can be performed from desktop computers, as well as in mobility using laptops and smartphones/tablets, thanks to the increasing deployment of cellular and Wi-Fi networks. 
From a privacy point of view, the set of websites frequently visited by a user is a sensitive information, since it can disclose her interests, social habits, religious belief, sexual preference, and political orientation.

In the field of traffic analysis, \textit{website fingerprinting} generally indicates the attempt to infer the website or even webpage \revOne{visited by a user surfing the Internet with her mobile device, by analyzing the network traffic generated by the mobile web browser.} 
This type of analysis has been extensively treated in the domain of personal computers, where machine learning techniques have been proved to be very effective~\cite{Liberatore2006,Herrmann2009,Panchenko2011}. 
Since we focus on mobile devices, in this section we survey the works that target users navigating through the web browser of their mobile devices.

Spreitzer et al. in~\cite{Spreitzer2016} develop an Android app being able to capture the data-usage statistics of the browser app, and leverage them to fingerprint \revOne{the mobile webpages} visited by the user of the mobile device on which that app is installed. 
This solution is not affected by encryption, since it only requires to know the amount of data transmitted and received by the browser app (which is easily obtainable in Android). 
\revOne{The proposed app is evaluated on a set of $500$ possible pages that the user can visit. 
%Overall, $97\%$ of $2{,}500$ page visits are correctly inferred. 
The authors also evaluate their fingerprinting app when the network traffic is protected by Tor (in particular, by using the Orbot proxy and the Orweb browser).} 
%Out of a set of $100$ possible pages that the user can visit, $95\%$ of $500$ page visits are correctly inferred.

\section{Points of Capturing for Traffic Analysis Targeting Mobile Devices}
\label{sec:points_of_capturing}

Another meaningful categorization of the work is based on the point where the network traces are captured in order to build the traffic dataset(s).
%Another meaningful categorization for the works in the field of traffic analysis is according to the point where the network traces are captured in order to build the traffic dataset. 
In Table~\ref{tab:points_of_capturing}, we report the point(s) of capturing for each surveyed work. 
\revOne{As shown in Figure~\ref{fig:papers_by_point_of_capturing}, the most common points of capturing are wired network equipments (twenty two works), followed by mobile devices themselves (twenty works), Wi-Fi access points (twelve works), and mobile devices emulators and Wi-Fi monitors (five works each).} 
In one work, the mobile traffic is simulated via software. 
As shown in Table~\ref{tab:points_of_capturing}, four works leverage two different types of point of capturing, and one work even three. 
\revOne{In the following sections, we provide a definition for and discuss each of the points of capturing we encountered in the surveyed work. 
For each point of capturing, we also point out pro, cons, and relevant aspects that have to be taken into account, as a guideline to properly design a network traffic collection environment for mobile devices.} 
%We also discuss the effect of encryption (e.g., IPsec, SSL/TLS) on the collected datasets and performed analyses.

\begin{table}
\caption{The surveyed works by where the mobile traffic is captured.}
\label{tab:points_of_capturing}
\centering
\begin{tabular}{|c|l|c|c|c|c|c|c|}
\hline
\multicolumn{1}{|c|}{\textbf{Year}} & \multicolumn{1}{c|}{\textbf{Paper}} & \rot{\textbf{Mobile Devices (Emulated)}} & \rot{\textbf{Mobile Devices (Real)}} & \rot{\textbf{Network Simulators}} & \rot{\textbf{Wired Network Equipments }} & \rot{\textbf{Wi-Fi Access Points}} & \rot{\textbf{Wi-Fi Monitors}} \\ \hline
% 2010
\multirow{5}{*}{2010} & Afanasyev et al. \cite{Afanasyev2010} & & & & \OKB & \OKB & \\ \cline{2-8}
& Falaki et al. \cite{Falaki2010} & & \OKB & & & & \\ \cline{2-8}
& Husted et al. \cite{Husted2010} & & & \OKB & & & \\ \cline{2-8}
& Maier et al. \cite{Maier2010} & & & & \OKB & & \\ \cline{2-8}
& Shepard et al. \cite{Shepard2010} & & \OKB & & & & \\ \hline
% 2011
\multirow{4}{*}{2011} & Finamore et al. \cite{Finamore2011} & & & & \OKB & & \\ \cline{2-8}
& Gember et al. \cite{Gember2011} & & & & & \OKB & \\ \cline{2-8}
& Lee et al. \cite{Lee2011} & & & & \OKB & & \\ \cline{2-8}
& {Rao et al. \cite{Rao2011}} & & & & {\OKB} & & \\ \hline
% 2012
\multirow{9}{*}{2012} & Baghel et al. \cite{Baghel2012} & & & & \OKB & & \\ \cline{2-8}
& Chen et al. \cite{Chen2012} & & & & \OKB & & \\ \cline{2-8}
& Ham et al. \cite{Ham2012} & & \OKB & & & & \\ \cline{2-8}
& Musa et al. \cite{Musa2012} & & & & & & \OKB \\ \cline{2-8}
& {Shabtai et al. \cite{Shabtai2012}} & & {\OKB} & & & & \\ \cline{2-8}
& {Stevens et al. \cite{Stevens2012}} & & & & & {\OKB} & \\ \cline{2-8}
& Su et al. \cite{Su2012} & & \OKB & & & & \\ \cline{2-8}
& Wei et al. \cite{Wei2012} & & & & \OKB & & \\ \cline{2-8}
& {Wei et al. \cite{Wei2012ProfileDroid}} & & {\OKB} & & & & \\ \hline
% 2013
\multirow{5}{*}{2013} & Barbera et al. \cite{Barbera2013} & & & & & & \OKB \\ \cline{2-8}
& Kuzuno et al. \cite{Kuzuno2013} & & \OKB & & & & \\ \cline{2-8}
& Qazi et al. \cite{Qazi2013} & & \OKB & & & \OKB & \\ \cline{2-8}
& {Rao et al. \cite{Rao2013}} & & & & {\OKB} & & \\ \cline{2-8}
& {Watkins et al. \cite{Watkins2013}} & & & & & {\OKB} & \\ \hline
% 2014
\multirow{6}{*}{2014} & Chen et al. \cite{Chen2014} & & & & \OKB & \OKB & \OKB \\ \cline{2-8}
& Coull et al. \cite{Coull2014} & & \OKB & & & & \\ \cline{2-8}
& Crussell et al. \cite{Crussell2014} & \OKB & & & & & \\ \cline{2-8}
& Lindorfer et al. \cite{Lindorfer2014} & \OKB & & & & & \\ \cline{2-8}
& Shabtai et al. \cite{Shabtai2014} & & \OKB & & & & \\ \cline{2-8}
& Verde et al. \cite{Verde2014} & & & & \OKB & & \\ \hline
% 2015
\multirow{9}{*}{2015} & Chen et al. \cite{Chen2015} & & & & \OKB & & \\ \cline{2-8}
& Fukuda et al. \cite{Fukuda2015} & & \OKB & & & & \\ \cline{2-8}
& Le et al. \cite{Le2015} & & \OKB & & & & \\ \cline{2-8}
& Park et al. \cite{Park2015} & & & & \OKB & & \\ \cline{2-8}
& Soikkeli et al. \cite{Soikkeli2015} & & \OKB & & & & \\ \cline{2-8}
& Song et al. \cite{Song2015} & & \OKB & & & & \\ \cline{2-8}
& Wang et al. \cite{Wang2015} & & & & & & \OKB \\ \cline{2-8}
& Yao et al. \cite{Yao2015} & \OKB & & & & \OKB & \\ \cline{2-8}
& Zaman et al. \cite{Zaman2015} & & \OKB & & & & \\ \hline
% 2016
\multirow{13}{*}{2016} & Alan et al. \cite{Alan2016} & & & & & \OKB & \\ \cline{2-8}
& Conti et al. \cite{Conti2016} & & & & \OKB & & \\ \cline{2-8}
& \revOne{Fu et al. \cite{Fu2016}} & & & & & \revOne{\OKB} & \\ \cline{2-8}
& {Mongkolluksamee et al. \cite{Mongkolluksamee2016}} & & {\OKB} & & & & \\ \cline{2-8}
& Narudin et al. \cite{Narudin2016} & \OKB & \OKB & & & & \\ \cline{2-8}
& {Nayam et al. \cite{Nayam2016}} & & & & {\OKB} & & \\ \cline{2-8}
& Ren et al. \cite{Ren2016} & & & & \OKB & & \\ \cline{2-8}
& Ruffing et al. \cite{Ruffing2016} & & & & & & \OKB \\ \cline{2-8}
& Saltaformaggio et al. \cite{Saltaformaggio2016} & & & & & \OKB & \\ \cline{2-8}
& Spreitzer et al. \cite{Spreitzer2016} & & \OKB & & & & \\ \cline{2-8}
& Tadrous et al. \cite{Tadrous2016} & & & & & \OKB & \\ \cline{2-8}
& {Vanrykel et al. \cite{Vanrykel2016}} & & & & {\OKB} & & \\ \cline{2-8}
& {Wang et al. \cite{Wang2016}} & & & & {\OKB} & & \\ \hline
% 2017
\multirow{8}{*}{2017} & {Arora et al. \cite{Arora2017}} & & {\OKB} & & & & \\ \cline{2-8}
& \revOne{Chen et al. \cite{Chen2017}} & \revOne{\OKB} & & & & & \\ \cline{2-8}
& \revOne{Cheng et al. \cite{Cheng2017}} & & & & \revOne{\OKB} & & \\ \cline{2-8}
& {Continella et al. \cite{Continella2017}} & & & & {\OKB} & & \\ \cline{2-8}
& {Espada et al. \cite{Espada2017}} & & {\OKB} & & & & \\ \cline{2-8}
& {Malik et al. \cite{Malik2017}} & & & & & {\OKB} & \\ \cline{2-8}
& Taylor et al. \cite{Taylor2017} & & & & \OKB & & \\ \cline{2-8}
& {Wei et al. \cite{Wei2017}} & & & & {\OKB} & & \\ \hline
\end{tabular}
\end{table}

\begin{figure}
\centering
\includegraphics[scale=0.70]{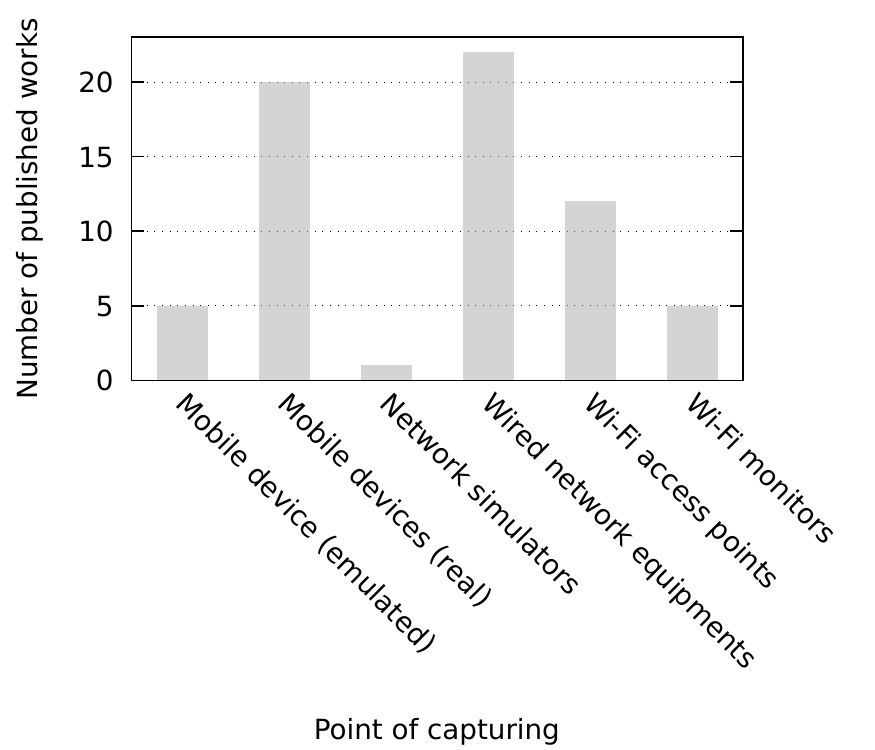}
\caption{Number of published works contributing traffic analysis methods targeting mobile devices, sorted by point of capturing.}
\label{fig:papers_by_point_of_capturing}
\end{figure}

\subsection{Wired Network Equipments}
\label{sec:wired_network_equipments}

In this section, we review the works in which the mobile traffic is captured at one or more wired network equipments. 
Such equipments can be deployed into two types of network:
\begin{itemize}
\item Small-scale networks, serving a reduced number of mobile users (from one single mobile device to a few tens). 
Researchers often deploy such networks to collect the traffic they need in a controlled environment. 
The equipments associated to this type of network are small Internet gateways~\cite{Verde2014},
VPN servers~\cite{Rao2013,Ren2016,Vanrykel2016}, and traditional desktop computers that log all the traffic traversing the wired link between the APs to which the mobile devices are associated and the Internet~\cite{Rao2011,Baghel2012,Wei2012,Chen2015,Park2015,Conti2016,Nayam2016,Wang2016,Cheng2017,Continella2017,Taylor2017}. 
The user population typically consists only of the targeted mobile devices (i.e., there is no need to filter out non-mobile traffic from the captured traces).
\item Large-scale networks, serving thousands of users. 
In such case, the considered network equipments are edge routers (i.e., routers that connect customers to their ISP's backbone)~\cite{Maier2010}, top-level routers~\cite{Lee2011,Chen2012,Verde2014}, Internet gateways~\cite{Afanasyev2010,Chen2014,Wei2017}, switches~\cite{Chen2014}, or generic points of presence within national ISPs and campus networks~\cite{Finamore2011}. 
The user population typically includes also non-mobile users (e.g., laptop users), and the network traffic they generate must be removed from the captured traces.
\end{itemize}
%\revOne{The works in which the mobile traffic is extracted from traces captured at one or more wired network equipments serving small-scale and large-scale networks.} % are presented in sections~\ref{sec:wired_network_equipments_small-scale} and~\ref{sec:wired_network_equipments_large-scale}, respectively.

%\subsubsection{Small-scale Networks}
\label{sec:wired_network_equipments_small-scale}

\begin{table*}
\caption{The surveyed works in which the mobile traffic is extracted from traces captured at one or more wired network equipments serving a small number of mobile devices.}
\label{tab:wired_network_equipments_small-scale}
\centering
%\begin{small}
\begin{tabular}{|
>{\hspace{\tabred}}c<{\hspace{\tabred}}|
>{\hspace{\tabred}}l<{\hspace{\tabred}}|
>{\hspace{\tabred}}c<{\hspace{\tabred}}|
m{0.15\textwidth}|m{0.20\textwidth}|m{0.15\textwidth}|}
\hline
\textbf{Year} & \multicolumn{1}{c|}{\textbf{Paper}} & \textbf{Network Equipment(s)} & \multicolumn{1}{c|}{\textbf{Target(s)}} & \multicolumn{1}{c|}{\textbf{Capturing Details}} & \textbf{Leveraged Information} \\ \hline
% 2011
{2011} & {Rao et al. \cite{Rao2011}} & {Forwarding server} & {Android and iOS clients for Netflix and YouTube} & {$180$ seconds for each video playback} & {HTTP messages} \\ \hline
% 2012
\multirow{3}{*}{2012} & \multirow{2}{*}{Baghel et al. \cite{Baghel2012}} & \multirow{2}{*}{Forwarding server} & Facebook Android app & $90$ minutes, no user interaction & \multirow{2}{*}{Layer-2+ data} \\ \cline{4-5}
& & & Skype Android app & Five hours, no user interaction & \\ \cline{2-6}
& Wei et al. \cite{Wei2012} & Forwarding server & $102$ malicious Android apps \revOne{from a public Android malware dataset, and popular Android apps from Google Play Store} & No details & DNS data \\ \hline
% 2013
\multirow{2}{*}{{2013}} & \multirow{2}{*}{{Rao et al. \cite{Rao2013}}} & \multirow{2}{*}{{VPN server}} & {The top-$100$ free Android apps in Google Play Store, and $209$ \revOne{free} iOS apps from Apple App Store} & {Up to ten minutes of manual interaction} & \multirow{2}{*}{{Layer-3+ data}} \\ \cline{4-5}
& & & {$732$ free Android apps from a third-party Android marketplace} & {Android Debug Bridge (ADB) scripting and Monkey are leveraged to execute $100{,}000$ actions for each app} & \\ \hline
% 2014
{2014} & {Verde et al.} \cite{Verde2014} &{Gateway router} & {$26$ mobile devices} & {One month of monitoring} & {NetFlow records} \\ \hline
% 2015
\multirow{2}{*}{2015} & Chen et al. \cite{Chen2015} & Forwarding server & $5560$ malicious Android apps \revOne{(from $177$ malware families)} & Each app is stimulated for five minutes using Monkeyrunner & Layer-2+ data \\ \cline{2-6}
& Park et al. \cite{Park2015} & Forwarding server & Eleven user actions from KakaoTalk Android app & Each user action is automatically executed $100$ times & IP headers, TCP headers \\ \hline
% 2016
\multirow{6}{*}{2016} & Conti et al. \cite{Conti2016} & Forwarding server & $58$ user actions from seven Android apps & Android Debug Bridge (ADB) scripting is leveraged to execute $220$ sequences of actions for each app & IP headers, TCP headers \\ \cline{2-6}
& {Nayam et al. \cite{Nayam2016}} & {Forwarding server} & {$63$ Android and $35$ iOS free apps, all belonging to the ``Health \& Fitness'' category} & {\revOne{Three $30$-minutes-long runs per app,} driven using automated scripts (Appium for Android and Silk Mobile for iOS)} & {HTTP messages} \\ \cline{2-6}
& \multirow{2}{*}{Ren et al. \cite{Ren2016}} & \multirow{2}{*}{VPN server} & The top-$100$ free apps \revOne{from Google Play Store (Android), Apple App Store (iOS), and Windows Phone Store (Windows Phone)} & Five minutes of manual interaction & \multirow{2}{*}{Layer-3+ data} \\ \cline{4-5}
& & & $850$ of the top $1{,}000$ free apps from a third-party Android marketplace & Android Debug Bridge (ADB) scripting and Monkey are leveraged to execute $10{,}000$ actions for each app & \\ \cline{2-6}
& {Vanrykel et al. \cite{Vanrykel2016}} & {Two VPN servers} & {$1{,}260$ Android apps \revOne{from Google Play Store}} & {User interactions are simulated using The Monkey} & {HTTP messages} \\ \cline{2-6}
& {Wang et al. \cite{Wang2016}} & {Forwarding server} & {$8{,}312$ \revOne{benign Android apps from Google Play Store}, and $5{,}560$ \revOne{malicious Android apps}} & {Each app is stimulated using Monkeyrunner} & {TCP- and HTTP-related data} \\ \hline
% 2017
\multirow{3}{*}{2017} & \revOne{Cheng et al. \cite{Cheng2017}} & \revOne{Forwarding server} & \revOne{Seven Android apps, plus a self-developed PII-leaking Android app} & \revOne{Manual interaction} & \revOne{IP headers, TCP headers} \\ \cline{2-6}
& {Continella et al. \cite{Continella2017}} & {Forwarding server} & {$1{,}004$ Android apps \revOne{from Google Play Store}} & {Each app is stimulated for ten minutes using Monkey} & {HTTP messages} \\ \cline{2-6}
& Taylor et al. \cite{Taylor2017} & Forwarding server & $110$ \revOne{of the top-$200$} Android apps \revOne{from Google Play Store} & Android Debug Bridge (ADB) scripting is leveraged to simulate user-app interactions & IP headers, TCP headers \\ \hline
\end{tabular}
%\end{small}
\end{table*}

In Table~\ref{tab:wired_network_equipments_small-scale}, we present the works in which the collected mobile traffic comes from one or more wired network equipments serving a small number of mobile devices. 
For each work, we report the network equipments at which the mobile traffic is logged, the targets of the analysis, additional details about the capturing process, and the information leveraged for the analysis. 
We use the term \textit{forwarding server} to indicate a device that logs all the traffic traversing the wired link between the APs to which the monitored mobile devices are connected and the Internet.

\label{sec:wired_network_equipments_large-scale}

In Table~\ref{tab:wired_network_equipments_large-scale}, we present the works in which the mobile traffic is extracted from traces captured at one or more wired network equipments serving large-scale networks with thousands of users. 
\revOne{In the following, we provide additional information about the mobile data extraction process for each work.} % and the effects of encryption in the considered scenario.

\begin{sidewaystable*}
\caption{The surveyed works in which the mobile traffic is extracted from traces captured at one or more wired network equipments serving thousands of users.}
\label{tab:wired_network_equipments_large-scale}
\centering
\scalebox{0.85}{
\begin{tabular}{|c|>{\centering\arraybackslash}m{0.09\textwidth}|>{\centering\arraybackslash}m{0.12\textwidth}|>{\centering\arraybackslash}m{0.08\textwidth}|>{\centering\arraybackslash}m{0.08\textwidth}|>{\centering\arraybackslash}m{0.21\textwidth}|>{\centering\arraybackslash}m{0.28\textwidth}|}
\hline
\textbf{Year} & \textbf{Paper} & \textbf{Targeted Network Equipment(s)} & \textbf{Number of Monitored Users} & \textbf{Capturing Period} & \textbf{Leveraged Information} & \textbf{Methodology Applied to Extract Mobile Data} \\ \hline
% 2010
\multirow{2}{*}{2010} & \multicolumn{1}{l|}{Afanasyev et al. \cite{Afanasyev2010}} & Central Internet gateway of an urban Wi-Fi network & Over $2500$ simultaneous & $5$ days & \multicolumn{1}{m{0.20\textwidth}|}{Layer-3+ headers (excluding DHCP data\footnote{The DHCP requests are handled by the APs of the network, thus they do not reach the central Internet gateway.}) of the first packet of each flow for the first {quarter} of each hour} & 
\multicolumn{1}{m{0.27\textwidth}|}{Leverage RADIUS logs from the APs {of the network} to map each IP address observed at the \aRM{Internet} gateway to a MAC address, then inspect the Organizationally Unique Identifier (OUI) of the MAC address} \\ \cline{2-7}

& \multicolumn{1}{l|}{Maier et al. \cite{Maier2010}} & Edge router of an ISP's network & Over $20{,}000$ & $4$ days over $11$ months & Anonymized DSL data & 
\multicolumn{1}{m{0.27\textwidth}|}{Inspect the User-Agent field of HTTP messages or, for non-HTTP traffic, inspect the Time-To-Live (TTL) field of IP packets} \\ \hline
% 2011
\multirow{2}{*}{2011} & \multicolumn{1}{l|}{Finamore et al. \cite{Finamore2011}} & Five {points of presence} within national ISPs and campus networks & Unspecified & $1$ week & IP packets & 
\multicolumn{1}{m{0.24\textwidth}|}{Inspect the URL {requests}, looking for \texttt{app=youtube\_{}gdata} or \texttt{app=youtube\_{}mobile}} \\ \cline{2-7}
& \multicolumn{1}{l|}{Lee et al. \cite{Lee2011}} & Top-level router of a campus network & Unspecified & $6$ days & IP packets & 
\multicolumn{1}{m{0.27\textwidth}|}{Inspect packet headers, looking for information related to mobile operating systems} \\ \hline
% 2012
2012 & \multicolumn{1}{l|}{Chen et al. \cite{Chen2012}} & Gateway router of a campus network & Unspecified & $3$ days, $1$ day & \multicolumn{1}{m{0.20\textwidth}|}{{Up to} $900$ bytes of each incoming/outgoing packet, including IP, TCP, and application-level headers} & 
\multicolumn{1}{m{0.27\textwidth}|}{Inspect the IP address (since separate IP pools are used for Ethernet and WLAN), then inspect the User-Agent field of HTTP {messages}} \\ \hline
% 2014
\multirow{3}{*}{2014} & \multicolumn{1}{l|}{\multirow{2}{*}{Chen et al. \cite{Chen2014}}} & Wired switch serving the APs of a Wi-Fi network & Unspecified & Unspecified & \multicolumn{1}{m{0.20\textwidth}|}{Layer-2+ headers, plus DHCP and DNS payloads (all data are anonymized)} & None \\ \cline{3-7}
& & Internet gateways of a campus Wi-Fi network & $12{,}600$ & $1$ week & IP packets & None \\ \cline{2-7}

& \multicolumn{1}{l|}{Verde et al. \cite{Verde2014}} & Tier-2 router of a metropolitan Wi-Fi network & $200{,}000$ & $1$ day & NetFlow records & None \\ \hline
% 2017
{2017} & {Wei et al. \cite{Wei2017}} & {Internet gateway of a campus Wi-Fi network} & {$6{,}482$} & {$1$ month} & {Layer-3+ data} & {Inspect the IP address to check whether it belongs to the WLAN IP address pool, leverage the DHCP logs from the DHCP server to map the IP address to a MAC address, and inspect the Organizationally Unique Identifier (OUI) of the MAC address} \\ \hline
\end{tabular}
}
\end{sidewaystable*}

%\paragraph{Mobile Data Extraction}
\label{sec:wired_network_equipments_large-scale_mobile_data_extraction}

Afanasyev et al. in~\cite{Afanasyev2010} leverage the Organizationally Unique Identifier (OUI) of the MAC address to discriminate between mobile and non-mobile devices. 
This approach has the disadvantage that an OUI can be associated to devices of both types (e.g., several OUIs belonging to Apple are associated to iPhones and MacBooks as well). 
Maier et al. in~\cite{Maier2010} and Chen et al. in~\cite{Chen2012} inspect the User-Agent field of HTTP messages, which can be misleading and it is not present in non-HTTP mobile traffic. 
In such case, the authors in~\cite{Maier2010} inspect the Time-To-Live (TTL) field of IP packets. 
Finamore et al. in~\cite{Finamore2011} leverage the peculiar characteristics of YouTube traffic from mobile devices, while Lee et al. in~\cite{Lee2011} inspect the packet headers, looking for information related to mobile operating systems (without clarifying the nature of such information). 
Chen et al. in~\cite{Chen2014} do not elicit mobile traffic from the gathered network traces because they simply merge such data with traffic from other sources (see Section~\ref{sec:Wi-Fi_monitors} for details), then simulate tethering by modifying the source IP address of packets. 
Verde et al. in~\cite{Verde2014} do not need to extract the traffic of mobile devices from the collected network data because identifying such traffic is just the goal of their user fingerprinting method. 
{Wei et al. in~\cite{Wei2017} inspect the IP address to check whether it belongs to the IP address pool of the target WLAN, then leverage the DHCP logs from the DHCP server of the network to map the IP address to a MAC address, and finally inspect the Organizationally Unique Identifier (OUI) of the MAC address.}

\subsection{Mobile Devices (Real)}
\label{sec:mobile_devices}

The most direct way to collect mobile traffic is to place the point of capturing \textit{within} the mobile devices, leveraging their modern operating systems to run a full-fledged logging app that is able to gather the required information. 
This approach has several implications:
\begin{itemize}
\item The covered set of mobile devices tends to be small compared to network-side measurements, since the logging app has to be installed on the mobile device of each volunteer.
\item \revOne{The strongest advantage of this point of capturing is that the traffic is logged directly on the mobile device, therefore} we are sure that everything is captured belongs to that mobile device. 
In case of network-side logging, instead, the mobile traffic needs to be separated from the transmissions generated by other kinds of device, such as laptops and desktop computers. 
This process is error-prone, since some network information could be potentially misclassified. 
Moreover, it lacks completeness, because traffic that is not classified for some reason will be discarded even if it belongs to a mobile device.
\item The logging app must have the proper permissions to capture traffic on the mobile device.
\item The logging app must be lightweight. 
This means that it has to: 
(i) impose a negligible computational burden on the mobile device's CPU; 
(ii) occupy as few memory as possible to store traffic logs (this problem is easy resolvable if the logging app is \revOne{allowed to periodically upload} the logs to a remote server); and 
(iii) cause minor battery consumption, which is a major concern for mobile users.
\item It is possible to focus on the traffic generated by specific mobile apps, or the one transiting through a specific network interface (i.e., Wi-Fi or cellular).
\end{itemize}

In Table \ref{tab:mobile_devices}, we present the works in which the mobile traffic is captured within one or more mobile devices. 
For each work, we show the targeted mobile platforms, the number of mobile devices employed, the tool used to capture the network traffic, and the information leveraged for the analysis.

\begin{table*}
\caption{The surveyed works in which the mobile traffic is captured within one or more \revOne{real} mobile devices.}
\label{tab:mobile_devices}
\centering
\scalebox{0.95}{
\begin{tabular}{|c|l|c|c|c|m{0.20\textwidth}|}
\hline
\textbf{Year} & \multicolumn{1}{c|}{\textbf{Paper}} & \textbf{Mobile Platform(s)} & \textbf{Number of Devices} & \textbf{Capturing Tool} & \multicolumn{1}{c|}{\textbf{Leveraged Information}} \\ \hline
% 2010
\multirow{4}{*}{2010} & \multirow{3}{*}{Falaki et al. \cite{Falaki2010}} & \multirow{2}{*}{Android} & $33$ & Custom logging app & Per-app Tx/Rx bytes \\ \cline{4-6}
& & & $2$ & tcpdump & \multirow{2}{*}{Layer-2+ data} \\ \cline{3-5}
& & Windows Mobile & $8$ & Netlog & \\ \cline{2-6}
& Shepard et al. \cite{Shepard2010} & iOS & $25$ & Custom logging app & Per-interface Tx/Rx bytes, IP headers/packets \\ \hline
% 2012
\multirow{4}{*}{2012} & Ham et al. \cite{Ham2012} & Android & $10$ & Custom logging app & Per-process/Per-interface Tx/Rx bytes/packets \\ \cline{2-6}
& {Shabtai et al. \cite{Shabtai2012}} & {Android} & {$2$} & {Custom logging app} & {Per-app cellular/Wi-Fi Tx/Rx bytes/packets} \\ \cline{2-6}
& Su et al. \cite{Su2012} & Android & Unspecified & tcpdump & Layer-2+ data \\ \cline{2-6}
& {Wei et al. \cite{Wei2012ProfileDroid}} & {Android} & {$2$} & {tcpdump} & {Layer-2+ data} \\ \hline
% 2013
\multirow{2}{*}{2013} & Kuzuno et al. \cite{Kuzuno2013} & Android & $1$ & tcpdump & Layer-2+ data \\ \cline{2-6}
& Qazi et al. \cite{Qazi2013} & Android & $5$ & netstat & netstat logs \\ \hline
% 2014
\multirow{2}{*}{2014} & Coull et al. \cite{Coull2014} & iOS & Unspecified & \parbox{0.20\textwidth}{Unspecified (on a Mac, by using the iOS Remote Virtual Interface)} & Packet sizes within iMessage's APNS connection \\ \cline{2-6}
& Shabtai et al. \cite{Shabtai2014} & Android & \revOne{$1$} & Custom logging app & Per-app Tx/Rx bytes and percent out of total Tx/Rx bytes \\ \hline
% 2015
\multirow{7}{*}{2015} & \multirow{2}{*}{Fukuda et al. \cite{Fukuda2015}} & Android & Over $800$ & \multirow{2}{*}{Custom logging app} & \multirow{2}{*}{\parbox{0.20\textwidth}{Per-app/Per-interface Tx/Rx bytes/packets}} \\ \cline{3-4}
& & iOS & Over $700$ & & \\ \cline{2-6}
& Le et al. \cite{Le2015} & Android & $9$ & Custom logging app & Per-app IP headers/packets \\ \cline{2-6}
& Soikkeli et al. \cite{Soikkeli2015} & Unspecified & $120$ & Custom logging app & Tx/Rx bytes \\ \cline{2-6}
& Song et al. \cite{Song2015} & Android & $1$ & Custom logging app & IP packets \\ \cline{2-6}
& \multirow{2}{*}{Zaman et al. \cite{Zaman2015}} & \multirow{2}{*}{Android} & \multirow{2}{*}{$1$} & Shark for Root & Layer-2+ data \\ \cline{5-6}
& & & & netstat & netstat logs \\ \hline
% 2016
\multirow{3}{*}{2016} & {Mongkolluksamee et al. \cite{Mongkolluksamee2016}} & {Android} & {$1$} & {tcpdump} & {Layer-2+ data} \\ \cline{2-6}
& Narudin et al. \cite{Narudin2016} & Android & $1$ & tPacketCapturePro & Per-app layer-2+ data \\ \cline{2-6}
& Spreitzer et al. \cite{Spreitzer2016} & Android & Unspecified & Custom logging app & Tx/Rx TCP bytes of the browser app \\ \hline
% 2017
\multirow{2}{*}{{2017}} & {Arora et al. \cite{Arora2017}} & {Android} & {$1$} & {Unspecified} & {IP packets} \\ \cline{2-6}
& {Espada et al. \cite{Espada2017}} & {Android} & {$1$} & {tcpdump} & {Layer-2+ data} \\ \hline
\end{tabular}
}
\end{table*}

Android is the most targeted mobile platform, mainly because its open nature makes it easy to develop a traffic logger from scratch, or simply port one of the available desktop solutions. 
Nevertheless, Fukuda et al. in~\cite{Fukuda2015} and Shepard et al. in~\cite{Shepard2010} show that effective traffic loggers can be successfully deployed also on iOS devices. 
Packet sniffing tools such as Shark for Root and the networking module of DELTA logging tool~\cite{spolaor2017delta} are based on tcpdump, while tPacketCapturePro leverages the \texttt{VPNService} class of the \revOne{Android libraries. 
This class is} also used in the custom logging apps by Le et al. in~\cite{Le2015}, and Song and Hengartner in~\cite{Song2015}. 

The number of targeted devices is a meaningful \revOne{information, especially for works} which carry out mobile traffic characterization~\cite{Falaki2010,Shepard2010,Wei2012ProfileDroid,Fukuda2015,Espada2017}, or study the usage habits of mobile users~\cite{Falaki2010,Ham2012,Fukuda2015,Soikkeli2015}. 
With regard to such works, we observe that only Fukuda et al. in~\cite{Fukuda2015} leverage a suitable population (over $1500$ mobile devices), while the others count from a few tens to slightly more than a hundred of mobile devices. 
We point out that the problem is less relevant for the works in~\cite{Wei2012ProfileDroid,Espada2017}, since the focus of the analysis is on the apps, rather than mobile devices.

\subsection{Wi-Fi Access Points}
\label{sec:Wi-Fi_access_points}

As reported in~\cite{Fukuda2015}, mobile users are increasingly offloading their traffic demands to Wi-Fi networks. 
This practice has become very common at home, where Wi-Fi modems are employed to make the wired Internet connection of the house available to laptops and mobile devices. 
Moreover, free Wi-Fi networks are often deployed in shops and public places (e.g., parks, malls, train stations), as well as at social events (e.g., meetings, conferences, concerts).

A Wi-Fi network typically consists of two types of hardware equipments: 
(i) access points (APs), which leverage the 802.11 standard to provide network connectivity to the associated wireless devices; 
and (ii) gateways, which forward the network traffic coming from the APs to the Internet (or to a higher-level gateway, in case of a hierarchical network infrastructure), and vice versa. 
It is worth to notice that these two categories are not mutually exclusive, since a hardware equipment can act as both access point and gateway (e.g., Wi-Fi modems). 
In this section, we survey the works that apply analysis methods to mobile traffic captured at one or more Wi-Fi access points (we dealt with traffic capturing at gateways in Section~\ref{sec:wired_network_equipments}).

From an analysis point of view, we make the following observations:
\begin{itemize}
\item Compared with cellular networks, Wi-Fi networks cover a smaller geographical area, as well as much less users. 
For this reason, the mobile traffic captured at the APs of a Wi-Fi network is representative of a more restricted user population (e.g., the customers of a shop, the students of a campus), enabling fine-grained analysis.
\item Since Wi-Fi networks are typically free of charge, mobile users can carry out intensive network activities (e.g., watching videos from a streaming platform). 
Such kind of activities are hard to observe in cellular networks due to the fees applied to Internet traffic by network providers.
\item Wi-Fi networks usually serve not only mobile devices, but also other kinds of device, such as desktop computers and laptops. Therefore, if the analysis targets mobile devices, the network traffic belonging to non-mobile devices must be properly filtered out from the collected network traces.
\item If the monitored Wi-Fi network employs several access points, the information gathered at each AP must be properly combined with the one from the other APs, in order to produce a comprehensive network trace. This process can be tricky whenever the APs (or the traffic sniffers deployed at the APs) are not perfectly synchronized, causing timestamps from different sources to be staggered.
\end{itemize}

\begin{table*}
\caption{The surveyed works in which the mobile traffic is captured at one or more Wi-Fi access points.}
\label{tab:Wi-Fi_access_points}
\centering
\begin{tabular}{|c|c|l|m{0.40\textwidth}|}
\hline
\textbf{Scale of the Targeted Wi-Fi Network(s)} & \textbf{Year} & \multicolumn{1}{c|}{\textbf{Paper}} & \multicolumn{1}{c|}{\textbf{Leveraged Information}} \\ \hline
% few APs, controlled environment
\multirow{10}{*}{\revOne{Few APs} in controlled environment} & {2012} & {Stevens et al. \cite{Stevens2012}} & {HTTP messages} \\ \cline{2-4}
& \multirow{2}{*}{2013} & Qazi et al. \cite{Qazi2013} & Network- and transport-layer information \\ \cline{3-4}
& & {Watkins et al. \cite{Watkins2013}} & {Inter-packet time of ICMP responses} \\ \cline{2-4}
& 2014 & Chen et al. \cite{Chen2014} & IP headers, TCP headers \\ \cline{2-4}
& 2015 & Yao et al. \cite{Yao2015} & HTTP messages \\ \cline{2-4}
& \multirow{4}{*}{2016} & Alan et al. \cite{Alan2016} & IP headers, TCP headers \\ \cline{3-4}
& & \revOne{Fu et al. \cite{Fu2016}} & \revOne{Size and timing of IP packets} \\ \cline{3-4}
& & Saltaformaggio et al. \cite{Saltaformaggio2016} & IP headers \\ \cline{3-4}
& & Tadrous et al. \cite{Tadrous2016} & 802.11 frames \\ \cline{2-4}
& {2017} & {Malik et al. \cite{Malik2017}} & {Inter-packet time of ICMP responses or IP packets related to video streaming} \\ \hline
% multiple APs, real deployment
\multirow{2}{*}{Multiple APs in real deployment} & 2010 & Afanasyev et al. \cite{Afanasyev2010} & {Data-link-} and {network-layer} information from RADIUS logs \\ \cline{2-4}
& 2011 & Gember et al. \cite{Gember2011} & Layer 2+ \revOne{data} \\ \hline
\end{tabular}
\end{table*}

In Table~\ref{tab:Wi-Fi_access_points}, we present the surveyed works in which the mobile traffic is captured at one or more Wi-Fi access points. 
\revOne{We can identify two distinct experimental settings:} 
\revOne{in the former,} the authors monitor \revOne{few access points}, deployed in a controlled environment to provide Internet connectivity to a small number of mobile devices~\cite{Stevens2012,Qazi2013,Watkins2013,Chen2014,Yao2015,Alan2016,Fu2016,Saltaformaggio2016,Tadrous2016,Malik2017}; % (more details in Section~\ref{sec:Wi-Fi_access_points_single_AP});
\revOne{in the latter,} several APs of a real Wi-Fi network serving a large number of users are monitored~\cite{Afanasyev2010,Gember2011}.

\revOne{In the case of multiple APs in a real deployment, it is fundamental to separate the network traffic related to mobile devices from the one related to other types of device (e.g., laptops).} 
%\revOne{While the from a single controlled AP uses the same principles of the methods relying on a forwarding server (discussed in Section~\ref{sec:wired_network_equipments_small-scale}, the works that rely on multiple APs in a real deployment needs to define alternative methodology for data collection.}
Afanasyev et al. in~\cite{Afanasyev2010} leverage {encryption-agnostic} data-link- and network-layer information from the RADIUS logs collected by the over $500$ APs of the Google Wi-Fi network in Mountain View (California, {USA}), and discriminate among desktop computers, laptops, and mobile devices by relying on the Organizationally Unique Identifier (OUI) of the MAC address. 
{Gember et al. in~\cite{Gember2011} carry out} a comparison between mobile and non-mobile devices with regard to network traffic properties and habits of users. 
To discriminate between the two types of device, they apply the following methodology: 
%Since the goal of their analysis is a comparison between mobile and non-mobile devices with regard to network traffic properties and habits of users, Gember et al. in~\cite{Gember2011} discriminate between the two types of device applying the following methodology: 
(i) for a device generating HTTP traffic, the User-Agent field of the HTTP messages is compared with a list of strings clearly related to mobile devices, in order to determine whether the device is a mobile one (match) or not (mismatch); 
(ii) for a device that does not generate HTTP traffic, the Organizationally Unique Identifier (OUI) of its MAC address is inspected to determine whether it is a mobile or non-mobile device. 
%Since the analysis is mainly focused on transport and application layers, most of the authors' findings are related to non-encrypted traffic.
%Whenever the encryption at network (IPsec) or transport layer (SSL/TLS) is employed, the HTTP information becomes inaccessible, thus making the discrimination process (if not the entire analysis) infeasible.

\subsection{Mobile Devices (Emulated)}
\label{sec:emulators}

As for their desktop counterparts, \aRM{mobile} apps must be properly tested not only throughout their development process, but also before their final submission to a marketplace. 
The simplest testing methodology consists of installing the mobile app on one or more physical mobile devices. 
However, this approach has two shortcomings: (i) the number of different test configurations (each consisting of a hardware device and a version of a compatible mobile operating system) is limited, while the number of configurations available on the market is large; 
and (ii) the difficulty in automating the tests within the mobile devices, due to lack of tools and resource constraints.

An alternative solution is to install the \aRM{mobile} app to be tested in a mobile device emulator, which is a virtual machine that is able to simulate the components and operations of a mobile operating system. This approach has several implications:
\begin{itemize}
\item The Software Development Kit (SDK) of a mobile platform typically provides a mobile device emulator for testing purposes. Therefore, the developers can cut the expense for physical mobile devices and simply buy a machine to run the emulator (or, even better, run the emulator directly on the machine where the code is written, without any additional expenditure).
\item If properly endowed with computational power and memory, the testing machine can run multiple mobile device emulators in parallel, speeding up the overall testing process or letting the developers increment the set of tests to be executed on the app.
\item A mobile device emulator can be quite easily controlled from the outside, helping test automation and thus reducing human intervention.
\item There are important limitations on which components and operations of a mobile operating system can be emulated. Such limitations reduce the types of test that can be actually executed on a {given} app.
\end{itemize}

Since a machine running a mobile device emulator is responsible for forwarding the network traffic from the emulator to the Internet and vice versa, it constitutes an ideal point of capturing for mobile traffic. 
This approach is particularly useful if the focus of the analysis is on the network traffic of a specific mobile app. 
In Table~\ref{tab:emulators}, we present the targeted mobile platforms and number of considered apps for the surveyed works in which the mobile traffic is captured at one or more machines running mobile device emulators.

\begin{table}
\caption{The surveyed works in which the mobile traffic is captured at one or more machines running mobile device emulators.}
\label{tab:emulators}
\centering
\begin{tabular}{|c|l|c|c|c|}
\hline
& & \multicolumn{3}{c|}{\textbf{Number of \aRM{Mobile} Apps Run in}} \\ 
\multirow{2}{*}{\textbf{Year}} & \multicolumn{1}{c|}{\multirow{2}{*}{\textbf{Paper}}} & \multicolumn{3}{c|}{\textbf{Mobile Device Emulator(s)}} \\ \cline{3-5}

& & \textbf{Android} & \textbf{iOS} & \textbf{Symbian} \\ \hline
% 2014
\multirow{2}{*}{2014} & Crussell et al. \cite{Crussell2014} & $165{,}426$ & None & None \\ \cline{2-5}
& Lindorfer et al. \cite{Lindorfer2014} & Over $1$M & None & None \\ \hline
% 2015
2015 & Yao et al. \cite{Yao2015} & $651{,}000$ & $68{,}000$ & $10{,}000$ \\ \hline
% 2016
2016 & Narudin et al. \cite{Narudin2016} & $1{,}030$ & None & None \\ \hline
% 2017
\revOne{2017} & \revOne{Chen et al. \cite{Chen2017}} & \revOne{$5{,}000$} & \revOne{None} & \revOne{None} \\ \hline
\end{tabular}
\end{table}

Crussell et al. in~\cite{Crussell2014} carry out ad fraud detection on two sets of Android apps: 
(i) $130{,}339$ apps {crawled} from nineteen {different} marketplaces{;} 
and (ii) $35{,}087$ apps that {probably are} malware {(}provided by {an unspecified} security company{)}. 
The authors apply the following modus operandi for capturing the network traffic of a given app: 
(i) the app is installed on a newly created Android emulator image; 
(ii) {a logger starts to capture the emulator's network traffic; 
and (iii)} the app is first run in the foreground for $60$ seconds, then it runs in the background for another $60$ seconds. 
%The proposed framework is not resilient to encryption, since it needs to inspect the HTTP and DNS data generated by apps{. 
%However, the authors' analysis covers most of the available ad libraries. This means that such libraries do not usually employ any form of encryption for their data transfers, and simply rely on plain HTTP}.

Lindorfer et al. in~\cite{Lindorfer2014} present ANDRUBIS, a publicly available system for the analysis of Android apps. 
For each submitted app, ANDRUBIS applies both static and dynamic analysis techniques in order to study how the app behaves. 
Moreover, during the $240$ seconds of the dynamic analysis, the network traffic generated by the app running in the sandbox is captured for a later analysis focused on high-level protocols (e.g., DNS, HTTP, IRC). 
%Unfortunately, such analysis is not feasible if the app encrypts its network traffic.

Yao et al. in~\cite{Yao2015} carry out app identification on three mobile platforms, namely Android, iOS, and Symbian. 
To capture network traffic from the selected \aRM{mobile} apps, the authors run them in mobile device emulators and trigger their network behavior using UI automation tools (the framework also supports the capturing {at a Wi-Fi access point}, see Section~{\ref{sec:Wi-Fi_access_points}}). 
%Since the system requires to inspect HTTP {messages}, it does not work if an app leverages HTTPS or lower-layers encryption.

Narudin et al. in~\cite{Narudin2016} leverage machine learning to build a classifier being able to detect malware on the Android platform. 
They consider two sets of Android apps: 
(i) the top twenty free (benign) apps available in the Google Play Store; 
and (ii) $1{,}000$ malicious apps from $49$ malware families, provided by the Android Malware Genome Project, as well as $30$ new {(in 2013)} malicious apps from fourteen malware families, collected by the authors. 
To capture {the traffic of the} malicious apps, two online dynamic analysis platforms, namely Anubis and SandDroid, are leveraged {(the traffic of the benign apps is logged on a real device, see Section~\ref{sec:mobile_devices})}. 
%To capture the traffic of the benign apps, the authors employ a traffic logger installed on a real device (see Section~\ref{sec:mobile_devices} for details). For the malicious apps, two dynamic online analysis platforms, namely Anubis and SandDroid, are leveraged. 
%The proposed analysis requires to inspect HTTP messages, hence it cannot take encrypted network traffic {as input}.

\revOne{
Chen et al. in~\cite{Chen2017} carry out app identification targeting the Android platform, and evaluate their solution on $2{,}500$ apps from Google Play Store and seven other third-party Android marketplaces, and $2{,}500$ malicious apps from VirusTotal. 
The considered apps are run in emulators for five minutes, and stimulated via an automatic UI exploration tool; such tool first randomly explore the possible interactions with the app, then heuristically generates new interactions from the ones that have been already explored.
}

\subsection{Wi-Fi Monitors}
\label{sec:Wi-Fi_monitors}

We define as \textit{Wi-Fi monitor} a hardware equipment that is able to scan the Wi-Fi radio bands (i.e., $2.4$ and $5$ GHz) in order to capture the transiting IEEE 802.11 frames. 
The most common configuration consists of a traditional Wi-Fi device (e.g., a Peripheral Component Interconnect (PCI) card in a desktop computer) set in monitor mode, i.e., the device passively listens the nearby Wi-Fi transmissions. 
To effectively eavesdrop the network traffic of a Wi-Fi device, the monitor must be within the target's range of transmission. 
Such range depends on many factors, including the selected radio band, the power of the Wi-Fi module, and the surrounding buildings.

Wi-Fi monitors can be easily deployed at a low cost, and let oversee a good number of Wi-Fi devices. 
However, {there are a few issues that have to be addressed in order to effectively use Wi-Fi monitors for eavesdropping}:
\begin{itemize}
\item In case more than one monitor is deployed, an IEEE 802.11 frame can be eavesdropped by multiple distinct monitors if they are too close to each other. 
When traffic traces provided by different monitors are merged, the duplicate captures must be properly deleted.
\item The timestamp of each eavesdropped IEEE 802.11 frame depends on the internal clock of the Wi-Fi monitor that captured it. 
Since the network data collected by distinct monitors are merged to build a comprehensive dataset, it is crucial to consider internal clock differences between monitors (unless they are synchronized in some way).
\end{itemize}

In Table \ref{tab:Wi-Fi_monitors}, we provide information about the capturing process carried out in the works that employ one or more Wi-Fi monitors to collect the network traffic of mobile devices. 
%{In what follows,} we discuss two additional issues, and how {the works reported in Table \ref{tab:Wi-Fi_monitors}} address them:
%\begin{itemize}
%\item 
\revOne{A typical concern related to the mobile traffic analysis is to filter out from the captured network traces the traffic generated by non-mobile devices. 
In~\cite{Musa2012} and~\cite{Barbera2013} the traffic capturing takes place in a location and time such that the collected traffic only belongs to mobile devices. 
In~\cite{Wang2015} and~\cite{Ruffing2016} the network data generated by non-mobile devices is filtered out since the MAC address of each targeted mobile device is known.}
%\item 
%802.11 frames content can be protected by encryption. 
%This is not a concern for Musa and Eriksson in~\cite{Musa2012} and Barbera et al. in~\cite{Barbera2013}, since their frameworks focus on probe requests, which are sent in clear. 
%The analyses carried out by Wang et al. in~\cite{Wang2015} and Ruffing et al. in~\cite{Ruffing2016} are encryption-agnostic, since they leverage only size and/or timing of the captured 802.11 frames. 
%However, this statement does not hold for Chen et al. in~\cite{Chen2014}, since their analysis requires to access IP payloads.
%\end{itemize}

\begin{table*}
\caption{The surveyed works in which the mobile traffic is captured using one or more Wi-Fi monitors.}
\label{tab:Wi-Fi_monitors}
\centering
\begin{tabular}{|c|l|c|c|m{0.29\textwidth}|c|}
\hline
\textbf{Year} & \multicolumn{1}{c|}{\textbf{Paper}} & \multicolumn{1}{m{0.065\textwidth}|}{\centering \textbf{Number of Wi-Fi Monitors}} & \textbf{Capturing Duration} & \multicolumn{1}{c|}{\textbf{Targeted Population}} & \textbf{Leveraged Information} \\ \hline
% 2012
\multirow{3}{*}{2012} & \multirow{3}{*}{Musa et al. \cite{Musa2012}} & $5$ & $9$ months & People along the streets near an university campus & \multirow{3}{*}{802.11 probe requests} \\ \cline{3-5}
& & $6$ & $12$ hours & People along {fairly} busy roads of a city & \\ \cline{3-5}
& & $7$ & $12$ hours & People along an arterial road of a city & \\ \hline
% 2013
\multirow{3}{*}{2013} & \multirow{3}{*}{Barbera et al. \cite{Barbera2013}} & $5$ & \multicolumn{1}{m{0.15\textwidth}|}{From $40$ minutes to $7$ hours} & People at two political meetings, two Pope's masses, a big mall, and a train station & \multirow{3}{*}{802.11 probe requests} \\ \cline{3-5}
& & $1$ & $6$ weeks & People at an university campus & \\ \cline{3-5}
& & $1$ & Unspecified & People at streets and aggregation places of a city & \\ \hline
% 2014
\multirow{2}{*}{2014} & \multirow{2}{*}{Chen et al. \cite{Chen2014}} & $9$ & $2$ days & People at OSDI 2006 & \multirow{2}{*}{Size and header of IP packets} \\ \cline{3-5}
& & $8$ & $5$ days & People at SIGCOMM 2008 & \\ \hline
% 2015
2015 & Wang et al. \cite{Wang2015} & $1$ & Unspecified & \revOne{One iOS device} & \multicolumn{1}{m{0.20\textwidth}|}{Size and timing of (possibly encrypted) 802.11 frames} \\ \hline
% 2016
2016 & Ruffing et al. \cite{Ruffing2016} & $1$ & $3$ months & Two Android devices, two iOS devices, a Windows Phone device, and a Symbian device & \multicolumn{1}{m{0.20\textwidth}|}{Timing of (possibly encrypted) 802.11 frames} \\ \hline
\end{tabular}
\end{table*}

\subsection{Network Simulators}
\label{sec:simulator}

\revOne{Among the surveyed works, one in particular does not capture the mobile traffic from real or emulated mobile devices, but instead generate it via a software simulator.}
%In this section, we deal with works in which the mobile traffic is not captured from real mobile devices or mobile device emulators, but instead it is generated by a software simulator.
This approach can be useful to study particular deployments of mobile devices that are not observable in a real-world scenario due to technical difficulties, economical constraints, or limits imposed by law. 
If the simulation is realistic, the resulting network traces will be really close to the ones that are collected on real or emulated mobile devices.

Network traffic simulation typically works as follows:
\begin{enumerate}
\item The information about the simulated environment (e.g., geographical extension, buildings, streets) is provided to the system.
\item The information about the actors (e.g., mobile devices, laptops, access points) is provided to the system. 
For each actor, such information includes its technical specifications, its position within the simulated environment, and its network behavior. 
If the actor is used by a {human user}, her sociological characteristics and behavioral patterns are also provided.
\item The points of capturing are positioned within the simulated environment.
\item The network transmissions of the actors are simulated according to realistic physical laws and social dynamics.
\end{enumerate}

Husted and Myers in~\cite{Husted2010} develop a 3D simulation of a large population of mobile devices deployed in a dense metropolis where no other Wi-Fi devices (e.g., access points) are present. 
A fraction of the mobile devices act as trackers (i.e., they are the points of capturing) and scan the air in order to capture Wi-Fi probe requests transmitted by the rest of the population (i.e., the trackees). 
The system properly simulates the propagation of probe requests (which are transmitted in clear) in the environment, and takes into account the diurnal behavior of mobile users (e.g., go to work in the morning, come home in the evening). 
The resulting network traffic dataset is leveraged for \revOne{position estimation} (more details in Section~\ref{sec:position_estimation}).

\section{Targeted Mobile Platforms in Traffic Analysis}
\label{sec:mobile_platforms}

The network traffic of a mobile device depends on its operating system. 
Since each mobile OS has its own implementation of the network protocol stack, it generates data transmissions with peculiar network properties. 
Exploiting such properties is fundamental to devise effective methods for the analysis of mobile traffic. 
For example, the TCP window size scale option (i.e., the value that is negotiated during the TCP three-way handshake to increase the TCP receiver window size beyond $65{,}535$ bytes) is always $16$ for iOS, while it can be either $2$, $4$, or $64$ for Android.
Chen et al. in~\cite{Chen2014} exploit this distinction (together with other differences with regard to network traffic) to successfully recognize whether a target mobile device is running one of those OSes.

In this section, we present the surveyed works according to the mobile platforms they target. 
As shown in Figure~\ref{fig:papers_by_platform}, only thirteen works propose analyses that are platform-independent, i.e., they do not take into account the platform which the targeted mobile devices belong to (it is worth to notice that two of them, namely Lee et al. in~\cite{Lee2011} and Chen et al. in~\cite{Chen2014}, also present other types of analysis that are instead tailored to specific mobile platforms). 
Among the other works, Android is the most targeted mobile platform (\revOne{$45$} works), followed by iOS (fifteen works), Windows Mobile/Phone (four works), and Symbian (two works). 
As shown in Table~\ref{tab:mobile_platforms}, nine works target two mobile platforms, three works target three platforms, and one work even four. 
%{In the remaining of this section,} for each mobile platform we provide an overview with regard to system architecture and applications, then we report the works that target that platform.
%Each of the following sections presents a mobile platform according to the number of works they are involved in. 
We present the targeted mobile platforms, sorted by the the number of works involved: Android in Section~\ref{sec:Android}, iOS in Section~\ref{sec:iOS}, Windows Mobile/Phone in Section~\ref{sec:Windows_Mobile/Phone}, and Symbian in Section~\ref{sec:Symbian}. 
Each of the above-mentioned sections is organized in three parts:
the first part provides an overview of the system architecture; 
the second part describes the apps specific for that platform; 
and the third part reviews the works that carry out traffic analysis targeting that platform.
Finally, we discuss the works that do not belong to any specific mobile platform in Section~\ref{sec:platform-independent_works}.

\begin{table}
\caption{Targeted mobile platforms in the surveyed works.}
\label{tab:mobile_platforms}
\centering
\begin{tabular}{|c|l|c|c|c|c|}
\hline
\textbf{Year} & \multicolumn{1}{c|}{\textbf{Paper}} & \rot{\textbf{Android }} & \rot{\textbf{iOS }} & \rot{\textbf{Symbian }} & \rot{\textbf{Windows Phone }} \\ \hline
% 2010
\multirow{2}{*}{2010} & Falaki et al. \cite{Falaki2010} & \OKB & & & \OKB \\ \cline{2-6}
& Shepard et al. \cite{Shepard2010} & & \OKB & & \\ \hline
% 2011
\multirow{2}{*}{2011} & Lee et al. \cite{Lee2011} & \OKB & \OKB & & \\ \cline{2-6}
& {Rao et al. \cite{Rao2011}} & {\OKB} & {\OKB} & & \\ \hline
% 2012
\multirow{7}{*}{2012} & Baghel et al. \cite{Baghel2012} & \OKB & & & \\ \cline{2-6}
& Ham et al. \cite{Ham2012} & \OKB & & & \\ \cline{2-6}
& {Shabtai et al. \cite{Shabtai2012}} & {\OKB} & & & \\ \cline{2-6}
& {Stevens et al. \cite{Stevens2012}} & {\OKB} & & & \\ \cline{2-6}
& Su et al. \cite{Su2012} & \OKB & & & \\ \cline{2-6}
& Wei et al. \cite{Wei2012} & \OKB & & & \\ \cline{2-6}
& {Wei et al. \cite{Wei2012ProfileDroid}} & {\OKB} & & & \\ \hline
% 2013
\multirow{4}{*}{2013} & Kuzuno et al. \cite{Kuzuno2013} & \OKB & & & \\ \cline{2-6}
& Qazi et al. \cite{Qazi2013} & \OKB & & & \\ \cline{2-6}
& {Rao et al. \cite{Rao2013}} & {\OKB} & {\OKB} & & \\ \cline{2-6}
& {Watkins et al. \cite{Watkins2013}} & {\OKB} & & & \\ \hline
% 2014
\multirow{5}{*}{2014} & Chen et al. \cite{Chen2014} & \OKB & \OKB & & \\ \cline{2-6}
& Coull et al. \cite{Coull2014} & & \OKB & & \\ \cline{2-6}
& Crussell et al. \cite{Crussell2014} & \OKB & & & \\ \cline{2-6}
& Lindorfer et al. \cite{Lindorfer2014} & \OKB & & & \\ \cline{2-6}
& Shabtai et al. \cite{Shabtai2014} & \OKB & & & \\ \hline
% 2015
\multirow{8}{*}{2015} & Chen et al. \cite{Chen2015} & \OKB & & & \\ \cline{2-6}
& Fukuda et al. \cite{Fukuda2015} & \OKB & \OKB & & \\ \cline{2-6}
& Le et al. \cite{Le2015} & \OKB & & & \\ \cline{2-6}
& Park et al. \cite{Park2015} & \OKB & & & \\ \cline{2-6}
& Song et al. \cite{Song2015} & \OKB & & & \\ \cline{2-6}
& Wang et al. \cite{Wang2015} & & \OKB & & \\ \cline{2-6}
& Yao et al. \cite{Yao2015} & \OKB & \OKB & \OKB & \\ \cline{2-6}
& Zaman et al. \cite{Zaman2015} & \OKB & & & \\ \hline
% 2016
\multirow{13}{*}{2016} & Alan et al. \cite{Alan2016} & \OKB & & & \\ \cline{2-6}
& Conti et al. \cite{Conti2016} & \OKB & & & \\ \cline{2-6}
& \revOne{Fu et al. \cite{Fu2016}} & \revOne{\OKB} & & & \\ \cline{2-6}
& {Mongkolluksamee et al. \cite{Mongkolluksamee2016}} & {\OKB} & & & \\ \cline{2-6}
& Narudin et al. \cite{Narudin2016} & \OKB & & & \\ \cline{2-6}
& Nayam et al. \cite{Nayam2016} & \OKB & \OKB & & \\ \cline{2-6}
& Ren et al. \cite{Ren2016} & \OKB & \OKB & & \OKB \\ \cline{2-6}
& Ruffing et al. \cite{Ruffing2016} & \OKB & \OKB & \OKB & \OKB \\ \cline{2-6}
& Saltaformaggio et al. \cite{Saltaformaggio2016} & \OKB & \OKB & & \\ \cline{2-6}
& Spreitzer et al. \cite{Spreitzer2016} & \OKB & & & \\ \cline{2-6}
& Tadrous et al. \cite{Tadrous2016} & \OKB & \OKB & & \\ \cline{2-6}
& {Vanrykel et al. \cite{Vanrykel2016}} & {\OKB} & & & \\ \cline{2-6}
& {Wang et al. \cite{Wang2016}} & {\OKB} & & & \\ \hline
% 2017
\multirow{7}{*}{2017} & {Arora et al. \cite{Arora2017}} & {\OKB} & & & \\ \cline{2-6}
& \revOne{Chen et al. \cite{Chen2017}} & \revOne{\OKB} & & & \\ \cline{2-6}
& \revOne{Cheng et al. \cite{Cheng2017}} & \revOne{\OKB} & & & \\ \cline{2-6}
& {Continella et al. \cite{Continella2017}} & {\OKB} & & & \\ \cline{2-6}
& {Espada et al. \cite{Espada2017}} & {\OKB} & & & \\ \cline{2-6}
& {Malik et al. \cite{Malik2017}} & {\OKB} & {\OKB} & & {\OKB} \\ \cline{2-6}
& Taylor et al. \cite{Taylor2017} & \OKB & & & \\ \hline
\end{tabular}
\end{table}

\begin{figure}
\centering
\includegraphics[scale=0.70]{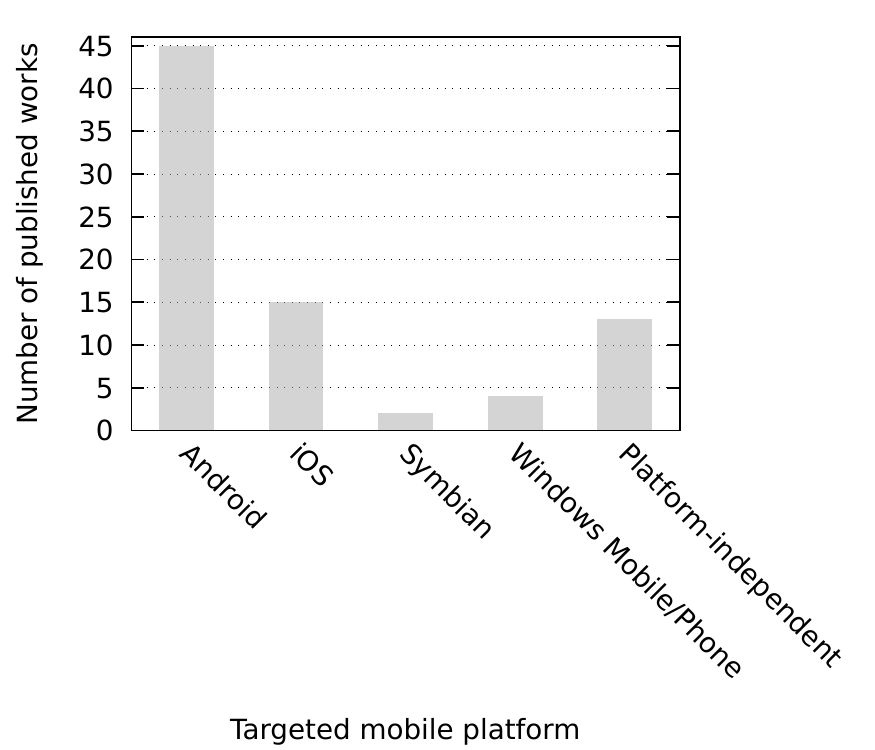}
\caption{Number of published works contributing traffic analysis methods targeting mobile devices, sorted by mobile platform.}
\label{fig:papers_by_platform}
\end{figure}

\subsection{Android}
\label{sec:Android}

Android is an open-source mobile operating system developed by Google. Android is also promoted by the Open Handset Alliance (OHA), a consortium of $84$ firms (including Google, as well as several important actors of the mobile market, like HTC, Samsung, and LG) which is devoted to the development of open standards for mobile devices. 
Android was unveiled at the end of 2007, and the first batch of commercial Android devices appeared a year later. 
Many mobile device manufacturers soon started deploying Android on their flagship products, and the popularity of the operating system rapidly increased. 
Nowadays, Android is the dominant mobile operating system, with a market share of $68.4\%$ in June 2016, according to the statistics reported in~\cite{StatistaMobileOSes2016}.

%The rest of the section is organized as follows. 
%In Section~\ref{sec:Android_architecture}, we provide an overview of Android's system architecture. 
%In Section~\ref{sec:Android_apps}, we deal with Android apps. 
%Finally, in Section~\ref{sec:Android_network_traffic_analysis}, we review the works that carry out traffic analysis targeting the Android platform.

\subsubsection{System Architecture}
\label{sec:Android_architecture}

\begin{figure*}[h]
\centering
 \subfloat[Android.\label{fig:Android_architecture}]{
      \includegraphics[scale=0.35]{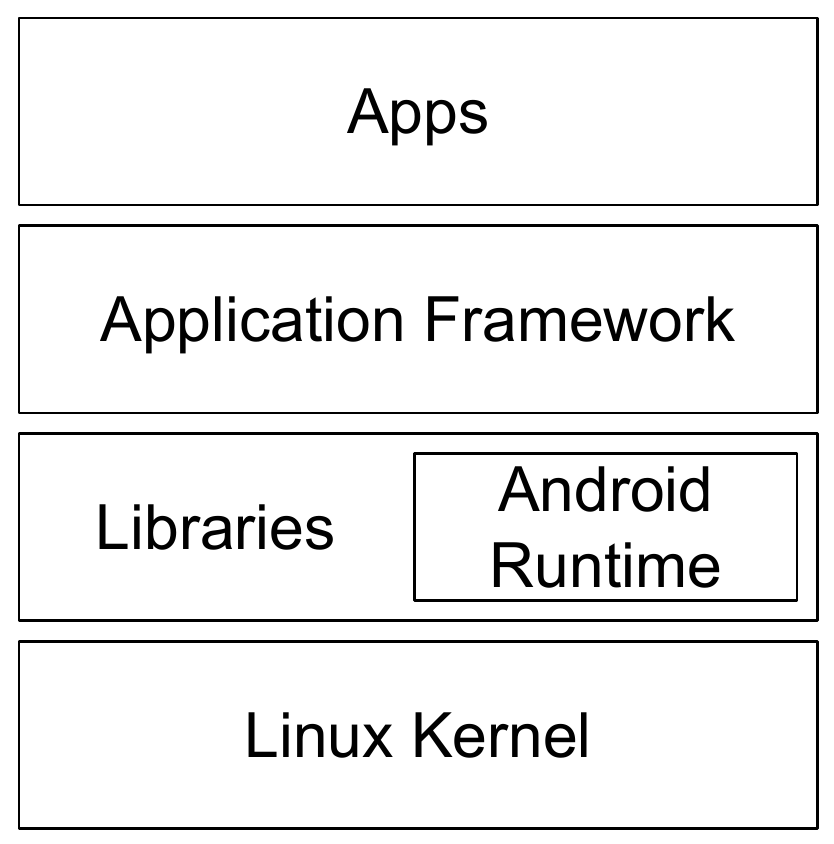}
    }
%\\ %, trim= 0.1cm 0.2cm 1.5cm 1cm]
 \hspace{0.1cm}
 \subfloat[iOS.\label{fig:iOS_architecture}]{
      \includegraphics[scale=0.35]{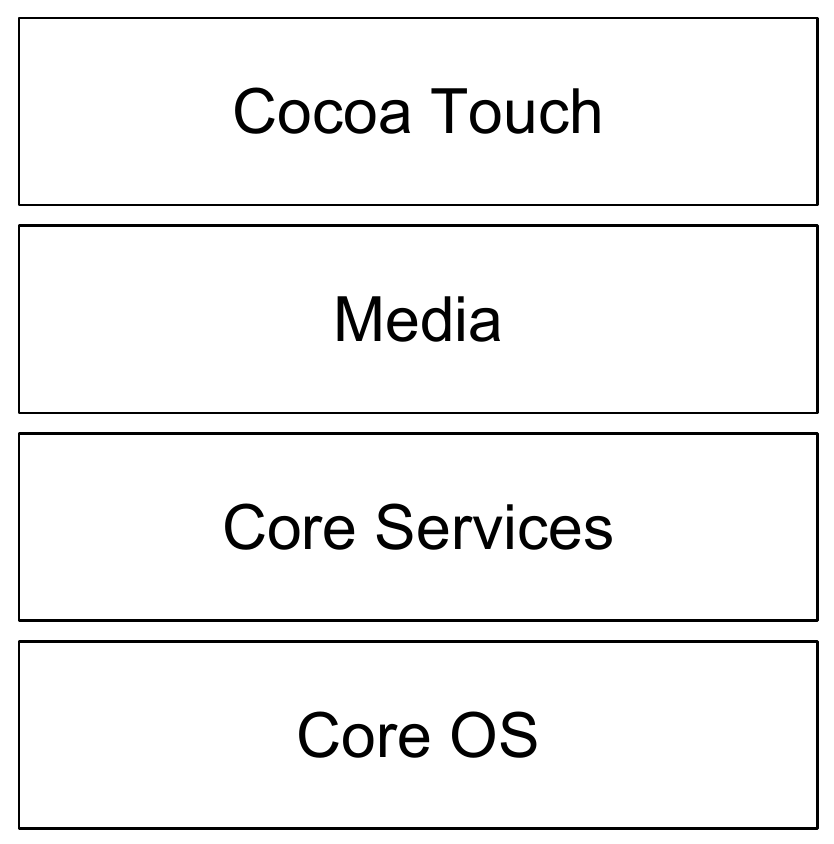}
    }
 \hspace{0.1cm}
 \subfloat[Windows Mobile.\label{fig:Windows_Mobile_architecture}]{
      \includegraphics[scale=0.35]{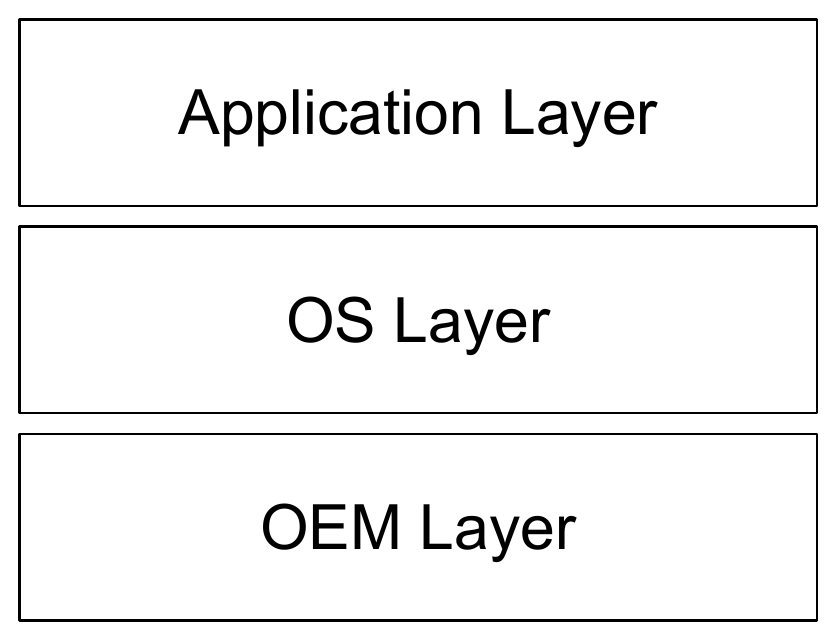}
    }   
    \hspace{0.1cm}
    \subfloat[Windows Phone.\label{fig:Windows_Phone_architecture}]{
    \includegraphics[scale=0.35]{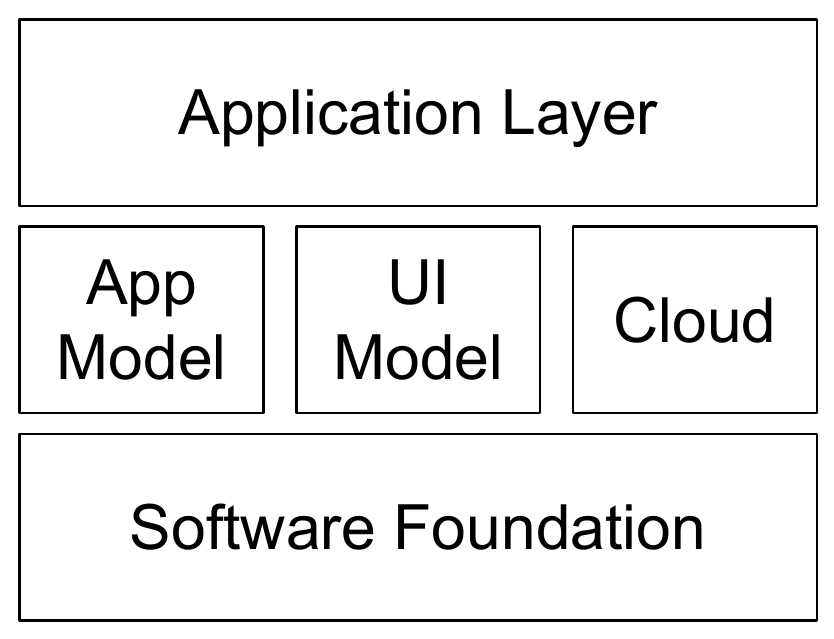}
    }   
 \hspace{0.1cm}
 \subfloat[Symbian.\label{fig:Symbian_architecture}]{
      \includegraphics[scale=0.35]{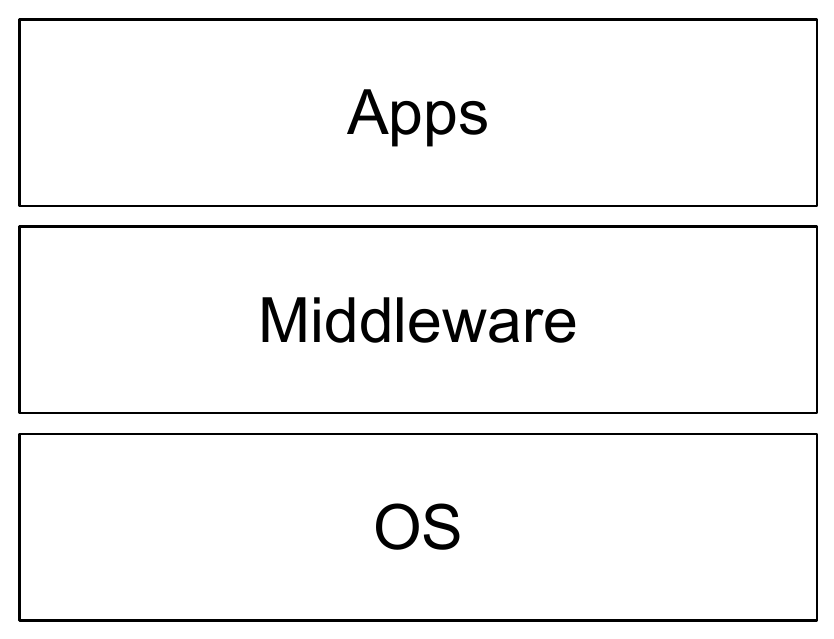}
    }

\caption{System architectures of mobile operating systems.}
\label{fig:sysOS}
\end{figure*}
\begin{comment}

\begin{figure}[h]
\centering
\includegraphics[scale=0.50]{Android_system_architecture.pdf}
\caption{Android's system architecture.}
\label{fig:Android_architecture}
\end{figure}
\end{comment}

As shown in Figure~\ref{fig:Android_architecture}, the architecture of the Android operating system consists of a stack of four abstraction layers:
\begin{enumerate}
\item At the first layer, a Linux kernel provides system services (e.g., memory, power, and process management), preemptive multitasking, a network stack, and drivers for hardware devices (e.g., display, camera).
\item The second layer contains the Android Runtime (ART), which is the application runtime environment. 
{Before Android 5.0 (Lollipop), the execution of Android apps is managed by the Dalvik virtual machine process. 
Android apps and services are typically written in Java and executed in a Dalvik Virtual Machine after being converted from Java Virtual Machine to Dalvik bytecode. %, compiled to bytecode for the Java Virtual Machine, then translated into Dalvik bytecode, which is designed for high efficiency in low resource environments. 
%The Dalvik Virtual Machine is able to run multiple apps simultaneously, providing security, isolation, memory management, and threading support. 
ART adopts a different approach: the Dalvik bytecode is translated into native instructions to be later executed on the runtime environment of the device. 
This solution increases efficiency and reduces power consumption. 
%Together with ART, the second layer 
This layer also includes native libraries that provide several functionalities (e.g., 2D/3D graphics, encryption, SQLite database management).
\item The third layer is the application framework, i.e., the environment that runs and manages Android apps. 
Among the available services that compose such environment, 
(i) the Activity Manager manages app lifecycle and activity stack; 
(ii) the Content Providers allow apps to share data with other apps; 
(iii) the Telephony Manager interfaces with telephony services available on the device; 
(iv) the Notifications Manager prompts the user with notification or alerts raised by apps; and (v) the Location Manager provides the apps with periodic updates regarding the location of the device.}
\item The fourth layer is constituted by the apps, which can be native (e.g., web browser, email client) or provided by a third party.
\end{enumerate}

\subsubsection{Apps}
\label{sec:Android_apps}

{Android apps run in a sandbox and their access to each system's resource is regulated by a specific permission that has to be given by the user.}
Before Android 6.0 (Marshmallow), an app's required permissions are presented to the user at the beginning of the installation process. The user must grant all the required permissions in order to install the app on her device. 
From Android 6.0 on, %In the subsequent versions of Android (i.e., 
permissions are managed individually, and users can grant or revoke each permission according to their usability and security needs.

A third-party Android app is shipped in an APK (Application Package Kit) file, which can be downloaded from the developer's website and manually installed on the device. 
To simplify the process, Android users typically rely on the app stores, or app marketplaces, which are programs that allow them to browse the available apps, as well as to install, update, and remove them. 
Google Play Store (formerly Android Market) is the primary app store installed on Android devices, and hosts over $2{,}500{,}000$ apps~\cite{AppBrain2016} distributed by both Google itself and third-party developers under Google's license and compatibility requirements. 
However, the openness of Android has allowed the birth of a number of other third-party app marketplaces (e.g., GetJar, F-Droid, the app store by Amazon), which release apps under policies different from Google's one.

\subsubsection{Traffic Analysis}
\label{sec:Android_network_traffic_analysis}

Since mobile apps are a key component of the success of the Android operating system, it is not surprising that most of the surveyed works focus their analysis on the network traffic generated by Android apps. 
The achieved results show that it is possible to successfully fingerprint an Android app {(or type of app)}~\cite{Lee2011,Qazi2013,Rao2013,Le2015,Yao2015,Alan2016,Mongkolluksamee2016,Chen2017,Taylor2017}, as well as an action performed by a mobile user on her Android device~\cite{Watkins2013,Park2015,Conti2016,Fu2016,Saltaformaggio2016}. 
In~\cite{Rao2013,Le2015,Song2015,Ren2016,Vanrykel2016,Cheng2017,Continella2017}, it is reported that Android apps extensively leak the PII of mobile users, and the works in~\cite{Stevens2012,Kuzuno2013} highlight that an important role in this phenomenon is played by the embedded ad libraries. 
Regarding mobile advertisement, Crussell et al. in~\cite{Crussell2014} prove that many Android apps trick the advertisement business model in order to let their developers illicitly earn money. 
{In~\cite{Vanrykel2016}, the authors exploit the sensitive identifiers that are present in mobile traffic to fingerprint Android users.} 
In~\cite{Spreitzer2016}, the network statistics of Android's default web browser are leveraged for website fingerprinting. 
Finally, several works aim at detecting malicious Android apps: in~\cite{Wei2012,Zaman2015,Narudin2016,Wang2016,Arora2017}, automated detection frameworks that can be employed by marketplaces and security companies are presented; instead, the authors in~\cite{Shabtai2012,Shabtai2014} present apps that can enable malware detection directly within the mobile devices of the end users.

In light of its market share, we argue that Android is the reference operating system for many mobile users, and Android devices are responsible for an important fraction of the worldwide mobile Internet traffic. 
For this reason, it is not surprising that many works aim at studying the properties of the network traffic generated by Android devices~\cite{Falaki2010,Rao2011,Baghel2012,Wei2012ProfileDroid,Lindorfer2014,Chen2015,Fukuda2015,Nayam2016,Tadrous2016,Espada2017}, as well as the usage habits of Android users~\cite{Falaki2010,Ham2012,Fukuda2015}.
Moreover, Android plays an important role in the works that deal with mobile OS identification~\cite{Chen2014,Ruffing2016,Malik2017}.

\subsection{iOS}
\label{sec:iOS}

{iOS %was formerly called iPhone OS and it 
is a proprietary mobile operating system developed by Apple. 
Such OS is exclusively deployed in Apple's mobile devices. % produced by that company. 
%After two years of development, 
iOS was officially released with the name iPhone OS in 2007. 
%as the operating system that will run in iPhone, Apple's first commercial smartphone. 
%At the end of the same year, iOS was extended to support other Apple devices (e.g., iPod Touch). 
Later, this mobile OS was extended to support other Apple's mobile devices: iPod Touch (Apple's multimedia player) in 2007 and iPad (Apple's tablet) in 2010.} 
According to the statistics reported in~\cite{StatistaMobileOSes2016}, iOS is the second most popular mobile operating system, with a market share of $20.32\%$ in June 2016.

% The rest of the section is organized as follows. In Section~\ref{sec:iOS_architecture}, we provide an overview of iOS's system architecture. In Section~\ref{sec:iOS_apps}, we deal with iOS apps. Finally, in Section~\ref{sec:iOS_network_traffic_analysis}, we review the works that carry out traffic analysis targeting the iOS platform.

\subsubsection{System Architecture}
\label{sec:iOS_architecture}

\begin{comment}
\begin{figure}
\centering
\includegraphics[scale=0.50]{iOS_system_architecture.pdf}
\caption{iOS's system architecture.}
\label{fig:iOS_architecture}
\end{figure}
\end{comment}

As shown in Figure \ref{fig:iOS_architecture}, the architecture of the iOS operating system consists of a stack of four abstraction layers, each providing different services and technologies:
\begin{itemize}
\item The Core OS layer contains: 
(i) the kernel; 
(ii) the device drivers; 
(iii) the interfaces to access the low-level features of the operating system (e.g., file system, memory, concurrency, networking); 
and (iv) the interfaces to access the frameworks that provide several core functionalities (e.g., support for external hardware, Bluetooth, authentication, cryptography, support for VPN tunnels).
\item The Core Services layer includes the mandatory system services for running apps. 
These services provide core functionalities (e.g., account management, location services, cellular network services), as well as high-level features (e.g., P2P, data protection, file sharing, SQLite, XML).
\item The Media layer contains technologies leveraged by developers to implement multimedia content in their apps (i.e., audio, video, and graphic).
\item The Cocoa Touch layer provides the key frameworks which define the appearance of apps and grant the access to high-level system services (e.g., push notifications, touch-based input, multi-tasking).
\end{itemize}

\subsubsection{Apps}
\label{sec:iOS_apps}

Apple distributes the iOS Software Development Kit (SDK), which contains the tools needed to develop, test, and deploy native iOS apps. 
Apps are written in Objective-C or Swift, and leverage the iOS system frameworks. 
Such frameworks provide the interfaces that developers need to write software for the iOS platform. Apps are physically installed on the devices, and run directly on their operating system.

Third-party iOS apps are available to users in the App Store, Apple's digital distribution platform, which was launched in 2008. 
The apps are developed with the iOS SDK and released after Apple's approval. 
The review process aims at assessing that the distributed apps fulfill precise usability and security requirements. 
According to the statistics reported in~\cite{StatistaAppleAppStore2016}, the App Store hosts about two million apps, available for various iOS devices (e.g., iPhone, iPad). 
It is worth to notice that there exist also unofficial marketplaces that distribute iOS apps (e.g., Cydia), but they all require a jailbroken iOS device. 
In a jailbroken iOS device, software vulnerabilities have been exploited to remove the restrictions imposed by Apple on iOS. 
This practice is required to allow the download and installation of apps, extensions, and themes that are unavailable through the official Apple App Store.

\subsubsection{Traffic Analysis}
\label{sec:iOS_network_traffic_analysis}

%Act    2 | OS  3          AP 3 | Dev 3 | Emu 1 | Mnt 2 | Wrd 3
%App    3 | TC  3
%LkgPII 1 | Usg 1

Mobile apps are a fundamental building block of the iOS user experience. 
For this reason, several solutions have been proposed to effectively fingerprint them~\cite{Lee2011,Rao2013,Wang2015,Yao2015}, as well as to detect the interactions between an iOS user and a specific app installed on her mobile device~\cite{Coull2014,Saltaformaggio2016}. 
The authors in~\cite{Rao2013,Ren2016} investigate the disclosure of sensitive information by iOS apps, discovering that many of them leak the PII of the user. 
Regarding OS identification, Coull et al. in~\cite{Coull2014} discriminates between iOS and OS X, while the frameworks presented in~\cite{Chen2014,Ruffing2016,Malik2017} consider iOS among the targeted mobile operating systems. 
Finally, a few works study the properties of the network traffic generated by iOS devices \cite{Shepard2010,Rao2011,Fukuda2015,Nayam2016,Tadrous2016}, and the usage habits of iOS users \cite{Fukuda2015}.

%Chen et al., 2014 \cite{Chen2014} OS - AP - also Android
%Coull and Dyer, 2014 \cite{Coull2014} Act, OS - Dev - iMessage
%Fukuda et al., 2015 \cite{Fukuda2015} TC, Usg - Dev - also Android - logging app
%Lee et al., 2011 \cite{Lee2011} App - Wrd - also Android
%Ren et al., 2016 \cite{Ren2016} LkgPII - Wrd - also Android and Windows Phone
%Ruffing et al., 2016 \cite{Ruffing2016} OS - Mnt - also Android, Windows Phone, and Symbian
%Saltaformaggio et al. \cite{Saltaformaggio2016} Act - AP - also Android
%Shepard et al., 2010 \cite{Shepard2010} TC - Dev - logging app
%Tadrous and Sabharwal, 2016 \cite{Tadrous2016} TC - AP - also Android
%Wang et al., 2015 \cite{Wang2015} App - Mnt
%Yao et al., 2015 \cite{Yao2015} App - Emu, Wrd - also Android and Symbian

%Ruffing et al., 2016 \cite{Ruffing2016} OS - Mnt - also Android, iOS, and Windows Phone
%Yao et al., 2015 \cite{Yao2015} App - Emu, Wrd - also Android and iOS

\subsection{Windows Mobile/Phone}
\label{sec:Windows_Mobile/Phone}

In the early 1990s, %while starting to assume a dominant position in the market of desktop operating systems, 
Microsoft began to develop a new operating system for minimalist computers and embedded systems. 
This OS, later called Windows CE and officially released in 1996, was the basis for the operating systems that make Microsoft enter into the mobile market at the beginning of 2000s. 
The first batch of mobile devices running a Microsoft's OS were Windows Mobile smartphones. 
They became available in 2003 and targeted business users at first. 
The lifecycle of Windows Mobile lasted for approximately seven years, ending in 2010 with the release of its successor, Windows Phone, which had a new user interface and aimed at the consumer market. 
The last iteration of this OS was Windows Phone 8.1, released in 2014 and succeeded by Windows 10 Mobile at the end of 2015. 
Overall, Microsoft's mobile OSes struggle to acquire a relevant market share and seem not to threaten the duopoly by Android and iOS (a trend confirmed by the fact that, according to the statistics reported in~\cite{StatistaMobileOSes2016}, only $1.94\%$ of mobile devices were Windows Phone ones in June 2016).

%In the rest of the section, we will focus on the mobile operating systems developed by Microsoft that appear in the surveyed works, namely Windows Mobile and Windows Phone. We describe their system architectures in Section~\ref{sec:Windows_Mobile/Phone_architecture}, then we treat their apps in Section~\ref{sec:Windows_Mobile/Phone_apps}. Finally, we review the works that carry out traffic analysis targeting these two mobile OSes in Section~\ref{sec:Windows_Mobile/Phone_network_traffic_analysis}.

\subsubsection{System Architectures}
\label{sec:Windows_Mobile/Phone_architecture}

\begin{comment}
\begin{figure}
\centering
\includegraphics[scale=0.50]{Windows_Mobile_system_architecture.pdf}
\caption{Windows Mobile's system architecture.}
\label{fig:Windows_Mobile_architecture}
\end{figure}
\end{comment}

As shown in Figure \ref{fig:Windows_Mobile_architecture}, the architecture of the Windows Mobile operating system follows a stack model, consisting of three abstraction layers:
\begin{itemize}
\item The Original Equipment Manufacturer (OEM) layer is positioned at the bottom of the stack. This layer directly communicates with the underlying hardware components (e.g., microprocessor, RAM, ROM, digital signal processors, input/output modules). 
%It includes the boot-loader, configuration files, drivers, and the OEM Adaptation Layer (OAL), which allows OEMs to adapt the system to a specific platform and consists of functions related to system start-up, interrupt management, profiling, power management, timer, and clock.
\item The Operating System (OS) layer includes the kernel, the core DLLs, the object store (which offers file system, registry, and database persistent storage), multimedia technologies, the device manager, communication and networking services, and the Graphic Windowing and Events Subsystem (GWES). 
The later one provides an interface between the OS, the app, and the user. %, and the operating system.
\item The Application layer consists of the apps, from either Microsoft itself or third parties.
\end{itemize}

\begin{comment}
\begin{figure}
\centering
\includegraphics[scale=0.50]{Windows_Phone_system_architecture.pdf}
\caption{Windows Phone's system architecture.}
\label{fig:Windows_Phone_architecture}
\end{figure}
\end{comment}

As shown in Figure \ref{fig:Windows_Mobile_architecture}, the architecture of the Windows Phone operating system is different, although it maintains the three-levels stack:
\begin{itemize}
\item At the bottom of the stack, the Software Foundation layer includes: (i) the kernel, which manages security, networking, and storage; and (ii) the interfaces that mediate the access to the underlying hardware components (e.g., sensors, camera).
\item The intermediate layer is composed by three elements: (i) the App Model, which is the component providing first-class access to several functionalities that are important for apps (e.g., isolation, licensing, software updates, data sharing); (ii) the UI Model, which manages the user interface of the operating system; and (iii) the components that enable the integration with Microsoft's cloud services.
\item The Application layer includes the frameworks available to developers for building the user interface and logic of their apps.
\end{itemize}

\subsubsection{Apps}
\label{sec:Windows_Mobile/Phone_apps}

Apps for Windows Mobile are developed using the official Software Development Kit (SDK) released by Microsoft, and can be written either in C++ (``native'' apps) or C\#{}/Basic (``managed'' apps). At the end of 2009, Microsoft set up a digital distribution platform, called Windows Marketplace for Mobile, to organize and centralize the release of apps for the Windows Mobile platform. With the advent of Windows Phone, Microsoft started to progressively abandon Windows Mobile, by ending support and closing Windows Marketplace for Mobile in 2012.

Although the SDK and libraries are different, the apps for Windows Phone are written with the same languages used for Windows Mobile apps (i.e., C++, C\#{}, and Basic), plus HTML5 and JavaScript for web-based apps. The official software distribution platform for Windows Phone, called Windows Phone Marketplace (and later renamed Windows Phone Store), was launched by Microsoft at the end of 2010, and subsequently merged into the Windows Store (i.e., Microsoft's universal software marketplace) in 2015.

\subsubsection{Traffic Analysis}
\label{sec:Windows_Mobile/Phone_network_traffic_analysis}

Only a few works we survey target Microsoft's mobile OSes. 
Falaki et al. in~\cite{Falaki2010} deploy a custom logging app on Windows Mobile devices to capture their network traffic and study its properties. 
{Ren et al. in~\cite{Ren2016} investigate PII leaks through network traffic generated by} devices running several mobile operating systems, including Windows Phone. 
Finally, the works in~\cite{Ruffing2016,Malik2017} deal with mobile OS identification, and Windows Phone is among the operating systems that the proposed frameworks are able to recognize.

%Falaki et al., 2010  TC, Usg - Dev - also Android - logging app
%Ren et al., 2016 \cite{Ren2016} LkgPII - Wrd - also Android and iOS
%Ruffing et al., 2016 \cite{Ruffing2016} OS - Mnt - also Android, iOS, and Symbian

\subsection{Symbian}
\label{sec:Symbian}

Symbian is a mobile operating system originally developed for PDAs in 1998, and subsequently moved into cellphones and smartphones in the following years. 
{Running exclusively on ARM processors, Symbian requires an additional middleware to form a complete operating system and to provide a user interface.} 
During the 2000s, Symbian became the most popular mobile OS, since many mobile manufacturers, particularly Nokia, chose it to power their devices. 
A non-profit organization, the Symbian Foundation, was created in 2008 to drive the development of the operating system and promote the adoption of Nokia's middleware, namely S60.
However, with the advent of Android and iOS, and Nokia adopting Windows Phone for its devices, the popularity of the Symbian platform rapidly decreased. 
The Symbian Foundation closed in 2010, and the development of the OS ended in that period. 
According to the statistics reported in~\cite{StatistaMobileOSes2016}, Symbian is almost disappeared, with a market share of only $2.22\%$ in June 2016.

%The rest of the section is organized as follows. In Section~\ref{sec:Symbian_architecture}, we provide an overview of Symbian's system architecture. In Section~\ref{sec:Symbian_apps}, we deal with Symbian apps. Finally, in Section~\ref{sec:Symbian_network_traffic_analysis}, we review the works that carry out traffic analysis targeting the Symbian platform.

\subsubsection{System Architecture}
\label{sec:Symbian_architecture}

\begin{comment}

\begin{figure}
\centering
\includegraphics[scale=0.50]{Symbian_system_architecture.pdf}
\caption{Symbian's system architecture.}
\label{fig:Symbian_architecture}
\end{figure}
\end{comment}

As shown in Figure \ref{fig:Symbian_architecture}, the architecture of the Symbian operating system consists of a stack of three abstraction layers:
\begin{itemize}
\item The OS layer is the core of a Symbian system, and contains the kernel, which provides the interfaces to access the underlying hardware, and several essential services (e.g., communications, text and data handling, graphics).
\item The Middleware layer provides a software platform which consists of higher-level generic APIs available to the apps of the upper layer. These APIs include the native UI frameworks, as well as frameworks for app lifecycle, higher-level protocols, and data handling. Different platforms are not compatible, i.e., apps developed for a platform cannot run on the others.
\item The Apps layer includes apps that interact with the user and background services that provide functionalities to the apps.
\end{itemize}

\subsubsection{Apps}
\label{sec:Symbian_apps}

As we already explained in Section~\ref{sec:Symbian_architecture}, all Symbian devices share a common core, on top of which different software platforms are built to provide an execution environment for user apps (actually implementing the Middleware layer shown in Figure \ref{fig:Symbian_architecture}). 
Backed by different groups of mobile device manufacturers, three software platforms were created for Symbian:
\begin{itemize}
\item S60 (Series 60) was the most popular Symbian platform, officially supported by the Symbian Foundation and deployed in the products of several mobile device manufacturers, including Nokia, Samsung, and LG. 
S60 was able to run apps developed in Java MIDP, C++, Python, and Adobe Flash. 
Third-party developers had to distribute their apps by either releasing them in the marketplaces (the most important stores were run by Nokia and Opera Software), or pre-installing them in the mobile devices of some manufacturers.
\item UIQ (User Interface Quartz) was developed by UIQ Technology, and supported by Sony Ericsson and Motorola. 
The platform was able to run native apps written in C++ using the Symbian/UIQ Software Development Kit (SDK), as well as Java apps. The development of UIQ stopped in 2008, when the Symbian Foundation was established and chose S60 as its reference Symbian platform.
\item MOAP (Mobile Oriented Applications Platform) was the platform chosen by NTT DoCoMo, a major Japanese cellular operator, for its FOMA (Freedom of Mobile Multimedia Access) service, which was a W-CDMA-based 3G telecommunications service. Supported by a few Japanese companies, like Fujitsu and Sharp, MOAP did not spread outside of Japan. It was not an open development platform, i.e., there were no third-party apps.
\end{itemize}

\subsubsection{Traffic Analysis}
\label{sec:Symbian_network_traffic_analysis}

\revOne{Only two works target the Symbian operating system.} 
The first work by Ruffing et al.~\cite{Ruffing2016} deals with the identification of the OS of mobile devices, and Symbian is among the operating systems that the proposed framework is able to recognize. 
The second work by Yao et al.~\cite{Yao2015} presents an app identification system which is trained and evaluate on, among others, $10{,}000$ Symbian apps from the Nokia OVI Store.

\subsection{Platform-independent Works}
\label{sec:platform-independent_works}

We survey several works in which the analysis performed on the network traffic is generic, which means that it is not specific for a particular mobile platform. 
Some of these works leverage the 802.11 probe requests that are sent by mobile devices of any platform to discover if an already known Wi-Fi access point is nearby: 
in~\cite{Barbera2013}, sociological information is inferred from the probe requests of a population of mobile users, while in~\cite{Chen2012,Musa2012} probe requests are exploited to estimate a mobile device's geographical position and movements, respectively. 
Besides, some works simply group together mobile devices of different platforms and consider them as a unique category. 
In~\cite{Chen2012,Wei2017}, the properties of the network traffic generated by mobile devices in a campus Wi-Fi network are studied. 
In~\cite{Afanasyev2010,Maier2010,Gember2011}, mobile and non-mobile devices are compared on network traffic properties and users' usage habits, and the same is done in~\cite{Finamore2011} but limited to the YouTube service. 
Finally, Verde et al. in~\cite{Verde2014} present a user fingerprinting method that is successfully used to recognize the presence of some target mobile users within a small test network and a large Wi-Fi one.

%Afanasyev et al., 2010 \cite{Afanasyev2010} TC, Usg - AP, Wrd - desktop computers, laptops, mobile devices
%Barbera et al., 2013 \cite{Barbera2013} Soc - Mnt - probe requests
%Chen et al., 2012 \cite{Chen2012} TC - Wrd - smart MHDs in campus Wi-Fi network
%Finamore et al., 2011 \cite{Finamore2011} TC, Usg - Wrd - YouTube
%Gember et al., 2011 \cite{Gember2011} TC, Usg - AP - handheld vs non-hanheld in campus Wi-Fi network
%Husted and Myers, 2010 \cite{Husted2010} Pos - Sim - probe requests
%Maier et al., 2010 \cite{Maier2010} TC, Usg - Wrd - DSL traces
%Musa and Eriksson, 2012 \cite{Musa2012} position estimation, wi-fi monitors, probe requests
%Verde et al., 2014 \cite{Verde2014} Usr - Wrd - NetFlow records

%\rikiTODO{!!!!!!Check plagiarism of previous part!!!!!!!!!!!!!!}

\section{Models and Methods in Traffic Analysis Targeting Mobile Devices}
\label{sec:models_methods}

In the literature, researchers leverage several different models and methods to carry out the analysis of the network traffic of mobile devices. 
%In the literature, researchers leverage many different methods and models in order to carry out their analysis of network traffic of mobile devices. 
The application of such instruments is strictly correlated with the point of listening (as described in Section~\ref{sec:points_of_capturing}), as well as the information extracted from the captured traffic. 
%The application of such methods is strictly correlated with the point of listening (as described in Section~\ref{sec:XXXX  }) and the information provided by the traffic capture.
Being able to capture packet-level unencrypted data (e.g., HTTP messages) constitutes the most optimistic scenario since all the information enclosed in the network packets (e.g., the URLs that has been contacted) is available in clear. 
%Capturing packet-level unencrypted traffic (e.g., HTTP) is the most optimistic scenario since all the information enclosed in a network packet is available in clear (e.g., URL, payload, IP addresses).
Under these conditions, it is possible to effectively apply Deep Packet Inspection (DPI) techniques. 
%Under this settings, it is possible to apply Deep Packet Inspection (DPI) techniques.
Unfortunately, such types of analysis cannot be carried out if the information available in the network traffic is affected by the following factors: 
%Unfortunately, this method cannot be applied whether two factors affect the information available:
(i) the presence of encrypted traffic at different layers (i.e., IPsec at network layer, SSL/TLS at transport layer); and
%(i) the presence of encrypted traffic at different layers (i.e., SSL/TLS on transport and IPsec on network-layers);
(ii) the application of traffic aggregation or sampling (e.g., NetFlow, IPFIX). 
%(ii) the traffic aggregation or sampling (e.g., Netflow, IPFIX).
For this reason, network traffic analysts have to rely on mathematical models and methods to cope with the lack of available information whenever such difficulties are in place. 
According to their goals, researchers can still apply techniques to analyze the mobile network traffic.
For example, it is possible to rely on statistical-based techniques to perform traffic characterization. 
%%For this reason, network analysts have to rely on mathematical models and methods and models to cope with the lack of information available.
Among those techniques, machine learning provides several approaches to classify and cluster network traffic.
%Among those methods, machine learning techniques play an important role in classifying and clustering network traffic.

In this section, we provide a deeper insight into the models and methods leveraged in the state of the art to devise solutions for the analysis of mobile devices' network traffic. 
In particular, we provide an overview of the methodology followed to perform traffic analysis with machine learning in Section~\ref{sub:elements}, describing each step in the procedure.
In Section~\ref{sec:machine_learning}, instead, we deal with application of machine learning to several types of traffic analysis targeting mobile devices.

\revOne{
\subsection{Overview and Elements on Machine Learning}
\label{sub:elements}

Machine learning (ML) is the branch of artificial intelligence that studies algorithms that can be used to learn from and make predictions on data. 
Such algorithms are typically adopted to solve problems for which a traditional algorithmic solution (i.e., a finite sequence of instructions) is hard, if not impossible to find. 

In this section, we provide an introduction on the basic concepts of machine-learning-based analysis.
This is also mean to be a guideline that reports the principal steps that have to be followed to properly train a machine learning model (summarized in Figure~\ref{fig:MLworkflow}). 
For each step, we define its purpose and describe the different methods used by the surveyed works. % to carry it out.

\begin{figure}[ht]
\centering
\includegraphics[width=0.49\textwidth]{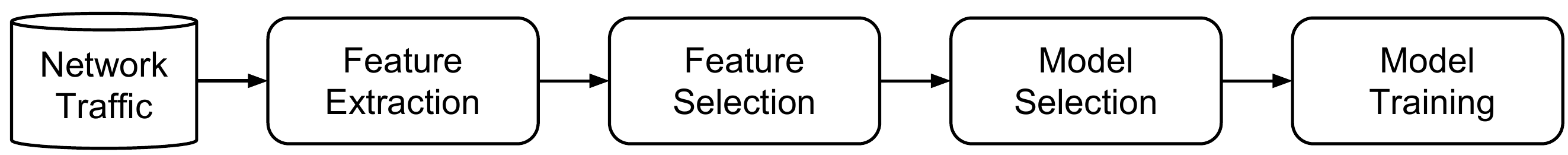}
 
\caption{Procedure for a machine learning analysis.}
\label{fig:MLworkflow}
\end{figure}

\subsubsection{Features extraction from network traffic}
\label{sub:feature_extraction}

As a first step, the network traffic collected from a capturing point (See Section~\ref{sec:points_of_capturing}) has to be transformed into a format that can be used by a machine learning technique.
This step is called \textit{feature extraction} and typically takes a network entity as an input (e.g., packets, flows) and provides a fixed-sized vector for features that represent the properties of such entity.
A good feature extraction has to enclose in such output vector as much as possible the information available from the network traffic collected according to the final target of the analysis.
Some examples of feature extraction methods used in the surveyed work are:
\begin{itemize}
\item Time Series (TS) are sequence of entities ordered by the time in which they occur~\cite{hamilton1994time}. 
As an example time series, a flow of packets can be represented as a sequence of packet sizes. 
Since time series related to network traffic do not have a fixed length, an additional transformation is necessary to obtain a fixed size vector. 
\item Statistical Feature Extraction (SFE) consists in applying statistical primitives (e.g., mean, standard deviation) along the data dimensions. 
As an example, from a sequence of packet sizes it is possible to extract a feature vector in which each element corresponds to a statistical primitive.
\item Histograms are used for feature extraction to represent the distribution of values, by aggregating them into a fixed number of bins. 
Each bin corresponds to a range of values and it counts the occurrences of values within that range. 
\item Bag-of-Word (BoW) is a model that first identify distinguishable entities (i.e., words) and then counts the frequency of occurrence of each specific entity in the input data. 
This model is broadly used in document classification. 
\end{itemize}
Moreover, in Section~\ref{sub:othermethods} we describe into detail two other feature extraction and analysis methods that rely on dictionaries and graphs. 
At the end of the feature extraction, we obtain a dataset that can be seen as a matrix in which each row corresponds to an entity (a.k.a., observation) and a column to a feature (a.k.a., dimension).
Optionally, it is possible to normalize the whole dataset by column to a specific interval of values (e.g., from 0 to 1). 
}

\revOne{
\subsubsection{Feature Selection}
\label{sub:feature_selection}
Given the dataset obtained from the previous step, it is often necessary to select a subset of features which are meaningful to describe the reality to be modeled.
Indeed, a high-dimensional dataset may present a phenomena called \textit{curse of dimensionality}.
The target of a model is to identify similarities within observations belonging to the same class, and such similarities need to be statistically significant to not mislead during the training of a model. 
%In order to prove , 
Hence, the amount of observations needed to prove that a similarity is not due to sparsity grows exponentially with the number of features of the dataset. 
Thus, a dataset with an unbalanced amount of observations among classes (e.g., the unbalance between the number of malicious and benign traffic traces in malware analysis) and a high-dimensionality may be not able to prove whether the features are statistically significant. %, thus avoiding the curse of dimensionality.
Moreover, the high amount of features also increase the computational cost to train a model.
In order to cope with that, it is possible to apply feature selection (or feature reduction) techniques, among which:

\begin{itemize}
\item Analysis of Variance (ANOVA)~\cite{fisher1925statistical} is a statistical method to analyze the difference between the mean values of features in a dataset.
ANOVA evaluates the statistical significance of features in the dataset running a statistical test (i.e., a $t$-test that aims to reject the null hypothesis) generalized to work on multiple means. 

\item Principal Components Analysis ({PCA})~\cite{dunteman1989principal} and Independent Component Analysis (ICA)~\cite{iversen1987analysis} are used in signal processing to identify the components of a signal, but they are also used for feature selection tasks. 
PCA aims to find and remove correlated features in order to reduce the dimensionality of a dataset. 
To do so, PCA decomposes the feature set into a set of orthogonal components that show a high variance (i.e., covariance matrix's eigenvectors of the original dataset). 
Similarly to PCA, ICA removes correlated features by finding statistically independent components.

\item Relative Mutual Information (RMI) measures the mutual dependence between couples of features. In practice, RMI quantifies the amount of information expressed by a feature evaluating the conditional entropy of that feature given another one~\cite{peng2005feature}.
\end{itemize}
}

\revOne{
\subsubsection{Model Selection and Training}
Once the features set is defined, the learning process can proceed with the actual training of a model.
A model has to be selected taking into account two main factors: (i) the goal of the analysis; and (ii) the availability of labeled data.
For this reason, it is necessary to carry out some preliminary analysis to understand which model is the most suitable for the available data. 
These factors also determine whether the outcome result has to be a binary or multi-labels classification, or a clustering task. 
Once the candidate models have been selected according the available dataset and the goal that has to be achieved, it is possible to evaluate the performance of such models varying their hyper-parameters. 
This procedure is called \textit{model selection}, and it is a good practice to carry it out before training a final model. 
It is possible to run this evaluation on a dataset portion (i.e., training set) leaving the remainder of the dataset (i.e., testing set) for the evaluation of the final model. 
For the sake of simplicity, here we only mention the so called \textit{holdout method}, but we describe in detail this and other dataset partitioning methods in Section~\ref{sec:dataset_partitioning}.

In what follows, we introduce two main machine learning approaches used in mobile traffic analysis: supervised and unsupervised learning.

\paragraph{Supervised Learning}
The most frequently used machine learning methods in traffic analysis follow the supervised learning paradigm.
This paradigm permit to extract knowledge from labeled datasets.
This category of learning also comprehends semi-supervised learning, which allow the presence of unlabeled data. 
\begin{itemize}

\item \textit{Support Vector Machines (SVM)} classifier is a method that aim at separate observations into classes relying on a hyperplane (which is identified by three observations, i.e., vectors). 
Since observations may not be separable by an hyperplane (i.e., \textit{linear SVM}), it is possible to apply kernel functions, such as \textit{Polynomial Kernel (PK)} and \textit{Radial Basis Function (RBF)}, to make the observations linearly separable by projecting them to a high dimensional feature.

\item \textit{Decision Trees} classifier is a simple method that relies on tree structures. 
The components of these structures are: (i) nodes, each representing a condition on a feature; (ii) leaves, each representing a feature vector.  
Among the possible algorithms to generate a decision tree, the most popular are \textit{ID3}, \textit{C4.5}, and \textit{J48}.

\item \textit{Ensemble Methods} combine the results of many weak classifiers into a more powerful one. 
This combination of results follows a strategy such as Boosting or Bootstrap aggregation. An example of ensemble methods is \textit{Random Forest (RF)} classifier which usually relies on decision trees as weak learners.

\item The \textit{Probabilistic Learning} methods leverage Directed Acyclic Graphs (DAGs) to represent the probability between variables, each represented by a node. 
For each edge that connects a node $A$ from to a node $B$, it is associated with a probability function that, given an input, outputs the probability to have a transition from $A$ to $B$. 

\textit{Bayesian network (BN)} is an example of method that assumes each variable is conditionally independent from the others, i.e., \textit{Na\"ive Bayes (NB)}.
As another example, \textit{Hidden Markov Model (HMM)} assumes a conditional probability distribution among hidden variables (i.e., unobservable states) respecting the Markov property. 
HMM is often applied in pattern and speech recognition.

\item \textit{Regression} methods consist of explaining through a function the relation between a dependent variable (i.e., class) given an independent variable (i.e., data). 
An example is the \textit{Linear Regression (LiReg)}, which aims to fit the data with a linear model.
Another example is the \textit{Logistic Regression (LoReg)} which is a linear regression that gives in output not a continuous value, but a binary value instead.

\item \textit{$k$-Nearest Neighbors (kNN)} method simply consists of mapping labeled observations into a space. 
During the classification of an unlabeled example $X$, the algorithm assigns the most recurrent class label among the $k$ nearest neighbors of $X$ (with $k$ fixed beforehand).

\item \textit{Artificial Neural Networks (NN)}
 are learning methods based on nodes (i.e., neurons) connected to each other and arranged into layers. 
 Each neuron applies a propagation function to its incoming connections and generates an output. 
 The connections between neurons are weighted, and during the training such weights are updated using a back-propagation algorithm (e.g., gradient descent). 
 Traditional neural networks have three kinds of neuron, according to the location (i.e., input, hidden, or output layer).
 In particular, a neural network without cycles (i.e., feed-forward) with more than one layer of hidden neurons is called \textit{Multilayer Perceptron (MLP)}. 
 Some other variants that allow cycles and multiple layers of hidden neurons are \textit{Convolutional (CNN)} and \textit{Recurrent (RNN)} neural networks.
 Another example of neural network is the \textit{Neural-Fuzzy classifier (NFz)}.
 Neural networks can be also used for unsupervised and reinforced learning.

\end{itemize}

\paragraph{Unsupervised Learning}
This machine learning approach aims to group observations into clusters according to their similarity. 
This task is called \textit{unsupervised} because it does not need any previous knowledge about observations.
Indeed, unsupervised learning is useful in goals of analysis that do not rely on a labeled datasets, such as traffic characterization. 

Clustering methods rely on distance metrics to measure the similarity between observations. 
Such metrics are particularly useful when network entities (e.g., packets, flows) and sets are involved. 
Examples of distance metrics used in the surveyed work are \textit{Euclidean} and \textit{Compression} distances, the optimal warping path of \textit{Dynamic Time Warping (DTW)}~\cite{Moeller2007}, and \textit{Jaccard's Index (JI)}~\cite{real1996probabilistic}.

\textit{Hierarchical Clustering (HC)} is a clustering method that aims to build a hierarchy of clusters according to a specific strategy. 
The \textit{agglomerative} strategy starts with a cluster for each observation and iteratively merges clusters according to their similarity (keeping trace of the hierarchy among clusters), until all observations are aggregated in a single cluster (a.k.a., Bottom-up strategy).
The hierarchy of clusters resulting from this process can be represented through a dendrogram.
By setting a \textit{cut-off} parameter, it is possible to cut the resulting hierarchy at a specific height to obtain a set of clusters.

\textit{$k$-Means Clustering (kMeans)} is a method that aims to group observations into $k$ clusters. 
Such method is initialized with a preliminary division of the observations according to their position in the feature space. 
Then, it iterates on each observation alternating two steps: ``assignation'' and ``update''.
The ``assignation'' step associates an observation to the cluster whose centroid (i.e., mean of all cluster elements) is the nearest (i.e., having the shortest distance centroid-observation).
The ``update'' step recomputes the centroids of clusters given the observation's new assignation.
The algorithm converges when the assignation step does not move any observation from the previous iteration.
}

\subsection{Machine Learning Applications by Goal of the Analysis}
\label{sec:machine_learning}

\revOne{Machine learning is effectively applied in most of the works we survey to perform such type of analysis targeting mobile devices. 
In the following sections, we review the applications of machine learning according to the popularity of the goal of the analysis (in the same order used in Section~\ref{sec:goals}).}
%In particular, we report the machine learning techniques applied to traffic characterization (Section~\ref{sec:machine_learning_traffic_characterization}), app identification (Section~\ref{sec:machine_learning_app_identification}), malware detection (Section~\ref{sec:machine_learning_malware_detection}), PII leakage detection (Section~\ref{sec:machine_learning_PII_leakage_detection}), user action identification (Section~\ref{sec:machine_learning_user_action_identification}), operating system identification (Section~\ref{sec:machine_learning_OS_identification}), position estimation (Section~\ref{sec:machine_learning_position_estimation}), user fingerprinting (Section~\ref{sec:machine_learning_user_fingerprinting}), ad fraud detection (Section~\ref{sec:machine_learning_ad_fraud_detection}), tethering detection (Section~\ref{sec:machine_learning_tethering_detection}), and website fingerprinting (Section~\ref{sec:machine_learning_website_fingerprinting})}.

\subsubsection{Traffic Characterization}
\label{sec:machine_learning_traffic_characterization}

Nayam et al. in~\cite{Nayam2016} study the network behavior of $63$ Android and $35$ iOS free apps. To find similarities between the apps, the authors apply the following methodology:
\begin{itemize}
\item The TCP and UDP traffic of each analyzed app is partitioned according to the type of domains to which it is related: (i) advertisement; (ii) tracking; (iii) popular services (e.g., Google, Facebook); and (iv) other domains.
\item For each app, the following attributes are computed: (i) total number of sessions; (ii) session rate for each type of domain; and (iii) percentage of sessions for each type of domain.
\item $k$-means clustering is applied to group together the apps that show a similar network behavior.
\end{itemize}
In Table \ref{tab:machine_learning_traffic_characterization}, we report the resulting app classification.

\begin{table*}
\caption{Classification of the apps according to their network behavior (Nayam et al.~\cite{Nayam2016}).}
\label{tab:machine_learning_traffic_characterization}
\centering
\begin{tabular}{|c|m{0.55\textwidth}|c|}
\hline
\textbf{Cluster} & \multicolumn{1}{c|}{\textbf{Network behavior}} & \textbf{Classification} \\ \hline
0 & Excessive ad-related traffic, and excessive number of sessions & \multirow{5}{*}{Suspicious} \\ \cline{1-2}
1 & Excessive ad- and tracking-related traffic, and excessive number of sessions & \\ \cline{1-2}
2 & Excessive ad-related traffic, excessive traffic related to other domains, and excessive number of sessions & \\ \cline{1-2}
3 & Excessive tracking-related traffic, and excessive number of sessions & \\ \cline{1-2}
4 & Excessive ad-related traffic, but very low network activity & \\ \hline
5 & High portion of traffic related to popular services, but very low use of them and very low network activity & \multirow{2}{*}{Innocuous} \\ \cline{1-2}
6 & High use of popular services, but very low network activity & \\ \hline
7 & High portion of traffic related to other domains, but very low tracking-related traffic and very low use of popular services & \multirow{3}{*}{Potentially suspicious} \\ \cline{1-2}
8 & High portion of traffic related to other domains, but very low use of them and very low portion of tracking-related traffic & \\ \cline{1-2}
9 & High portion of traffic related to other domains, but very low use of them & \\ \hline
\end{tabular}
\end{table*}

\subsubsection{App Identification}
\label{sec:machine_learning_app_identification}

Supervised learning is applied for app identification in~\cite{Qazi2013,Le2015,Wang2015,Alan2016,Mongkolluksamee2016,Taylor2017}. 
The methodology followed to build the app classifier is the same in all such works: 
(i) the network traffic of the selected mobile apps is captured; 
(ii) for each mobile app, feature vectors are extracted from its network traces and labeled with the name{/type} of that app; 
and (iii) the chosen classifier is trained on the labeled feature vectors. 
In Table~\ref{tab:machine_learning_app_identification}, \revOne{we report the leveraged features and employed classifiers for each of the surveyed works in which machine learning is applied for app identification}. 
Moreover, in the following we provide additional information about the reinforced-learning-based method that Taylor et al. propose in~\cite{Taylor2017} to cope with the problem of \textit{ambiguous} networks flows, i.e., network flows that are not useful in order to uniquely identify an app. 
Generated by third-party libraries (e.g., ad libraries) which can be embedded in different apps, such flows hinder the training of a classifier. 
To tackle the problem, the authors propose a method composed of four stages: (i) a preliminary classifier is trained using a preliminary training set; (ii) the preliminary classifier is evaluated using a preliminary testing set; (iii) samples which are wrongly labeled by the preliminary classifier are re-labeled as ``ambiguous''; and (iv) a reinforced classifier is trained using the re-labeled dataset, including ``ambiguous'' as a new class.
%As an additional contribution for the extended version of their paper in~\cite{Taylor2016}, Taylor et al. in~\cite{Taylor2017} proposed a reinforced learning-based method to cope with the problem of ambiguous flows. Network flows are ambiguous if they are not useful to uniquely identify an app i.e., they are generated by third-party libraries (e.g., ad libraries) and shared by different apps. Due to their nature, such flows hinder the training of a model, thus the authors propose a method composed of four stages to tackle such problem: (i) training of a preliminary model with a preliminary training set; (ii) evaluate the preliminary model with a preliminary testing set (iii) re-label as \textit{ambiguous} the samples which are labeled wrongly by the preliminary classifier; and (iv) training of a reinforced classifier with the re-labeled dataset (including \textit{ambiguous} as a new class).

\begin{table*}
\caption{The surveyed works in which machine learning is applied for app identification.}
\label{tab:machine_learning_app_identification}
\centering
\begin{tabular}{|>{\hspace{\tabred}}c<{\hspace{\tabred}}
|>{\hspace{\tabred}}l<{\hspace{\tabred}}
|m{0.5\textwidth}
|>{\hspace{\tabred}}l<{\hspace{\tabred}}
|}
\hline
\textbf{Year} & \multicolumn{1}{c|}{\textbf{Paper}} & \multicolumn{1}{c|}{\textbf{Features}} & \multicolumn{1}{c|}{\textbf{Classifier}} \\ \hline
% 2013
2013 & Qazi et al. \cite{Qazi2013} & \revOne{{\tiny N/A}} & C5.0 decision tree \\ \hline
% 2015
\multirow{2}{*}{2015} & Le et al. \cite{Le2015} & $84$ network-level features \revOne{belonging to five typologies (packet length statistics, payload length statistics, inter-arrival time statistics, bursts timing, overall flow statistics, and TCP flags)} & Linear SVM \\ \cline{2-4}
& Wang et al. \cite{Wang2015} & \revOne{Average and standard deviation of the size/time of all the transmitted/received 802.11 frames, and average size/time of the low 20\%, mid 60\%, and high 20\%} & Random forest \\ \hline
% 2016
\multirow{4}{*}{2016} & \multirow{3}{*}{Alan et al. \cite{Alan2016}} & Burst sizes (rounded to the nearest $32$ bytes) of the first $64$ IP packets & Based on Jaccard index \\ \cline{3-4}
& & Sizes of the first $64$ IP packets (using the minus sign for incoming packets) & Gaussian NB \\ \cline{3-4}
& & Sizes of the first $64$ IP packets (using the minus sign for incoming packets), modified by term frequency -- inverse document frequency transformation and normalization & Multinomial NB \\ \cline{2-4}
& Mongkolluksamee et al. \cite{Mongkolluksamee2016} & $35$ graphlet- and $24$ histogram-based features extracted from network information \revOne{(source/destination IP address, protocol, source/destination port, size)} & Random forest \\ \hline
% 2017
2017 & Taylor et al. \cite{Taylor2017} & \revOne{$18$} statistics \revOne{(minimum, maximum, mean, median, absolute deviation, standard deviation, variance, skew, kurtosis, percentiles from 10\% to 90\%, and number of values)} computed on the \revOne{transmitted/received/both} IP packet sizes within TCP flows & Random forest \\ \hline
\end{tabular}
\end{table*}

\subsubsection{PII Leakage Detection}
\label{sec:machine_learning_PII_leakage_detection}

Machine learning is applied for PII leakage detection in~\cite{Kuzuno2013,Ren2016,Cheng2017}. \revOne{In particular,} the authors \revOne{of such works} leverage hierarchical clustering~\revOne{\cite{Kuzuno2013}}, C4.5 decision tree~\revOne{\cite{Ren2016}, and random forest~\cite{Cheng2017}}.

Kuzuno and Tonami in~\cite{Kuzuno2013} investigate the leakage of sensitive information due to ad libraries embedded into free Android apps. 
The proposed framework works as follows:
\begin{itemize}
\item The HTTP traffic of the target mobile apps is captured.
\item The payloads of HTTP messages are inspected, and each message is labeled according to the fact that it contains sensitive information or not.
\item The HTTP messages containing sensitive information are clustered using hierarchical clustering. The following metrics are employed:
\begin{itemize}
\item The HTTP message destination distance $d_{dst}$, which is defined as:
\begin{align}
&d_{dst}(p_{x},p_{y}) = d_{ip}(p_{x},p_{y}) \nonumber \\
&\quad + d_{port}(p_{x},p_{y}) + d_{host}(p_{x},p_{y})
\end{align}
%$$$d_{dst}(p_{x},p_{y}) = \sum_{k\in\{ip, port, host\}}d_{k}(p_{x},p_{y})$$
where $p_{n} = \{ ip_{n}, port_{n}, host_{n} \}$ with $ip_{n}$ a destination IPv4 address, $port_{n}$ a port number, $host_{n}$ a HTTP host, and the distances are defined as:
\begin{align}
d_{ip}(p_{x},p_{y}) &= lmatch(ip_{x},ip_{y})/32 \nonumber \\
%&\in \left[0,1\right]\\
d_{port}(p_{x},p_{y}) &= match(port_{x},port_{y}) \nonumber \\
%&\in \left[0,1\right]\\
d_{host}(p_{x},p_{y}) &= \frac{ed(host_{x},host_{y})}{max(len(host_{x}),len(host_{y}))} \nonumber \\
%&\in \left[0,1\right]
\end{align}
where $lmatch()$ returns the number of common upper bits in two IP addresses, $match()$ returns 1 on matching ports and 0 otherwise, $ed()$ returns an edit distance, $len()$ returns the length of a character string, and $max()$ returns the greater of its two arguments.
{In particular, the values of distances $d_{ip}$, $d_{port}$ and $d_{host}$ are within an interval $\left[0,1\right]$.} 
\item The HTTP message content distance $d_{header}$, which is defined as:
\begin{align}
&d_{header}(p_{x},p_{y}) = d_{rline}(p_{x},p_{y}) \nonumber \\
&\quad + d_{cookie}(p_{x},p_{y}) + d_{body}(p_{x},p_{y})
\end{align}
where $p_{n} = \{ rline_{n}, cookie_{n}, body_{n} \}$ with $rline_{n}$ a request line, $cookie_{n}$ a cookie, $body_{n}$ a message body, and the distance is defined as:
\begin{equation}
d_{i}(p_{x},p_{y}) = ncd(i_{x},i_{y}) \in \left[0,1\right]
\end{equation}
where $i \in \{ rline, cookie, body \}$ and $ncd(k,z)$ is the normalized compression distance of the strings $k$ and $z$.
\item Given $C_{x}$ and $C_{y}$ two clusters of HTTP messages, the linkage criterion is the following:
\begin{equation}
d(C_{x},C_{y}) = \displaystyle \frac{\displaystyle \sum_{\substack{p_{x} \in C_{x} \\ p_{y} \in C_{y}}} d_{msg}(p_{x},p_{y})}{|C_{x}|*|C_{y}|}
\end{equation}
where:
\begin{align}
d_{msg}(p_{x},p_{y}) &= d_{dst}(p_{x},p_{y}) \nonumber \\
&\quad + d_{header}(p_{x},p_{y})
\end{align}
\end{itemize}
\item The conjunction signature set resulting from the clustering is employed to detect sensitive information leakage in mobile HTTP traffic.
\end{itemize}

Ren et al. in~\cite{Ren2016} focus on the mobile apps {for 
%of the following operating systems: 
Android, iOS, and Windows Phone.} 
The presented framework {is composed of three steps:}
%works as follows:
\begin{itemize}
\item The collected network traffic (which consists of HTTP/HTTPS flows) is inspected looking for the PII related to the target mobile devices. 
Each flow is labeled according to the fact that it leaked PII or not.
\item In the feature extraction phase, a bag-of-words model is used, with the flows being the documents and the structured data being the words. 
More in detail, each flow is partitioned into words (using tokens), then it becomes a vector of binary values. 
In such vector, each word is set to $1$ if it appears in the flow, otherwise it is set to $0$.
\item For each destination domain (identified using the \texttt{Host} field of the HTTP header), {the framework selects the features (i.e., the words) that are more suitable for classification. Finally, a C4.5 decision tree is trained on the feature vectors associated to that destination domain.}
\end{itemize}

\revOne{
The framework presented by Cheng et al. in~\cite{Cheng2017} for detecting the PII leaks of Android apps consists of three phases:
\begin{itemize}
\item In the pre-processing phase, the network traffic of the targeted apps is partitioned into flows according to the information available in IP and TCP headers. Moreover, among the flows of each app, the flow with the minimum overall distance (computed using Dynamic Time Warping) from the other flows of the app is elected as leader. Finally, the flows are converted into time series of packet sizes.
\item In the feature extraction phase, each time series is converted into a feature vector by computing the following features: the distance (computed using Dynamic Time Warping) of the series from the nearest leader, the weight of the behavior that generated the series (i.e., click, swipe, or other), the series duration, the number of packets in the series, the average packet size of the series, and the average packet interval of the series.
\item In the classification phase, a random forest classifier is trained using the feature vectors.
\end{itemize}
}

\subsubsection{Malware Detection}
\label{sec:machine_learning_malware_detection}

Researchers \revOne{have} effectively \revOne{employed} machine learning techniques to detect mobile malware from \revOne{the} network traffic \revOne{of mobile devices}. 
In particular, they \revOne{have} applied both supervised learning~\cite{Shabtai2012,Su2012,Shabtai2014,Narudin2016,Wang2016,Arora2017} and unsupervised learning~\cite{Wei2012}.

\paragraph{\revOne{Malware Detection via} Supervised Learning}

\begin{table*}
\caption{The surveyed works in which supervised learning is applied for malware detection.}
\label{tab:machine_learning_malware_detection}
\centering
\begin{tabular}{|c|l|m{0.5\textwidth}|m{0.24\textwidth}|}
\hline
\textbf{Year} & \multicolumn{1}{c|}{\textbf{Paper}} & \multicolumn{1}{c|}{\textbf{Features}} & \multicolumn{1}{c|}{\textbf{Classifier}} \\ \hline
% 2012
\multirow{8}{*}{2012} & \multirow{6}{*}{Shabtai et al. \cite{Shabtai2012}} & \multirow{6}{*}{Cellular/Wi-Fi sent/received bytes/packets} & Bayesian networks \\ \cline{4-4}
& & & \revOne{J48} decision tree \\ \cline{4-4}
& & & Histograms \\ \cline{4-4}
& & & $k$-means \\ \cline{4-4}
& & & Logistic regression \\ \cline{4-4}
& & & Naive Bayes \\ \cline{2-4}
& \multirow{2}{*}{Su et al. \cite{Su2012}} & \multirow{2}{*}{\parbox{0.40\textwidth}{Average and standard deviation of the number of sent/received packets, average and standard deviation of the number of sent/received bytes, and average session duration}} & \parbox{0.20\textwidth}{\mbox{} \\[0.10cm] \revOne{J48} decision tree \\[-0.05cm]} \\ \cline{4-4}
& & & \parbox{0.20\textwidth}{\mbox{} \\[0.10cm] Random forest \\[-0.05cm]} \\ \hline
% 2014
2014 & Shabtai et al. \cite{Shabtai2014} & Average sent/received bytes, average received bytes in percent out of total amount of transmitted bytes, inner/outer average send/receive interval, and average sent/received data in percent out of total transmitted data & \revOne{Decision tree} \\ \hline
% 2016
\multirow{7}{*}{2016} & \multirow{5}{*}{Narudin et al. \cite{Narudin2016}} & \multirow{5}{*}{\parbox{0.40\textwidth}{Source/Destination IP address, source/destination port, frame length/number, HTTP request type, number of frames received by unique source/destination in the last $t$ seconds from the same destination/source, and number of packets flowing from source to destination and vice versa}} & BN with\revOne{/without} feature selection \\ \cline{4-4}
& & & MLP with\revOne{/without} feature selection \\ \cline{4-4}
& & & \revOne{J48} with\revOne{/without} feature selection \\ \cline{4-4}
& & & $k$NN with\revOne{/without} feature selection \\ \cline{4-4}
& & & RF with\revOne{/without} feature selection \\ \cline{2-4}
& \multirow{2}{*}{Wang et al. \cite{Wang2016}} & Per-TCP-flow sent/received bytes, sent/received packets, and average sent/received packet size & \multirow{2}{*}{C4.5 decision tree} \\ \cline{3-3}
& & Per-HTTP-message Host, Request-URI, Request-Method, and User-Agent & \\ \hline
% 2017
\multirow{2}{*}{2017} & \multirow{2}{*}{Arora et al. \cite{Arora2017}} & Sent/Received packets per second/flow, ratio of incoming to outgoing bytes, maximum/average packet size, and minimum time interval between sent/received packets & \multirow{2}{*}{Naive Bayes} \\ \cline{3-3}
& & Sent/Received bytes/packets per second/flow, ratio of incoming to outgoing bytes/packets, first sent/received packet size, maximum/average packet size, minimum/maximum/average time interval between sent/received packets, average flow duration, and ratio of number of connections to number of destination IPs & \\ \hline
\end{tabular}
\end{table*}

In Table~\ref{tab:machine_learning_malware_detection}, for each of the surveyed works in which supervised learning is applied for malware detection, we report the leveraged features and employed classifiers.

Regarding the work by Shabtai et al. in~\cite{Shabtai2012}, the features reported in Table~\ref{tab:machine_learning_malware_detection} are only the ones related to network traffic.

Regarding the work by Arora and Peddoju in~\cite{Arora2017}, the authors present a feature selection algorithm to find the minimal set of features that achieves the best detection performance \revOne{(the first that is reported in Table~\ref{tab:machine_learning_malware_detection})}. Given a set of features $\{ F_{1}, \dots{}, F_{n} \}$, the proposed algorithm is composed of the following steps:
\begin{itemize}
\item Rank the features according to different metrics. Each metric produces a different ranking. The authors uses the following metrics: (i) the \textit{information gain} of a feature $F$, which is the reduction of entropy after observing $F$; and (ii) the chi-squared test, which expresses the difference between the expected and observed values.
\item For $k = 1$ up to $n$:
\begin{itemize}
\item Extract the top-$k$ features from each ranking, and keep only the ones that are present in all the rankings;
\item Use the selected features to perform a classification using a naive Bayes classifier, and compute the achieved F-measure;
\item If the F-measure computed above is greater than the one achieved in the previous steps, update the minimal set of features with the currently selected features.
\end{itemize}
\end{itemize}

\paragraph{\revOne{Malware Detection via} Unsupervised Learning}

The framework proposed by Wei et al. in~\cite{Wei2012} works as follows:
\begin{itemize}
\item A monitor collects DNS response messages.
\item The IP addresses within the answer and additional sections of the DNS response messages are mapped to geographical coordinates.
\item Independent Component Analysis (ICA) is used to compute the spatial uniform distribution of hosts (i.e., uniformity degree in the geographic distribution of hosts) and their spatial service relationship (which describes the relationship between a provider and a consumer by a service distance, and tends to be zero for a benign domain). 
{Both of these metrics} are leveraged to label mobile apps as benign or malicious.
\end{itemize}

\subsubsection{User Action Identification}
\label{sec:machine_learning_user_action_identification}

Machine learning is applied for user action identification in~\cite{Watkins2013,Coull2014,Park2015,Conti2016,Fu2016,Saltaformaggio2016}. In \revOne{these} works, the authors leverage several techniques: in the field of unsupervised learning, agglomerative hierarchical clustering and $k$-means clustering; in the field of supervised learning, linear regression, naive Bayes, random forest, and support vector machine.

Watkins et al. in [27] \revOne{employ a neural-fuzzy classifier that} exploits the inter-packet time of responses to ICMP packets (i.e., pings) to infer the type of action that the target user is performing on her mobile device. 
In particular, the authors focus on three types of user action: (i) CPU intensive; (ii) I/O intensive; and (iii) non-CPU intensive.

Coull and Dyer in~\cite{Coull2014} try to infer the language (among six possible choices: Chinese, English, French, German, Russian, and Spanish) and length of the messages exchanged between iMessage clients (on both iOS and OS X) and Apple's servers. For the language, a multinomial naive Bayes classifier is used with the count of each length/direction pair observed (direction indicates whether the data is going to or coming from Apple's servers). For the length, linear regression (with least squares estimation) is employed using the payload length as the explanatory variable and the message size as the dependent variable.

Conti et al. in~\cite{Conti2016} combine unsupervised and supervised learning to fingerprint several user actions of popular Android apps. The proposed framework works as follows:
\begin{itemize}
\item The network traffic generated by each user action is partitioned into flows (each flow is a time-ordered sequence of TCP segments exchanged during a single TCP session). Each flow is converted into three time series of packet sizes (with negative sizes for incoming traffic). One series is for incoming traffic, one is for outgoing traffic, and one combines traffic in both directions.
\item The flows are clustered using the agglomerative hierarchical clustering with the following linkage criterion:
\begin{equation}
d(u,v) = \displaystyle\sum_{\substack{1 \leq i \leq n \\ 1 \leq j \leq m}} \frac{distance(u[i], v[j])}{|u|*|v|}
\end{equation}
where $distance()$ is a distance function, and $u$ and $v$ are clusters of $n$ and $m$ elements, respectively. The distance function is defined as follows:
\begin{equation}
distance(f_{i},f_{j}) = \displaystyle\sum_{k = 1}^{n} w_{k} \times DTW(T_{k}^{i},T_{k}^{j})
\end{equation}
where $f_{i}$ is a flow consisting of a set of $n$ time series $\{T_{1}^{i}, \dots, T_{n}^{i}\}$, $w_{k}$ is a weight assigned to the $k^{th}$ time series, and $DTW(x,y)$ is the optimal warping path between the time series $x$ and $y$.
\item For every user action, each flow $f$ is assigned to the cluster that minimizes the distance between $f$ and the leader of the cluster (which is the flow that has the minimum overall distance from the other flows of the cluster). The $k^{th}$ feature indicates the number of flows that have been assigned to the $k^{th}$ cluster after the execution of that user action.
\item The final classification is performed using a random forest classifier.
\end{itemize}

The systems for user action identification developed by Park and Kim in~\cite{Park2015} and Saltaformaggio et al. in~\cite{Saltaformaggio2016} are similar to the one by Conti et al. \cite{Conti2016}. However, there are a few differences:
\begin{itemize}
\item In~\cite{Park2015}, each flow is represented by a single time series including both incoming and outgoing traffic. As a consequence, the clustering is applied to time series (in~\cite{Conti2016}, the authors consider sets of time series). Moreover, the distance function in the formula of the linkage criterion is simply $DTW()$.
\item In~\cite{Saltaformaggio2016}, the IP traffic is partitioned into server transactions, each containing the IP headers (ordered by time) of the packets exchanged with a specific remote host. The server transactions are converted into feature vectors by the following $26$ features: (i) send/receive average inter-packet time; (ii) the ratio of the number of packets sent/received to/from the server over the total number of packets exchanged with the server; (iii) the ratio of the size of the data sent/received to/from the server over the size of the total data exchanged with the server; and (iv) the number of packets sent/received within each of ten size ranges, normalized by the total number of sent/received packets. The feature vectors are then clustered using $k$-means clustering (following an incremental approach to find a suitable value for $k$), and the final classification is performed using a multi-class support vector machine.
\end{itemize}

\revOne{The framework proposed by Fu et al. in~\cite{Fu2016} is based on supervised learning and consists of the following steps:
\begin{itemize}
\item The captured network flows are partitioned into sessions, then hierarchical clustering is applied to group together the sessions that are related to the same user action (the authors call such groups of sessions as \textit{dialogs}).
\item The dialogs are converted into feature vectors by extracting features that are related to the size (e.g., median, standard deviation, percentage of packets in a given range) and timing (e.g., median, standard deviation) of IP packets. Each feature vector is labeled with the user action that generated the corresponding dialog.
\item A classifier is trained and evaluated on the feature vectors. The authors consider the following classifiers: random forest, gradient boosted trees, support vector machine, naive Bayes, and $k$-nearest neighbors.
\item The dialogs related to multiple user actions, which have been classified as ``unknown'' in the previous step, are split into sub-dialogs that are then classified using a trained hidden Markov model.
\end{itemize}
}

\subsubsection{Operating System Identification}
\label{sec:machine_learning_OS_identification}

Supervised learning techniques are applied for operating system identification in~\cite{Chen2014,Coull2014,Ruffing2016,Malik2017}.

Chen et al. in~\cite{Chen2014} present a naive Bayes classifier that leverages the following binary features:
\begin{itemize}
\item $TTL = 128$ (Windows) and $TTL \neq 128$ (Windows, Android, or iOS), where $TTL$ is the Time-To-Live (TTL) field of the IP header.
\item $ID_{mvr} < 0.05$ (mostly Windows, rarely Android), $ID_{mvr} \in \left[ 0.05, 0.40 \right]$ (mostly Android, rarely Windows), and $ID_{mvr} > 0.40$ (mostly iOS, rarely Android), 
%where $ID_{mvr}$ is the violation ratio of the monotonicity of the identification (ID) field of the IP header.
{where $ID_{mvr}$ is \revOne{the} monotonicity violation ratio of \revOne{the} identification (ID) field in \revOne{the} IP headers.}
\item $TS_{ratio} < 0.05$ (Windows) and $TS_{ratio} \geq 0.05$ (Android or iOS), where $TS_{ratio}$ is the ratio of segments with TCP timestamp option.
\item $WS = 4$ (mostly Windows, rarely Android), $WS = 16$ (iOS), $WS = 64$ (mostly Android, rarely Windows), and $WS = 256$ (Windows), where $WS$ is the TCP window size scale option.
\item $clock_{SD} \leq 3$ (mostly Android, rarely iOS and Windows) and $clock_{SD} > 3$ (iOS), where $clock_{SD}$ is the standard deviation of a device clock frequency, estimated using packets coming from its IP address.
\end{itemize}

Coull and Dyer in~\cite{Coull2014} leverage the sizes of encrypted packets exchanged between a target iMessage user and Apple's servers.
{The aim of this analysis is to
%in order to 
determine whether %she is running 
the iMessage client is running on iOS or OS X.} 
The authors use a binomial naive Bayes classifier with one class for each of the four possible (OS, direction) pairs, with direction indicating whether the packet is going to or coming from Apple's servers. 
The classifier operates on a binary feature vector of (size, direction) pairs, where the value for a given feature is $1$ if the corresponding pair is observed and $0$ otherwise.

The framework presented by Ruffing et al. in~\cite{Ruffing2016} combines together supervised learning and analysis of the frequency spectrum of packet timing. 
{The proposed methodology is composed of %consists of 
two phases:} 
%the following phases:
\begin{itemize}
\item Training phase:
\begin{itemize}
\item Each traffic trace, which is labeled with the operating system that generated it, is converted into a frequency spectrum.
\item Frequency components are extracted from the frequency spectra generated in the previous step.
\item A genetic algorithm is applied to separate the frequency components that are related to OS features from those that bring noise. The former are promoted features.
\end{itemize}
\item Identification phase:
\begin{itemize}
\item A new traffic trace $x$ is converted into a feature-extracted frequency spectrum $F^{x}$.
\item The identified operating system is provided by the following formula:
\begin{equation}
\displaystyle \argmax_{os \in OS} \frac{1}{n_{os}} \sum_{i = 1}^{n_{os}} corr(F^{x},F_{i}^{os})
\end{equation}
where $OS$ is the set of the considered mobile OSes, $n_{os}$ is the number of feature-extracted frequency spectra of the mobile operating system $os$, $F_{i}^{os}$ is the $i^{th}$ feature-extracted frequency spectrum of the mobile operating system $os$, and $corr(X,Y)$ is a function that computes the correlation between the frequency spectra $X$ and $Y$.
\end{itemize}
\end{itemize}

Malik et al. in~\cite{Malik2017} carry out OS identification by exploiting the inter-packet time of packets coming from the target mobile device. 
In particular, the authors focus on two types of packet: (i) the response to an ICMP packet sent to the target mobile device (active measurement); and (ii) an IP packet related to a video stream involving the target mobile device (passive measurement). 
\revOne{The presented framework consists of} a random forest classifier \revOne{that} is trained and evaluated on the inter-packet times of three mobile devices running Android, iOS, and Windows Phone, respectively.

\subsubsection{\revOne{Position Estimation}}
\label{sec:machine_learning_position_estimation}

Musa and Eriksson in~\cite{Musa2012} present a system for converting the detections of the Wi-Fi probe requests periodically transmitted by a target mobile device into a highly likely spatiotemporal trajectory within the monitored area.

The \revOne{position} estimation problem is formulated using a hidden Markov model (HMM): (i) each street of the covered area is partitioned into segments, and each segment represents a rectangular area in which a mobile device may be located; and (ii) a state of the HMM is assigned to each segment, and transition probabilities are used to model the behavior of mobile devices at intersections (i.e., go straight, turn left, or turn right). For each detection $det_{i}$ of the target mobile device and each state $s_{i}$ of the HMM, the emission probability $p(det_{i}|s_{i})$, which represents the probability of making the detection $det_{i}$ if the current state is $s_{i}$, is computed. Finally, the Viterbi's map-matching algorithm is applied to find the maximum-probability path, which is represented by a sequence of hidden states visited in the Markov model.

\subsubsection{User Fingerprinting}
\label{sec:machine_learning_user_fingerprinting}

Verde et al. in~\cite{Verde2014} present a framework for fingerprinting mobile users from NetFlow records. The proposed solution is based on supervised learning and hidden Markov model (HMM):
\begin{itemize}
\item Feature vectors are extracted from the NetFlow records of the target user, and then partitioned into subsets. Each subset is related to a specific network service. For each service, several HMMs are created by varying the number of states and the subset of features, and setting up their parameters via the $k$-means algorithm. The HMMs are trained in parallel, and subsequently converted into binary classifiers using a probability threshold $t$ (i.e., if the observation has a probability lower than $t$, it will be classified as $0$, otherwise it will be classified as $1$). Finally, the best performing HMM is selected for that network service.
\item Feature vectors are extracted from the NetFlow records captured at the monitored network, and subsequently partitioned according to the network service they belong to. Each feature vector is classifier with the HMM corresponding to its service. For each time interval, the results are aggregated into a new record that contains, for each HMM $h$, the number of feature vectors recognized by $h$ as belonging to the target user during the interval, times the weight of $h$ (which is computed during the training phase). Finally, a machine learning classifier is used to determine whether, during each of the time intervals, the network traffic contains data transmissions from the target user. The framework supports the following classifiers: support vector machine, random forest, RIPPER, multi-layer perceptron, and naive Bayes.
\end{itemize}

\subsubsection{Ad Fraud Detection}
\label{sec:machine_learning_ad_fraud_detection}

Crussell et al. in~\cite{Crussell2014} present a system being able to automatically run Android apps in emulators and analyze their application-layer traffic in order to detect \revOne{whether} they: 
(i) request ads while being in the background (i.e., ads are not displayed to the user); 
and (ii) click on ads without user interaction (i.e., false user clicks are simulated). The framework is based on supervised learning:
\begin{itemize}
\item The HTTP and DNS traffic of each analyzed app is extracted from its network traces, then causally related HTTP requests are linked to form request trees.
\item For each request page (identified by the host and path names of its URL), all the related HTTP requests are aggregated. After that, {the authors extract $33$ features} as follows:
\begin{itemize}
\item Ten features derive from query parameters:
\begin{itemize}
\item For each query parameter, {it is computed} the ratio of distinct values found for that parameter over the total number of times the parameter appeared in a request, as well as the ratio of distinct values found for that parameter over the \revOne{total} number of distinct apps. 
Each ratio is segmented into several intervals and the number of query parameters whose ratio is in each interval is counted. These counts contribute six features.
\item {It is also computed the entropy of each query parameter.}
%, its entropy is computed. 
The entropy is considered high if it is greater than $216$ bits, low otherwise. 
The number of query parameters that have, respectively, high and low entropy contribute two features.
\item The last two features are the average and the total number of query parameters.
\end{itemize}
\item Sixteen features derive from the request trees. 
Such features are related to their structure {(e.g., 
average height and depth of trees containing the page), 
%the average subtree height below the page, 
%the average depth of the page in the trees
%as well as 
number of children, their MIME types, and types of edge that connect the children to their parent.}
\item Seven features derive from HTTP headers {(e.g., status codes, requests' length, replies' length).}
\end{itemize}
\item The authors train a random forest classifier to classify each request page as ad-related (ARQ) or not (NARQ).
\item The ad request pages (i.e., the ARQ pages) and the HTTP request trees are leveraged to extract and verify ad impressions (i.e., displaying) and clicks.
\end{itemize}

\subsubsection{Tethering Detection}
\label{sec:machine_learning_tethering_detection}

Chen et al. in~\cite{Chen2014} apply supervised learning to detect whether a target mobile device is tethering its Internet connection to other devices. The proposed probabilistic classifier leverages the following binary features:
\begin{itemize}
\item $n_{OS} = 1$ (no tethering) and $n_{OS} > 1$ (tethering), where $n_{OS}$ is the number of operating systems identified from the packets coming from the same IP address (also the OS identification framework is based on machine learning, see Section~\ref{sec:machine_learning_OS_identification} for details).
\item $n_{TTL} = 1$ (no tethering with high probability) and $n_{TTL} > 1$ (tethering with high probability), where $n_{TTL}$ is the number of distinct TTLs in the packets coming from the same IP address.
\item $ts_{mvr} \leq 0$ and $ts_{mvr} > 0$, where $ts_{mvr}$ is the violation ratio of the TCP timestamp monotonicity of the segments coming from the same device (the idea is to exploit the fact that segments generated by the same device tend to monotonically increase TCP timestamp values, whereas segments from different devices tend to have mixed TCP timestamp values).
\item $clock_{SD} \leq 35$ and $clock_{SD} > 35$, where $clock_{SD}$ is the standard deviation of the clock frequency estimated using the packets coming from the same IP address (a large standard deviation is likely due to tethering).
\item $boot_{SD} \leq 1455$ and $boot_{SD} > 1455$, where $boot_{SD}$ is the standard deviation of the boot time inferred from the TCP timestamp values in the segments coming from the same device (the idea is to exploit the fact that different devices have distinct boot times and distinct initial TCP timestamp values).
\end{itemize}

\subsubsection{Website Fingerprinting}
\label{sec:machine_learning_website_fingerprinting}

Spreitzer et al. in~\cite{Spreitzer2016} fingerprint the websites visited by an Android user via the web browser of her mobile device, by leveraging the data-usage statistics of the browser app. The proposed framework is based on supervised learning:
\begin{itemize}
\item Each considered website $w_{i}$ is opened in the web browser of an Android device. At the same time, the TCP bytes transmitted and received by the browser app are sampled at a frequency $f$ for a period of $t$ seconds. The readings constitute a sample of the website $w_{i}$. The sampling process stops after collecting $n$ samples, which constitute the signature of the website $w_{i}$. All the generated signatures are included in a database $T$.
\item $T$ is loaded into an unprivileged Android app that is installed on the target mobile device. When the user opens a website within the web browser of the device, the app samples the transmitted and received TCP bytes of the browser app, and builds a signature $s$. For each signature $s_{i}$ in $T$, the similarity (which is a function based on the Jaccard index) between $s$ and $s_{i}$ is computed; the app returns the website corresponding to the signature $s_{i}$ that maximizes the similarity with $s$.
\end{itemize}

\subsection{Other Analysis Methods}
\label{sub:othermethods}
\revOne{Some traffic analyses does not require machine learning techniques to achieve their goals.
 In this section, we report two alternative methods.}
In Section~\ref{sec:dictionary}, we describe a way to turn a 
dictionary\footnote{We use the term \textit{dictionary} to indicate a collection of (key, value) pairs, such that each key appears at most once in the collection, and a value can be either a single value or an unordered set of values.} 
into an effective classifier. 
In Section~\ref{sec:graph_analysis}, we report two methodologies that rely on graphs. 
%%%%%%%%%%%%%%%TODO

\subsubsection{Dictionary}
\label{sec:dictionary}

A dictionary can be used as a one-feature classifier whether: 
(i) the keys are the values that the feature can take; and (ii) a set of class labels is associated to each key. 
This solution is suitable to solve classification problems in which there is a single feature, which takes a limited set of values.

Coull and Dyer in~\cite{Coull2014} target iMessage, Apple's instant messaging service, which is available as a mobile app for iOS or a traditional computer program for OS X. 
The objective is to fingerprint five distinct user actions (i.e., start typing, stop typing, send text, send attachment, and read receipt) by leveraging the size of the packets exchanged between the target iMessage client and Apple's servers. 
The authors study the packet sizes corresponding to the considered user actions, and notice that each user action has two distinctive packet sizes: 
{(i) one when a message is sent to Apple's servers; 
and (ii) one when a message is received from Apple's servers.} 
This property can be exploited to build a classifier that takes the form of a dictionary in which one or more user action labels are associated to each packet size observed in the training data. 
When a new packet arrives, the dictionary is queried to retrieve the user action label(s) for its payload length: if only one label is found, the packet is given that label; if two or more labels are returned, the user action most frequently associated to that payload size during training is chosen.

\subsubsection{Graph Analysis}
\label{sec:graph_analysis}

{In this section, we highlight a few works in which graph theory is leveraged for mobile traffic analysis.}

Barbera et al. in~\cite{Barbera2013} combine {graph theory} and traffic analysis to carry out sociological inference targeting mobile users (their findings are reported in Section~\ref{sec:sociological_inference}). 
More precisely, they present a methodology to build the social network\footnote{In a social network, the nodes correspond to the individuals, while the edges model the relationships between them.} of a group of mobile users from the probe requests sent by their mobile devices. 
The proposed procedure consists of the following steps:
\begin{itemize}
\item The dataset of collected probe requests is turned into an \textit{affiliation network}. 
An affiliation network $G = (V_{1}, V_{2}, E)$ is a bipartite graph in which $V_{1}$ is a set of \textit{actors}, $V_{2}$ is a set of \textit{groups} the actors belong to, and each edge $e \in E$ connecting an actor $v_{1} \in V_{1}$ to a group $v_{2} \in V_{2}$ represents a group membership. 
In the work by Barbera et al.~\cite{Barbera2013}, $V_{1}$ is the set of mobile devices (identified by their MAC address) that sent at least one probe request, $V_{2}$ is the set of SSIDs contained in the collected probe requests, and an edge $e \in E$ connecting a mobile device $v_{1} \in V_{1}$ to an SSID $v_{2} \in V_{2}$ represents $v_{1}$ having $v_{2}$ in its Preferred Network List (PNL), which is the list of the SSIDs of the Wi-Fi networks $v_{1}$ connected to in the past.
\item A \textit{similarity measure} $f : V_{1} \times V_{1} \rightarrow R$ is chosen to represent the strength of the social relationship between the users of each pair of mobile devices $u$ and $v$. Based on the Adamic-Adar similarity measure~\cite{Adamic2003}, {Barbera et al.~\cite{Barbera2013} define the $f$ function as:}
\begin{equation}
f(u,v) = \displaystyle\sum_{w \in N \left( u \right) \cap N \left( v \right)} \frac{1}{\log_{2} \left( | M \left( w \right) | \right)}
\end{equation}
where $N \left( u \right)$ is the PNL of the mobile device $u$, and $M \left( w \right)$ is the set of mobile devices that have the SSID $w$ in their PNL.
\item The affiliation network $G$ is turned into a social network $G' = (V_{1}, E')$ by applying the following rule:
\begin{equation}
\forall u, v \in V_{1}: \left( u, v \right) \in E' \Leftrightarrow f \left( u, v \right) > t
\end{equation}
where $t$ is a minimum similarity threshold.
\end{itemize}

Vanrykel et al. in~\cite{Vanrykel2016} present a graph building technique that processes a mobile traffic dataset in order to partition the network traces it contains by user. 
The idea is to exploit the sensitive identifiers that are typically present in the network traffic generated by mobile devices. 
The proposed methodology consists of the following steps:
\begin{itemize}
\item Through the analysis of TCP timestamps, the packets that belong to the same app session (therefore to the same mobile user) are grouped together into a node. Alternatively, it is possible to partition the packets by TCP session.
\item For each node, the sensitive identifiers present in HTTP messages are extracted according to host-specific rules.
\item To cluster the nodes into components, each one representing the network traffic related to a specific mobile user, the following rules are iteratively applied to all nodes:
\begin{itemize}
\item If the node's identifiers (or their hashed/encoded values) match the identifiers of an existing component, add the node to that component and merge their identifiers.
\item If the node's identifiers (or their hashed/encoded values) match the identifiers of multiple existing components, merge those components together, add the node to the resulting component, and merge their identifiers.
\item If the node's identifiers (or their hashed/encoded values) do not match the identifiers of any existing component, make the node a component on its own.
\end{itemize}
\end{itemize}

\revOne{
\section{Validation Methods and Results for Traffic Analysis Targeting Mobile Devices}
\label{sec:validation}

In this section, we critically evaluate and compare the results achieved by the surveyed works in the field of traffic analysis targeting mobile devices. 
Such results are linked to the analyzed datasets and presented by the goal of the analysis. 
To help the reader, we also devote an initial section (\ref{sec:validation_preliminaries}) to describe: (i) the evaluation metrics for the analyses that can be considered as a \textit{classification} problem, as well as those metrics that are specific of a particular goal; (ii) the dataset partitioning techniques for the analyses that follow the supervised learning paradigm; and (iii) the conventions that we will follow in the tables.

To offer a clear comparison among works, for each goal of the analysis (sorted by popularity as in Section~\ref{sec:goals}) we summarize with tables the information about validation in terms of datasets, methods, and results. 
Moreover, we discuss interesting aspects of validation and we comment the obtained results. 

\subsection{Preliminaries}
\label{sec:validation_preliminaries}

\subsubsection{Evaluation Metrics for Classification Problems}
\label{sec:metrics}

In a classification problem, given a set of $n \geq 2$ classes (or labels) $C = \{ C_{1}, \dots{}, C_{n} \}$ and a new instance $x$, the goal is to infer which class $x$ belongs to (i.e., which is the index $i$ such that $x \in C_{i}$). 
We distinguish between two types of classification: \textit{binary} and \textit{multi-class}.

In a binary classification problem, we have only two classes (i.e., $n = 2$): a \textit{positive} one, which represents a given property $P$, and a \textit{negative} one, which represents the negation of that property ($\bar{P}$). 
The goal is to infer whether $P$ holds for the new instance $x$. 
Under such conditions, we define as \textit{true positives} the correctly classified positive instances, as \textit{false positives} the negative instances incorrectly classified as positive, as \textit{true negatives} the correctly classified negative instances, and as \textit{false negatives} the positive instances incorrectly classified as negative.

In a multi-class classification problem, instead, we have at least three classes (i.e., $n \geq 3$). 
In such scenario, given two (possibly equal) class indexes $i$ and $j$ ($1 \leq i, j \leq n$), we define as $c_{i,j}$ the number of instances belonging to the class $C_{i}$ that are classified as belonging to the class $C_{j}$. 
As a consequence of the above definition, $c_{i,i}$ represents the number of instances belonging to the class $C_{i}$ that are correctly classified.

To evaluate the results achieved in the surveyed works that present traffic analyses we can consider as classification problems, we use four performance metrics: accuracy, precision, recall, and F-measure.

\textit{Accuracy} ---
%\label{sec:accuracy}
%In a binary classification problem, w
We define as \textit{accuracy} the number of correctly classified instances (i.e., the number of true positives $TP$ plus the number of true negatives $TN$) over the total number of classified instances (i.e., the sum of the number of true positives $TP$, the number of false positives $FP$, the number of true negatives $TN$, and the number of false negatives $FN$): 
\begin{equation}\label{eq:accuracy_binary}
accuracy = \frac{TP + TN}{TP + FP + TN + FN}
\end{equation}
Accuracy measures how much the considered instances have been correctly classified, and ranges from $0.0$ (none of the instances has been correctly classified) to $1.0$ (all the instances have been correctly classified).

%We can generalize the Equation~\ref{eq:accuracy_binary} to compute the accuracy for a multi-class classification problem: 
%\begin{equation}
%accuracy = \frac{\sum_{i = 1}^{n} c_{i,i}}{\sum_{i = 1}^{n} \sum_{j = 1}^{n} c_{i,j}}
%\end{equation}

\textit{Precision} ---
%\label{sec:precision}
%In a binary classification problem, w
We define as \textit{precision} the number of true positives $TP$ over the total number of instances that have been classified as positive (i.e., the number of true positives $TP$ plus the number of false positives $FP$): 
\begin{equation}\label{eq:precision_binary}
precision = \frac{TP}{TP + FP}
\end{equation}
Precision measures how much the instances classified as positive have been correctly classified, without any insight into the positive instances that could have been missed. 
Precision ranges from $0.0$ (none of the instances classified as positive is positive) to $1.0$ (all the instances classified as positive are positive).

%We can generalize the Equation~\ref{eq:precision_binary} to compute the precision for a given class $C_{i}$ in a multi-class classification problem: 
%\begin{equation}
%precision_{i} = \frac{c_{i,i}}{\sum_{j = 1}^{n} c_{j,i}}
%\end{equation}
%To compute an overall precision, the precisions of all the classes must be aggregated in some way (e.g., average, median).

\textit{Recall} ---
%\label{sec:recall}
%In a binary classification problem, w
We define as \textit{recall} the number of true positives $TP$ over the total number of instances that belong to the positive class (i.e., the number of true positives $TP$ plus the number of false negatives $FN$): 
\begin{equation}\label{eq:recall_binary}
recall = \frac{TP}{TP + FN}
\end{equation}
Recall measures how much the instances belonging to the positive class have been correctly classified, without any insight into the negative instances that could have been classified as positive. 
Recall ranges from $0.0$ (none of the instances belonging to the positive class have been correctly classified) to $1.0$ (all the instances belonging to the positive class have been correctly classified).

%We can generalize the Equation~\ref{eq:recall_binary} to compute the recall for a given class $C_{i}$ in a multi-class classification problem: 
%\begin{equation}
%recall_{i} = \frac{c_{i,i}}{\sum_{j = 1}^{n} c_{i,j}}
%\end{equation}
%To compute an overall recall, the recalls of all the classes must be aggregated in some way (e.g., average, median).

\textit{F-measure} ---
%\label{sec:f-measure}
%In a binary classification problem, w
We define as \textit{F-measure} the harmonic average of precision and recall: 
\begin{equation}\label{eq:f-measure_binary}
\text{\textit{F-measure}} = 2 \times \frac{precision \times recall}{precision + recall}
\end{equation}
The F-measure provides a single score to express the performance in a classification task, and ranges from $0.0$ (precision and recall are both $0.0$) to $1.0$ (precision and recall are both $1.0$).

%The F-measure for a given class $C_{i}$ in a multi-class classification problem can be obtained from the Equation~\ref{eq:f-measure_binary} by replacing precision and recall with those of class $C_{i}$: 
%\begin{equation}
%\text{\textit{F-measure}}_{i} = 2 \times \frac{precision_{i} \times recall_{i}}{precision_{i} + recall_{i}}
%\end{equation}
%To compute an overall F-measure, the F-measures of all the classes must be aggregated in some way (e.g., average, median).

\subsubsection{Goal-specific Evaluation Metrics}
\label{sec:goal-specific_metrics}

To evaluate the results achieved in the surveyed works that deal with app identification and user fingerprinting, we use additional performance metrics that are specific of those types of analysis: app matching rate and app identification rate for app identification; data aggregation rate for user fingerprinting.

\textit{App Matching Rate (AMR)} ---
%\label{sec:app_matching_rate}
%In an app identification problem, w
We define as \textit{app matching rate} the ratio between the number of network transmissions that are successfully matched with one of the targeted apps ($n_{\text{\textit{match}}}$), over the total number of captured network transmissions ($n_{\text{\textit{total}}}$):
\begin{equation}\label{eq:app_matching_rate}
\text{\textit{AMR}} = \frac{n_{\text{\textit{match}}}}{n_{\text{\textit{total}}}}
\end{equation}
The app matching rate measures how much the captured network transmissions belong to the targeted apps, and ranges from $0.0$ (none of the transmissions belong to the targeted apps) to $1.0$ (all transmissions belong to the targeted apps). 
The app matching rate is often used when no ground truth is available, i.e., the apps that actually generated the captured network transmissions are unknown.

\textit{App Identification Rate (AIR)} ---
%\label{sec:app_identification_rate}
%
%In an app identification problem, w
We define as \textit{app identification rate} the ratio between the number of identified apps ($n_{\text{\textit{identify}}}$), over the total number of targeted apps ($n_{\text{\textit{total}}}$):
\begin{equation}\label{eq:app_identification_rate}
\text{\textit{AIR}} = \frac{n_{\text{\textit{identify}}}}{n_{\text{\textit{total}}}}
\end{equation}
The app identification rate measures how many of the targeted apps have been successfully identified, and ranges from $0.0$ (none of the targeted apps have been identified) to $1.0$ (all the targeted apps have been identified).

\textit{Data Aggregation Rate (DAR)} ---
We define as \textit{data aggregation rate} the percentage of network data related to a mobile user that is correctly linked to that mobile user. 
The data aggregation rate ranges from $0.0$ (none of the network data related to the mobile user has been linked to that mobile user) to $1.0$ (all the network data related to the mobile user have been linked to that mobile user).

\subsubsection{Dataset Partitioning Techniques for Supervised Learning}
\label{sec:dataset_partitioning}
To evaluate the performance of a supervised-learning-based analysis, it is necessary to partition the dataset.
The \textit{holdout} method is the fundamental method to partition the dataset in order to get the data for training and testing, respectively" 
This method first creates two empty sets: training set and testing set. 
Then it randomly assigns each observation in a dataset to one of those sets according to a given pre-determined parameter $S_{training}$. 
This parameter indicates the proportion in terms of the number of elements that the training set include compared to the total elements of the dataset. 
As an example, with $S_{training} = 0.7$ the training set will include the $70\%$ of the total elements of the dataset, while the testing set will include the remaining $30\%$. 
Such proportion can also be expressed using two integers $n$ and $m$ that stand as the number of equipotent parts in which the dataset is partitioned and the number of parts used as testing set, respectively. 
As an example, with $n=5$ and $m=2$ a dataset is partitioned into five equipotent parts of which three used as training set and two as testing set.

All model selection and training processes must be done exclusively using the training set, without involving the testing set in any case.
Indeed, it is possible to further partition a training set and use part of it as a reduced testing set (i.e., validation set) to perform model selection and hyper-parameter tuning. 
The testing set has to be used for the only purpose of evaluating the performance of the trained model.

In multi-class classification, it is useful to ensure that training and testing sets hold the same proportion of observations belonging to a class. 
This option is called \textit{stratification} and is used to avoid imbalance of class representation between training and testing sets.

%\paragraph{Cross-validation}
%\label{sec:cross-validation}

Another method to evaluate a learning model is to perform a \textit{cross-validation}.
The cross-validation method consists in partitioning a dataset into $k$ folds (i.e., complementary and equipotent subsets) and iteratively training and testing a given model for $k$ times. 
For each run, one fold is considered as a testing set and the remaining $k-1$ folds as training set.
The overall results are obtained by aggregating the results of each run.
As for multi-class holdout method, the dataset can be partitioned in folds using the stratification option. 
Cross-validation can also be used on the training set only to perform model selection and hyper-parameter tuning.
A variant of cross-validation called \textit{leave-$p$-out cross-validation} allows specifying the exact number of observations $p$ to holdout as a testing set at each run (the case in which $p=1$ is called \textit{leave-one-out cross-validation}).

\subsubsection{Table Conventions}
\label{sec:table_conventions}

To uniquely identify the datasets that have no name in a given work, we will use arbitrary names of the form ``[yy].[author]\_{}[n]'', where [yy] is the last two digits of the publication year of the work, [author] is the surname of the first author of the work, and [n] is a progressive number, starting from zero.

Moreover, since we are space-constrained, we will shrink the content of the tables by using the following shortenings:
\begin{itemize}
\item We will use the character ``\#{}'' to shorten the word ``number''.
%\item We shorten the following words: hours (h), minutes (min), bytes (B), packets (pkt), headers (hdr), messages (msg), repetitions (rep), and percentage (\%). 
%\item We refer to ``machine learning'' methods as ML and the acronyms of specific methods are the one we introduced in Section~\ref{sec:models_methods}. 
\item Given a number $n$, we will write ``$n+$'' for ``over $n$''.
\item We will write ``Tx'' and ``Rx'' for ``transmitted'' and ``received'', respectively.
\item To reference a given OSI layer, we will use the corresponding ordering number (e.g., ``layer 2'' for the data-link layer, ``layer 3'' for the network layer, ``layer 4'' for the transport layer, ``layer 7'' for the application layer). To reference a given OSI layer and the layers above it, we will add the character ``+'' to the ordering number of the bottom layer (e.g., ``layer 3+'' for layers from network to application).
\item Given two decimal numbers $S_{\text{training}}$ and $S_{\text{test}}$ such that $S_{\text{training}} + S_{\text{test}} = 1$, we will write ``$S_{\text{training}}/S_{\text{test}}$'' to indicate the holdout dataset partitioning technique with $S_{\text{training}}$ of the data reserved to the training set, and $S_{\text{test}}$ of the data reserved to the test set (e.g., ``$0.7/0.3$'').
\item Given two integer numbers $n$ and $m$ such that $m < n$, we will write ``out($n,m$)'' to indicate the holdout dataset partitioning technique with $m$ parts out of $n$ reserved to the test set, and the remaining ones reserved to the training set (e.g., ``out($10,2$)'').
\item Given an integer number $k$, we will write ``kCross($k$)'' to indicate the $k$-fold cross-validation.
\end{itemize}
}

\revOne{
\subsection{Traffic Characterization}
\label{sec:validation_traffic_characterization}

In this section, we focus on the datasets used for traffic characterization works, since we have already summarized the findings of these works in Section~\ref{sec:traffic_characterization}. 

In Table~\ref{tab:validation_traffic_characterization_apps/services} we provide information about the datasets analyzed in the surveyed works that deal with traffic characterization targeting specific apps and/or mobile services.  
As we can notice, most of these works rely on data from the data-link layer and above.
Four out of nine works consider only one or two apps in their analysis (e.g., Skype, Youtube).
Two works focus on a specific category of apps: Chen et al. in~\cite{Chen2015} on malware, and Nayam et al. in~\cite{Nayam2016} on apps related to health and fitness. 
Moreover, Lindorfer et al. in~\cite{Lindorfer2014} study a set of one million apps.

In Table~\ref{tab:validation_traffic_characterization_population}, we report the information about the datasets of the works that focus their analysis on a population of mobile devices.
Most of the works consider more than two hundred devices (five out of nine), among which three works on more than two thousand devices.
Unfortunately, Chen et al. in~\cite{Chen2012} and Lee et al. in~\cite{Lee2011} do not specify the number of devices involved in their analysis.
}

\begin{table*}
%%\color{blue}
\caption{The datasets of the surveyed works that deal with traffic characterization targeting specific apps and/or mobile services.}
\label{tab:validation_traffic_characterization_apps/services}
\centering
\begin{tabular}{|c|l|c|m{0.17\textwidth}|c|m{0.17\textwidth}|m{0.07\textwidth}|}
\hline
\textbf{Year} & \multicolumn{1}{c|}{\textbf{Paper}} & \textbf{Mobile Platform(s)} & \multicolumn{1}{c|}{\textbf{App(s)/Mobile Service(s)}} & \textbf{Dataset} & \multicolumn{1}{c|}{\textbf{Point of Capturing}} & \multicolumn{1}{c|}{\textbf{Content}} \\ \hline
% 2011
\multirow{7}{*}{2011} & \multirow{5}{*}{Finamore et al. \cite{Finamore2011}} & \multirow{5}{*}{Platform-independent} & \multirow{5}{*}{YouTube} & US-Campus & \multirow{2}{*}{\parbox{0.17\textwidth}{Wired (within a campus network, one week)}} & \multirow{5}{*}{\parbox{0.07\textwidth}{IP packets}} \\ \cline{5-5}
& & & & EU1-Campus & & \\ \cline{5-6}
& & & & EU1-ADSL & \multirow{3}{*}{\parbox{0.17\textwidth}{Wired (within an ISP's network, one week)}} & \\ \cline{5-5}
& & & & EU1-FTTH & & \\ \cline{5-5}
& & & & EU2-ADSL & & \\ \cline{2-7}
& \multirow{2}{*}{Rao et al. \cite{Rao2011}} & \multirow{2}{*}{Android, iOS} & Netflix & NetMob & \multirow{2}{*}{\parbox{0.17\textwidth}{Wired (forwarding server, $180$ seconds per video)}} & \multirow{2}{*}{\parbox{0.07\textwidth}{Layer-2+ data}} \\ \cline{4-5}
& & & YouTube & YouMob & & \\ \hline
% 2012
\multirow{3}{*}{2012} & \multirow{2}{*}{Baghel et al. \cite{Baghel2012}} & \multirow{2}{*}{Android} & Facebook & 12.Baghel\_{}0 & Wired (forwarding server, $90$ minutes) & \multirow{2}{*}{\parbox{0.07\textwidth}{Layer-2+ data}} \\ \cline{4-6}
& & & Skype & 12.Baghel\_{}1 & Wired (forwarding server, five hours) & \\ \cline{2-7}
& Wei et al. \cite{Wei2012ProfileDroid} & Android & $19$ free and $8$ paid apps & 12.Wei\_{}0 & Devices (two devices) & Layer-2+ data \\ \hline
% 2014
2014 & Lindorfer et al. \cite{Lindorfer2014} & Android & Over $1{,}000{,}000$ apps & 14.Lindorfer\_{}0 & Emulators ($240$ seconds per app) & Layer-2+ data \\ \hline
% 2015
2015 & Chen et al. \cite{Chen2015} & Android & $5560$ malicious apps (from $177$ malware families) & 15.Chen\_{}0 & Wired (forwarding server, five minutes per app) & Layer-2+ data \\ \hline
% 2016
\multirow{3}{*}{2016} & \multirow{2}{*}{Nayam et al. \cite{Nayam2016}} & Android & $63$ free apps (``Health \& Fitness'' category) & \multirow{2}{*}{16.Nayam\_{}0} & \multirow{2}{*}{\parbox{0.17\textwidth}{Wired (forwarding server, three $30$-minutes-long runs per app)}} & \multirow{2}{*}{\parbox{0.07\textwidth}{HTTP messages}} \\ \cline{3-4}
& & iOS & $35$ free apps (``Health \& Fitness'' category) & & & \\ \cline{2-7}
& Tadrous et al. \cite{Tadrous2016} & Android, iOS & Five apps (common to the targeted mobile platforms) & 16.Tadrous\_{}0 & APs (one AP, $300$ sessions per app) & 802.11 frames \\ \hline
% 2017
2017 & Espada et al. \cite{Espada2017} & Android & Spotify & 17.Espada\_{}0 & Devices (one device) & Layer-2+ data \\ \hline
\end{tabular}
\end{table*}

\begin{table*}
%%\color{blue}
\caption{The datasets of the surveyed works that deal with traffic characterization targeting a population of mobile devices.}
\label{tab:validation_traffic_characterization_population}
\centering
\begin{tabular}{|c|l|c|m{0.17\textwidth}|m{0.17\textwidth}|m{0.20\textwidth}|}
\hline
\textbf{Year} & \multicolumn{1}{c|}{\textbf{Paper}} & \textbf{Dataset} & \multicolumn{1}{c|}{\textbf{Point of Capturing}} & \multicolumn{1}{c|}{\textbf{\# Mobile Devices}} & \multicolumn{1}{c|}{\textbf{Content}} \\ \hline
% 2010
\multirow{9}{*}{2010} & \multirow{2}{*}{Afanasyev et al. \cite{Afanasyev2010}} & 10.Afanasyev\_{}0 & APs ($28$ days) & \multirow{2}{*}{$2500$ simultaneous} & Layer-2 and -3 data from RADIUS logs \\ \cline{3-4} \cline{6-6}
& & 10.Afanasyev\_{}1 & Wired (central Internet gateway, five days) & & Layer-3+ headers (no DHCP data) of the first packet of each flow for the first quarter of each hour \\ \cline{2-6}
& \multirow{2}{*}{Falaki et al. \cite{Falaki2010}} & Dataset1 & \multirow{2}{*}{Devices} & Two Android, eight Windows Mobile & Layer-2+ data \\ \cline{3-3} \cline{5-6}
& & Dataset2 & & $33$ Android & Per-app Tx/Rx bytes \\ \cline{2-6}
& \multirow{4}{*}{Maier et al. \cite{Maier2010}} & SEP08 & \multirow{4}{*}{\parbox{0.17\textwidth}{Wired (an ISP's edge router, one day)}} & $200+$ & \multirow{4}{*}{Anonymized DSL data} \\ \cline{3-3} \cline{5-5}
& & APR09 & & $400+$ & \\ \cline{3-3} \cline{5-5}
& & AUG09a & & $500+$ & \\ \cline{3-3} \cline{5-5}
& & AUG09b & & $500+$ & \\ \cline{2-6}
& Shepard et al. \cite{Shepard2010} & 10.Shepard\_{}0 & Devices & $25$ iOS & IP packets \\ \hline
% 2011
\multirow{3}{*}{2011} & \multirow{2}{*}{Gember et al. \cite{Gember2011}} & Net1 & \multirow{2}{*}{\parbox{0.17\textwidth}{APs (campus network, three days)}} & $32{,}166$ & \multirow{2}{*}{Layer-2+ data} \\ \cline{3-3} \cline{5-5}
& & Net2 & & $112$ & \\ \cline{2-6}
& Lee et al. \cite{Lee2011} & 11.Lee\_{}0 & Wired (top-level router of a campus network, six days) & {\tiny N/A} & IP packets \\ \hline
% 2012
\multirow{2}{*}{2012} & \multirow{2}{*}{Chen et al. \cite{Chen2012}} & 12.Chen\_{}0 & Wired (gateway router of a campus Wi-Fi network, three days) & {\tiny N/A} & \multirow{2}{*}{\parbox{0.20\textwidth}{Up to $900$ bytes of each incoming/outgoing packet (including IP, TCP, and application-level headers)}} \\ \cline{3-5}
& & 12.Chen\_{}1 & Wired (gateway router of a campus Wi-Fi network, one day) & {\tiny N/A} & \\ \hline
% 2015
\multirow{3}{*}{2015} & \multirow{3}{*}{Fukuda et al. \cite{Fukuda2015}} & 2013 & \multirow{3}{*}{Devices} & \multirow{3}{*}{\parbox{0.17\textwidth}{$800+$ Android, $700+$ iOS}} & \multirow{3}{*}{\parbox{0.20\textwidth}{Per-app/Per-interface Tx/Rx bytes/packets}} \\ \cline{3-3}
& & 2014 & & & \\ \cline{3-3}
& & 2015 & & & \\ \hline
% 2017
2017 & Wei et al. \cite{Wei2017} & Traffic-May & Wired (Internet gateway of a campus network, one month) & $10{,}756$ Android, $11{,}328$ iOS, $618$ BlackBerry & Layer-3+ data \\ \hline
\end{tabular}
\end{table*}

\revOne{
\subsection{App Identification}
\label{sec:validation_app_identification}

% [Notes for the paragraph]

In Table~\ref{tab:validation_app_identification}, we summarize the information about validation for works that deal with app identification. 
Since this goal involves multi-label classification, we also provide results of traffic analysis proposals.
Eight works out of ten consider a reasonable sample of apps (i.e., more than 40), while Wang et al. in~\cite{Wang2015} and Mongkolluksamee et al. in~\cite{Mongkolluksamee2016} only consider thirteen and five apps, respectively.
Most of the works provide accuracy as a metric to evaluate their proposal, but recall and precision would have provided a better understanding of the performance.

In what follows, we provide some additional observations about these works.
We observe that Lee et al. in~\cite{Lee2011} and the \textit{mobUser} dataset by Rao et al. \cite{Rao2013} have no ground truth, therefore only app matching rate is provided as a result.
Alan et al. in~\cite{Alan2016} provide two additional findings: (i) a performance drop when trained classifiers are tested on updated versions of the same apps; 
%(-> app identification classifiers need to be often retrained).
(ii) a performance drop when training and testing involve different devices, which introduces a possible bias towards a specific OS or vendor. %the device used for training is different from the one used for testing (-> risk of bias ).
Although it is not clearly stated, the results reported in the Table~\ref{tab:validation_app_identification} for this work seem to be the ones using network traffic from the same app versions and installed on the same devices.

Regarding the work by Mongkolluksamee et al. in~\cite{Mongkolluksamee2016}, we underline that "16.Mongkolluksamee\_{}0" dataset (which the reported metrics refer to) is filtered out of background network traffic of Android OS and other apps. 
The authors observe a performance drop if such background traffic is kept, and mitigate that drop by removing short-lived flows.

Taylor et al. in~\cite{Taylor2017} further investigate other aspects with additional experiments (not reported in the table): 
(i) the effect of training on a dataset older than the testing one; 
(ii) the effect of performing training and testing on datasets from different devices, with different app versions, and both;
and (iii) the effect of the proposed reinforced learning approach to deal with ambiguous flows.
}

\begin{sidewaystable*}
%%\color{blue}
\caption{The achieved results in the surveyed works that deal with app identification.}
\label{tab:validation_app_identification}
\centering
\scalebox{0.65}{
\begin{tabular}{|c|l|m{0.21\textwidth}|c|m{0.16\textwidth}|m{0.10\textwidth}|m{0.16\textwidth}|c|c|c|c|c|c|}
\hline
\textbf{Year} & \multicolumn{1}{c|}{\textbf{Paper}} & \multicolumn{1}{c|}{\textbf{Targeted Apps}} & \textbf{Dataset} & \multicolumn{1}{c|}{\textbf{Point of Capturing}} & \multicolumn{1}{c|}{\textbf{Content}} & \multicolumn{1}{c|}{\textbf{Analysis Technique}} & \rot{\textbf{Accuracy }} & \rot{\textbf{Precision }} & \rot{\textbf{Recall }} & \rot{\textbf{F-measure }} & \rot{\textbf{
%App Matching Rate 
AMR
}} & \rot{\textbf{
%App Identification Rate 
AIR
}} \\ \hline
% 2011
2011 & Lee et al. \cite{Lee2011} & Top-$50$ apps in Google Play Store (Android) and Apple App Store (iOS) & 11.Lee\_{}0 & Wired (top-level router of a campus network, six days) & IP packets & Payload signature matching & {\tiny N/A} & {\tiny N/A} & {\tiny N/A} & {\tiny N/A} & $0.154$ & {\tiny N/A} \\ \hline
% 2013
\multirow{5}{*}{2013} & \multirow{2}{*}{Qazi et al. \cite{Qazi2013}} & \multirow{2}{*}{\parbox{0.21\textwidth}{Top-$40$ apps in Google Play Store (Android)}} & 13.Qazi\_{}0 & APs (one access point running OpenFlow) & Layer-3 and -4 data & \multirow{2}{*}{\parbox{0.16\textwidth}{Machine learning (C5.0 decision tree)}} & \multirow{2}{*}{{\tiny N/A}} & \multirow{2}{*}{{\tiny N/A}} & \multirow{2}{*}{{\tiny N/A}} & \multirow{2}{*}{$0.940$} & \multirow{2}{*}{{\tiny N/A}} & \multirow{2}{*}{{\tiny N/A}} \\ \cline{4-6}
& & & 13.Qazi\_{}1 & Devices (five Android devices) & netstat logs & & & & & & & \\ \cline{2-13}
& \multirow{3}{*}{Rao et al. \cite{Rao2013}} & The top-$100$ free apps from Google Play Store (Android), and $732$ apps from a free third-party marketplace (Android) & 13.Rao\_{}0 & \multirow{2}{*}{\parbox{0.16\textwidth}{Wired (VPN server, lab deployment, two Android devices and one iOS device)}} & \multirow{3}{*}{Layer-3+ data} & \multirow{3}{*}{\parbox{0.16\textwidth}{Inspection of Host and User-Agent fields of HTTP messages}} & {\tiny N/A} & {\tiny N/A} & {\tiny N/A} & {\tiny N/A} & {\tiny N/A} & 0.640 \\ \cline{3-4} \cline{8-13}
& & $209$ free apps from Apple App Store (iOS) & 13.Rao\_{}1 & & & & {\tiny N/A} & {\tiny N/A} & {\tiny N/A} & {\tiny N/A} & {\tiny N/A} & 0.890 \\ \cline{3-5} \cline{8-13}
& & The top-$100$ free apps from Google Play Store (Android), $732$ apps from a free third-party marketplace (Android), and $209$ free apps from Apple App Store (iOS) & mobUser & Wired (VPN server, real deployment, eleven Android and fifteen iOS devices) & & & {\tiny N/A} & {\tiny N/A} & {\tiny N/A} & {\tiny N/A} & 0.920 & {\tiny N/A} \\ \hline
% 2015
\multirow{3}{*}{2015} & Le et al. \cite{Le2015} & $70$ Android apps & 15.Le\_{}0 & Devices (nine Android devices) & Per-app IP headers/packets & Machine learning ($84$ network-level features, kCross($10$), linear support vector machine) & {\tiny N/A} & {\tiny N/A} & {\tiny N/A} & $0.701$ & {\tiny N/A} & {\tiny N/A} \\ \cline{2-13}
& Wang et al. \cite{Wang2015} & Thirteen popular free iOS apps & 15.Wang\_{}0 & Monitors (one monitor, one iOS device) & Size/Time of 802.11 frames & Machine learning (twenty statistical features, $0.6/0.4$, random forest) & $0.900+$ & {\tiny N/A} & {\tiny N/A} & {\tiny N/A} & {\tiny N/A} & {\tiny N/A} \\ \cline{2-13}
& Yao et al. \cite{Yao2015} & $651{,}000$ free apps from Google Play Store (Android), $68{,}000$ free apps from Apple App Store (iOS), and $10{,}000$ free apps from Nokia OVI Store (Symbian) & 15.Yao\_{}0 & APs (one access point), emulators & HTTP messages & Model the presence of app identifiers in the HTTP messages as conjunctive rules & $0.990$ & {\tiny N/A} & {\tiny N/A} & {\tiny N/A} & {\tiny N/A} & $0.900+$ \\ \hline
% 2016
\multirow{4}{*}{2016} & \multirow{3}{*}{Alan et al. \cite{Alan2016}} & \multirow{3}{*}{\parbox{0.21\textwidth}{$1{,}595$ of the top-$2{,}000$ free apps from Google Play Store (Android)}} & \multirow{3}{*}{16.Alan\_{}0} & \multirow{3}{*}{\parbox{0.16\textwidth}{APs (one access point, four Android devices)}} & \multirow{3}{*}{\parbox{0.10\textwidth}{IP and TCP headers}} & Machine learning (out($7,1$), classifier based on Jaccard index) & $0.740$ & {\tiny N/A} & {\tiny N/A} & {\tiny N/A} & {\tiny N/A} & {\tiny N/A} \\ \cline{7-13}
& & & & & & Machine learning (out($7,1$), Gaussian naive Bayes) & $0.820$ & {\tiny N/A} & {\tiny N/A} & {\tiny N/A} & {\tiny N/A} & {\tiny N/A} \\ \cline{7-13}
& & & & & & Machine learning (out($7,1$), multinomial naive Bayes) & $0.870$ & {\tiny N/A} & {\tiny N/A} & {\tiny N/A} & {\tiny N/A} & {\tiny N/A} \\ \cline{2-13}
& Mongkolluksamee et al. \cite{Mongkolluksamee2016} & Five popular free Android apps & 16.Mongkolluksamee\_{}0 & Devices (one Android device) & Layer-2+ data & Machine learning ($35$ graphlet- and $14$ histogram-based features, kCross($10$), random forest) & {\tiny N/A} & {\tiny N/A} & {\tiny N/A} & $0.960$ & {\tiny N/A} & {\tiny N/A} \\ \hline
% 2017
\multirow{9}{*}{2017} & \multirow{8}{*}{Taylor et al. \cite{Taylor2017}} & \multirow{3}{*}{\parbox{0.21\textwidth}{$110$ of the top-$200$ free apps in Google Play Store (Android)}} & Dataset-1 & \multirow{8}{*}{\parbox{0.16\textwidth}{Wired (forwarding server, one device)}} & \multirow{8}{*}{\parbox{0.10\textwidth}{IP and TCP headers}} & \multirow{8}{*}{\parbox{0.16\textwidth}{Machine learning ($54$ features based on packet-size statistics, random forest, random $0.75/0.25$ with $50$ averaged repetitions)}} & $0.731$ & $0.745$ & $0.728$ & $0.731$ & {\tiny N/A} & {\tiny N/A} \\ \cline{4-4} \cline{8-13}
& & & Dataset-4 & & & & $0.674$ & $0.686$ & $0.666$ & $0.669$ & {\tiny N/A} & {\tiny N/A} \\ \cline{4-4} \cline{8-13}
& & & Dataset-5 & & & & $0.696$ & $0.697$ & $0.681$ & $0.683$ & {\tiny N/A} & {\tiny N/A} \\ \cline{3-4} \cline{8-13}
& & \multirow{5}{*}{\parbox{0.21\textwidth}{$65$ of the top-$200$ free apps in Google Play Store (Android)}} & Dataset-1a & & & & $0.737$ & $0.744$ & $0.732$ & $0.734$ & {\tiny N/A} & {\tiny N/A} \\ \cline{4-4} \cline{8-13}
& & & Dataset-2 & & & & $0.655$ & $0.688$ & $0.676$ & $0.677$ & {\tiny N/A} & {\tiny N/A} \\ \cline{4-4} \cline{8-13}
& & & Dataset-3 & & & & $0.704$ & $0.713$ & $0.695$ & $0.698$ & {\tiny N/A} & {\tiny N/A} \\ \cline{4-4} \cline{8-13}
& & & Dataset-4a & & & & $0.669$ & $0.685$ & $0.668$ & $0.671$ & {\tiny N/A} & {\tiny N/A} \\ \cline{4-4} \cline{8-13}
& & & Dataset-5a & & & & $0.675$ & $0.677$ & $0.655$ & $0.660$ & {\tiny N/A} & {\tiny N/A} \\ \cline{2-13}
& Chen et al. \cite{Chen2017} & $2{,}500$ Android apps from Google Play Store and seven other third-party marketplaces, and $2{,}500$ malicious Android apps from VirusTotal & 17.Chen\_{}0 & Emulators (automatic UI exploration) & Layer-2+ data & Deep packet inspection & {\tiny N/A} & {\tiny N/A} & {\tiny N/A} & {\tiny N/A} & {\tiny N/A} & $0.761$ \\ \hline
\end{tabular}
}
\end{sidewaystable*}

\revOne{
\subsection{Usage Study}
\label{sec:validation_usage_study}

In Table~\ref{tab:validation_usage_study} we provide information about the datasets analyzed in the surveyed works that deal with usage study.
We can notice that eight works out of ten that deal with usage study also deal with traffic characterization (Tables~\ref{tab:validation_traffic_characterization_apps/services} and~\ref{tab:validation_traffic_characterization_population}).
Nonetheless, Table~\ref{tab:validation_usage_study} is meant to offer a proper comparison also with works that purely deal with usage study. 
An evident difference is that pure usage studies employ fewer devices when compared with other works. 
Indeed, while most of the other works employ more than two hundred devices, Ham et al. in~\cite{Ham2012} and Soikkeli et al. in~\cite{Soikkeli2015} validate their finding on only ten Android and $120$ (unspecified) devices, respectively. 
Moreover, it is interesting to notice that both these works just rely on transmitted and received bytes. 
%As we can notice, we already report validation information of, since 
}

\begin{table*}
%%\color{blue}
\caption{The datasets of the surveyed works that deal with usage study.}
\label{tab:validation_usage_study}
\centering
\scalebox{0.99}{
\begin{tabular}{|c|l|c|m{0.17\textwidth}|m{0.17\textwidth}|m{0.25\textwidth}|}
\hline
\textbf{Year} & \multicolumn{1}{c|}{\textbf{Paper}} & \textbf{Dataset} & \multicolumn{1}{c|}{\textbf{Point of Capturing}} & \multicolumn{1}{c|}{\textbf{\# Mobile Devices}} & \multicolumn{1}{c|}{\textbf{Content}} \\ \hline
% 2010
\multirow{8}{*}{2010} & \multirow{2}{*}{Afanasyev et al. \cite{Afanasyev2010}} & 10.Afanasyev\_{}0 & APs ($28$ days) & \multirow{2}{*}{$2500$ simultaneous} & Layer-2 and -3 data from RADIUS logs \\ \cline{3-4} \cline{6-6}
& & 10.Afanasyev\_{}1 & Wired (central Internet gateway, five days) & & Layer-3+ headers (no DHCP data) of the first packet of each flow for the first quarter of each hour \\ \cline{2-6}
& \multirow{2}{*}{Falaki et al. \cite{Falaki2010}} & Dataset1 & \multirow{2}{*}{Devices} & Two Android, eight Windows Mobile & Layer-2+ data \\ \cline{3-3} \cline{5-6}
& & Dataset2 & & $33$ Android & Per-app Tx/Rx bytes \\ \cline{2-6}
& \multirow{4}{*}{Maier et al. \cite{Maier2010}} & SEP08 & \multirow{4}{*}{\parbox{0.17\textwidth}{Wired (an ISP's edge router, one day)}} & $200+$ & \multirow{4}{*}{Anonymized DSL data} \\ \cline{3-3} \cline{5-5}
& & APR09 & & $400+$ & \\ \cline{3-3} \cline{5-5}
& & AUG09a & & $500+$ & \\ \cline{3-3} \cline{5-5}
& & AUG09b & & $500+$ & \\ \hline
% 2011
\multirow{8}{*}{2011} & \multirow{5}{*}{Finamore et al. \cite{Finamore2011}} & US-Campus & \multirow{2}{*}{\parbox{0.17\textwidth}{Wired (within a campus network, one week)}} & \multirow{5}{*}{\parbox{0.17\textwidth}{{\tiny N/A}}} & \multirow{5}{*}{\parbox{0.17\textwidth}{IP packets}} \\ \cline{3-3}
& & EU1-Campus & & & \\ \cline{3-4}
& & EU1-ADSL & \multirow{3}{*}{\parbox{0.17\textwidth}{Wired (within an ISP's network, one week)}} & & \\ \cline{3-3}
& & EU1-FTTH & & & \\ \cline{3-3}
& & EU2-ADSL & & & \\ \cline{2-6}
& \multirow{2}{*}{Gember et al. \cite{Gember2011}} & Net1 & \multirow{2}{*}{\parbox{0.17\textwidth}{APs (campus network, three days)}} & $32{,}166$ & \multirow{2}{*}{Layer-2+ data} \\ \cline{3-3} \cline{5-5}
& & Net2 & & $112$ & \\ \cline{2-6}
& Lee et al. \cite{Lee2011} & 11.Lee\_{}0 & Wired (top-level router of a campus network, six days) & {\tiny N/A} & IP packets \\ \hline
% 2012
2012 & Ham et al. \cite{Ham2012} & 12.Ham\_{}0 & Devices & Ten Android & Per-process/Per-interface Tx/Rx bytes/packets \\ \hline
% 2015
\multirow{4}{*}{2015} & \multirow{3}{*}{Fukuda et al. \cite{Fukuda2015}} & 2013 & \multirow{3}{*}{Devices} & \multirow{3}{*}{\parbox{0.17\textwidth}{$800+$ Android, $700+$ iOS}} & \multirow{3}{*}{\parbox{0.17\textwidth}{Per-app/Per-interface Tx/Rx bytes/packets}} \\ \cline{3-3}
& & 2014 & & & \\ \cline{3-3}
& & 2015 & & & \\ \cline{2-6}
& Soikkeli et al. \cite{Soikkeli2015} & 15.Soikkeli\_{}0 & Devices & $120$ & Tx/Rx bytes \\ \hline
% 2017
2017 & Wei et al. \cite{Wei2017} & Traffic-May & Wired (Internet gateway of a campus network, one month) & $10{,}756$ Android, $11{,}328$ iOS, $618$ BlackBerry & Layer-3+ data \\ \hline
\end{tabular}
}
\end{table*}

\revOne{
\subsection{PII Leakage Detection}
\label{sec:validation_PII_leakage_detection}

% [Notes for the paragraph]

In Table~\ref{tab:validation_PII_leakage_detection}, we summarize the validation methods used by works that deal with PII leakage detection.
%As for works that deal with traffic characterization, 
Most of the works on PII leakage detection focus on providing findings and observations about user private information transmission (in clear) to third-party services (summarized in detail in Section~\ref{sec:PII_leakage_detection}), rather than provide actual results from a classification task.
Despite this, Kuzuno et al. in~\cite{Kuzuno2013} and Ren et al. in~\cite{Ren2016} report results in terms of accuracy since they are the only ones that rely on machine learning methods. 
It worths also noticing that Kuzuno et al. in~\cite{Kuzuno2013} only report classification accuracy without providing any additional finding. 

As an overall consideration about datasets, five works out of eight carry out an analysis on more than one thousand apps. 
Besides, Song et al. in~\cite{Song2015} consider only 53 apps and Le et al. in~\cite{Le2015} do not specify the number of considered apps.
Moreover, Stevens et al. in~\cite{Stevens2012} carry out their analysis on a custom app that includes thirteen popular Android ad libraries.
}

\begin{sidewaystable*}
%%\color{blue}
\caption{The achieved results in the surveyed works that deal with PII leakage detection.}
\label{tab:validation_PII_leakage_detection}
\centering
\scalebox{0.80}{
\begin{tabular}{|c|l|m{0.21\textwidth}|c|m{0.16\textwidth}|m{0.10\textwidth}|m{0.16\textwidth}|c|c|c|c|}
\hline
\textbf{Year} & \multicolumn{1}{c|}{\textbf{Paper}} & \multicolumn{1}{c|}{\textbf{Selected Apps}} & \textbf{Dataset} & \multicolumn{1}{c|}{\textbf{Point of Capturing}} & \multicolumn{1}{c|}{\textbf{Content}} & \multicolumn{1}{c|}{\textbf{Analysis Technique}} & \rot{\textbf{Accuracy }} & \rot{\textbf{Precision }} & \rot{\textbf{Recall }} & \rot{\textbf{F-measure }} \\ \hline
% 2012
2012 & Stevens et al. \cite{Stevens2012} & One custom Android app with embedded ad libraries & 12.Stevens\_{}0 & APs (one access point, one Android device) & HTTP messages & Deep packet inspection & {\tiny N/A} & {\tiny N/A} & {\tiny N/A} & {\tiny N/A} \\ \hline
% 2013
\multirow{5}{*}{2013} & Kuzuno et al. \cite{Kuzuno2013} & $1{,}188$ free Android apps & 13.Kuzuno\_{}0 & Devices (one Android device) & Layer-2+ data & Machine learning (hierarchical clustering) & $0.940$ & {\tiny N/A} & {\tiny N/A} & {\tiny N/A} \\ \cline{2-11}
& \multirow{4}{*}{Rao et al. \cite{Rao2013}} & The top-$100$ free Android apps from Google Play Store, and $732$ Android apps from a free third-party marketplace & 13.Rao\_{}0 & \multirow{3}{*}{\parbox{0.16\textwidth}{Wired (VPN server, lab deployment, two Android devices and one iOS device)}} & \multirow{4}{*}{Layer-3+ data} & \multirow{4}{*}{\parbox{0.16\textwidth}{Deep packet inspection}} & {\tiny N/A} & {\tiny N/A} & {\tiny N/A} & {\tiny N/A} \\ \cline{3-4} \cline{8-11}
& & $209$ free iOS apps from Apple App Store & 13.Rao\_{}1 & & & & {\tiny N/A} & {\tiny N/A} & {\tiny N/A} & {\tiny N/A} \\ \cline{3-4} \cline{8-11}
& & $99$ malicious Android apps & 13.Rao\_{}2 & & & & {\tiny N/A} & {\tiny N/A} & {\tiny N/A} & {\tiny N/A} \\ \cline{3-5} \cline{8-11}
& & The top-$100$ free apps from Google Play Store (Android), $732$ apps from a free third-party marketplace (Android), and $209$ free apps from Apple App Store (iOS) & mobUser & Wired (VPN server, real deployment, eleven Android and fifteen iOS devices) & & & {\tiny N/A} & {\tiny N/A} & {\tiny N/A} & {\tiny N/A} \\ \hline
% 2015
\multirow{2}{*}{2015} & Le et al. \cite{Le2015} & Unspecified set of Android apps & 15.Le\_{}0 & Devices (nine Android devices) & Per-app IP headers/packets & Deep packet inspection & {\tiny N/A} & {\tiny N/A} & {\tiny N/A} & {\tiny N/A} \\ \cline{2-11}
& Song et al. \cite{Song2015} & $53$ Android apps from Google Play Store & 15.Song\_{}0 & Devices (one Android device) & IP packets & Deep packet inspection & {\tiny N/A} & {\tiny N/A} & {\tiny N/A} & {\tiny N/A} \\ \hline
% 2016
\multirow{2}{*}{2016} & Ren et al. \cite{Ren2016} & Top-$100$ free Android apps from Google Play Store, $850$ of the top-$1{,}000$ free Android apps from a third-party Android marketplace, top-$100$ free iOS apps from Apple App Store, and top-$100$ free Windows Phone apps from Windows Phone Store & 16.Ren\_{}0 & Wired (VPN server) & Layer-3+ data & Machine learning (bag-of-word model, C4.5 decision tree) & $0.981$ & {\tiny N/A} & {\tiny N/A} & {\tiny N/A} \\ \cline{2-11}
& Vanrykel et al. \cite{Vanrykel2016} & $1260$ Android apps from Google Play Store & 16.Vanrykel\_{}0 & Wired (two VPN servers) & HTTP messages & Deep packet inspection & {\tiny N/A} & {\tiny N/A} & {\tiny N/A} & {\tiny N/A} \\ \hline
% 2017
\multirow{4}{*}{2017} & \multirow{3}{*}{\revOne{Cheng et al. \cite{Cheng2017}}} & \multirow{3}{*}{\parbox{0.21\textwidth}{Seven Android apps, plus a self-developed PII-leaking malicious Android app}} & \multirow{3}{*}{17.Cheng\_{}0} & \multirow{3}{*}{\parbox{0.16\textwidth}{Wired (forwarding server, one Android device)}} & \multirow{3}{*}{\parbox{0.10\textwidth}{IP and TCP headers}} & Machine learning (random forest, both incoming and outgoing traffic) & {\tiny N/A} & $0.952$ & $0.947$ & $0.948$ \\ \cline{7-11}
& & & & & & Machine learning (random forest, incoming traffic only) & {\tiny N/A} & $0.981$ & $0.981$ & $0.980$ \\ \cline{7-11}
& & & & & & Machine learning (random forest, outgoing traffic only) & {\tiny N/A} & $0.991$ & $0.990$ & $0.990$ \\ \cline{2-11}
& Continella et al. \cite{Continella2017} & $1{,}004$ Android apps from Google Play Store & 17.Continella\_{}0 & Wired (forwarding server, six Android devices) & HTTP messages & Deep packet inspection, differential analysis & {\tiny N/A} & {\tiny N/A} & {\tiny N/A} & {\tiny N/A} \\ \hline
\end{tabular}
}
\end{sidewaystable*}

\revOne{
\subsection{Malware Detection}
\label{sec:validation_malware_detection}

% [Notes for the paragraph]
We present the validation methods and results for works that deal with malware detection in Table~\ref{tab:validation_malware_detection}.
It worths noticing that five works out of eight validate their proposals on datasets collected from less than two devices. 
Moreover, only Wei et al. in~\cite{Wei2012} and Wang et al. in~\cite{Wang2016} rely on a forwarding server, while the other datasets are collected directly from devices. 
Surprisingly, Narudin et al. in~\cite{Narudin2016} are the only that rely on emulators to collect malware traffic, despite the use of emulators-based sandboxes is a common practice in mobile malware analysis.

In what follows, we provide some additional observations regarding the validation of malware detection works.
The results of the work by Shabtai et al. in~\cite{Shabtai2012} are referred to their experiment I. 
Since experiment I provides two sub-experiments (i.e., gaming apps and tool apps), we averaged the results for each metric.
Moreover, the authors carry out three other experiments: 
in experiment II, the authors evaluate the effect of testing on apps not used in training; in experiment III, the authors evaluate the effect of performing training and testing on different devices; and in experiment IV, the authors evaluate the combination of both.
The authors also evaluate three metrics for feature selection: chi-square, Fisher score, and information gain.

% The maxi-table refers to the experiment I; each metric value we provide is the average of the two provided by the authors (one for games, the other for tools).

% See pages 16-18 for details on the content of the datasets, as well as on the other experiments (II, IV).

% In experiment II, the authors evaluate the effect of testing on apps not used in training; in experiment III, the authors evaluate the effect of performing training and testing on different devices; in experiment IV, the authors evaluate the combination of both.

% The authors evaluate the use of three metrics for feature selection: chi-square, Fisher score, and information gain.
% The "14.Shabtai\_{}0" dataset refers to the experiments with the infected versions of the benign apps, while the "14.Shabtai\_{}1" and "14.Shabtai\_{}2" ones refer to the experiments with the self-updating malware.

% [Shabtai et al. \cite{Shabtai2014}]
Regarding the datasets used by Shabtai et al. in~\cite{Shabtai2014}, the ``14.Shabtai\_{}0" dataset refers to the experiments with the infected versions of the benign apps, while ``14.Shabtai\_{}1" and ``14.Shabtai\_{}2" refer to the experiments with the self-updating malware.

Despite Zaman et al. in~\cite{Zaman2015} rely on DPI on HTTP messages and only focus on two malware samples, their proposal is able to identify only one malware out of two (i.e., DroidKungFu), thus having an accuracy of $0.5$. 
Moreover, their dataset involves traffic collected from a single device.  
Clearly, this proposal cannot be considered effective and well validated.

% [Zaman et al. \cite{Zaman2015}]

% Only DroidKungFu has been detected.

% [Narudin et al. \cite{Narudin2016}]
Narudin et al. in~\cite{Narudin2016} validate their proposal on two datasets, namely ``Ds1000'' and ``Priv''.
In particular, the authors argue that applying feature selection on ``Ds1000'' dataset slightly improves classifiers performance (except for multi-layer perceptron which has a performance drop). %Among the works that deal with malware detection
% Regarding the "Ds1000" dataset, the authors show that the feature selection slightly improves the performance of the classifiers, except for the multi-layer perceptron that instead has a slight drop in classification performance.
}

\begin{sidewaystable*}
%%\color{blue}
\caption{The achieved results in the surveyed works that deal with malware detection.}
\label{tab:validation_malware_detection}
\centering
\scalebox{0.70}{
\begin{tabular}{|c|l|m{0.21\textwidth}|c|m{0.16\textwidth}|m{0.10\textwidth}|m{0.16\textwidth}|m{0.12\textwidth}|c|c|c|c|}
\hline
\textbf{Year} & \multicolumn{1}{c|}{\textbf{Paper}} & \multicolumn{1}{c|}{\textbf{Selected Apps}} & \textbf{Dataset} & \multicolumn{1}{c|}{\textbf{Point of Capturing}} & \multicolumn{1}{c|}{\textbf{Content}} & \multicolumn{1}{c|}{\textbf{Analysis Technique}} & \multicolumn{1}{c|}{\textbf{Analysis Details}} & \rot{\textbf{Accuracy }} & \rot{\textbf{Precision }} & \rot{\textbf{Recall }} & \rot{\textbf{F-measure }} \\ \hline
% 2012
\multirow{8}{*}{2012} & \multirow{6}{*}{Shabtai et al. \cite{Shabtai2012}} & \multirow{6}{*}{\parbox{0.21\textwidth}{$40$ benign ($20$ games and $20$ tools) and $4$ malicious Android apps}} & \multirow{6}{*}{\parbox{0.12\textwidth}{\centering From 12.Shabtai\_{}0 to 12.Shabtai\_{}9 (ten datasets)}} & \multirow{6}{*}{\parbox{0.16\textwidth}{Devices (two Android devices)}} & \multirow{6}{*}{\parbox{0.10\textwidth}{Per-app cellular/Wi-Fi Tx/Rx bytes/packets}} & \multirow{6}{*}{\parbox{0.16\textwidth}{Machine learning (random $0.8/0.2$ on each dataset, one classifier run on each dataset per device, average)}} & Bayesian networks & $0.992$ & {\tiny N/A} & $0.984$ & {\tiny N/A} \\ \cline{8-12}
& & & & & & & J48 decision tree & $0.999$ & {\tiny N/A} & $0.999$ & {\tiny N/A} \\ \cline{8-12}
& & & & & & & Histograms & $0.931$ & {\tiny N/A} & $0.949$ & {\tiny N/A} \\ \cline{8-12}
& & & & & & & $k$-means & $0.793$ & {\tiny N/A} & $0.689$ & {\tiny N/A} \\ \cline{8-12}
& & & & & & & Logistic regression & $0.999$ & {\tiny N/A} & $0.998$ & {\tiny N/A} \\ \cline{8-12}
& & & & & & & Naive Bayes & $0.954$ & {\tiny N/A} & $0.913$ & {\tiny N/A} \\ \cline{2-12}
& \multirow{2}{*}{Su et al. \cite{Su2012}} & \multirow{2}{*}{\parbox{0.21\textwidth}{(Android) 49 malicious apps (from 22 malware families) and 60 benign apps for training, 50 malicious apps (from 22 malware families) and 70 benign apps for testing}} & \multirow{2}{*}{\parbox{0.12\textwidth}{\centering 12.Su\_{}0 (training) and 12.Su\_{}1 (testing)}} & \multirow{2}{*}{Devices} & \multirow{2}{*}{\parbox{0.16\textwidth}{Layer-2+ data}} & \multirow{2}{*}{\parbox{0.16\textwidth}{Machine learning (supervised learning)}} & \parbox{0.16\textwidth}{\rule{0.0cm}{0.4cm}\\ J48 decision tree\\ \rule{0.0cm}{0.5cm}} & $0.916$ & {\tiny N/A} & {\tiny N/A} & {\tiny N/A} \\ \cline{8-12}
& & & & & & & \parbox{0.16\textwidth}{\rule{0.0cm}{0.4cm}\\ Random forest\\ \rule{0.0cm}{0.5cm}} & $0.967$ & {\tiny N/A} & {\tiny N/A} & {\tiny N/A} \\ \cline{2-12}
& Wei et al. \cite{Wei2012} & $102$ malicious Android apps from a public Android malware dataset, and popular Android apps from Google Play Store & 12.Wei\_{}0 & Wired (forwarding server) & DNS data & Machine learning (unsupervised learning, kCross($10$)) & Indipendent Component Analysis (ICA) & $1.000$ & $1.000$ & {\tiny N/A} & {\tiny N/A} \\ \hline
% 2014
\multirow{3}{*}{2014} & \multirow{3}{*}{Shabtai et al. \cite{Shabtai2014}} & \multirow{3}{*}{\parbox{0.21\textwidth}{Several benign Android apps, ten self-updating malicious Android apps, and the infected version of five of the chosen benign apps}} & \parbox{0.12\textwidth}{\rule{0.0cm}{0.075cm}\\ \centering 14.Shabtai\_{}0\\ \rule{0.0cm}{0.0cm}} & \multirow{3}{*}{\parbox{0.16\textwidth}{Devices (one Android device)}} & \multirow{3}{*}{\parbox{0.10\textwidth}{Per-app Tx/Rx bytes and percent out of total Tx/Rx bytes}} & \multirow{3}{*}{\parbox{0.16\textwidth}{Machine learning (supervised learning)}} & \multirow{2}{*}{\parbox{0.12\textwidth}{Decision tree}} & {\tiny N/A} & {\tiny N/A} & $0.926$ & {\tiny N/A} \\ \cline{4-4} \cline{9-12}
& & & \parbox{0.12\textwidth}{\rule{0.0cm}{0.075cm}\\ \centering 14.Shabtai\_{}1\\ \rule{0.0cm}{0.025cm}} & & & & & {\tiny N/A} & {\tiny N/A} & $0.954$ & {\tiny N/A} \\ \cline{4-4} \cline{9-12}
& & & \parbox{0.12\textwidth}{\rule{0.0cm}{0.0cm}\\ \centering 14.Shabtai\_{}2\\ \rule{0.0cm}{0.0cm}} & & & & & {\tiny N/A} & {\tiny N/A} & $0.824$ & {\tiny N/A} \\ \hline
% 2015
2015 & Zaman et al. \cite{Zaman2015} & Two Android malware samples (DroidKungFu and AnserverBot) & 15.Zaman\_{}0 & Devices (one Android device) & Layer-2+ data and netstat logs & Malicious domain blacklisting & Inspect HTTP messages for the URLs of malicious domains & $0.500$ & {\tiny N/A} & {\tiny N/A} & {\tiny N/A} \\ \hline
% 2016
\multirow{12}{*}{2016} & \multirow{10}{*}{Narudin et al. \cite{Narudin2016}} & \multirow{5}{*}{\parbox{0.21\textwidth}{The top-twenty free (benign) Android apps from Google Play Store, and $1{,}000$ malicious Android apps (from $49$ malware families)}} & \multirow{5}{*}{Ds1000} & \multirow{10}{*}{\parbox{0.16\textwidth}{Devices (one Android device, for benign apps), emulators (for malicious apps)}} & \multirow{10}{*}{\parbox{0.10\textwidth}{Per-app layer-2+ data}} & \multirow{5}{*}{\parbox{0.16\textwidth}{Machine learning (supervised learning, feature selection, kCross($10$))}} & Bayesian network & {\tiny N/A} & $1.000$ & $1.000$ & $1.000$ \\ \cline{8-12}
& & & & & & & Multi-layer perceptron & {\tiny N/A} & $0.887$ & $0.883$ & $0.880$ \\ \cline{8-12}
& & & & & & & J48 decision tree & {\tiny N/A} & $1.000$ & $1.000$ & $1.000$ \\ \cline{8-12}
& & & & & & & $k$-nearest neighbors & {\tiny N/A} & $0.997$ & $0.997$ & $0.997$ \\ \cline{8-12}
& & & & & & & Random forest & {\tiny N/A} & $1.000$ & $1.000$ & $1.000$ \\ \cline{3-4} \cline{7-12}
& & \multirow{5}{*}{\parbox{0.21\textwidth}{The top-twenty free (benign) Android apps from Google Play Store, and $30$ new (in 2013) malicious Android apps (from fourteen malware families)}} & \multirow{5}{*}{Priv} & & & \multirow{5}{*}{\parbox{0.16\textwidth}{Machine learning (supervised learning, unspecified holdout)}} & Bayesian network & {\tiny N/A} & $0.841$ & $0.756$ & $0.750$ \\ \cline{8-12}
& & & & & & & Multi-layer perceptron & {\tiny N/A} & $0.880$ & $0.840$ & $0.839$ \\ \cline{8-12}
& & & & & & & J48 decision tree & {\tiny N/A} & $0.836$ & $0.738$ & $0.730$ \\ \cline{8-12}
& & & & & & & $k$-nearest neighbors & {\tiny N/A} & $0.884$ & $0.846$ & $0.845$ \\ \cline{8-12}
& & & & & & & Random forest & {\tiny N/A} & $0.837$ & $0.741$ & $0.734$ \\ \cline{2-12}
& \multirow{2}{*}{Wang et al. \cite{Wang2016}} & \multirow{2}{*}{\parbox{0.21\textwidth}{$8{,}312$ benign Android apps from Google Play Store, and $5{,}560$ malicious Android apps}} & \multirow{2}{*}{16Wang\_{}0} & \multirow{2}{*}{\parbox{0.16\textwidth}{Wired (forwarding server)}} & \multirow{2}{*}{\parbox{0.10\textwidth}{TCP- and HTTP-related data}} & \multirow{2}{*}{\parbox{0.16\textwidth}{Machine learning (supervised learning)}} & C4.5 decision tree on TCP-related features & {\tiny N/A} & {\tiny N/A} & $0.982$ & {\tiny N/A} \\ \cline{8-12}
& & & & & & & C4.5 decision tree on HTTP-related features & {\tiny N/A} & {\tiny N/A} & $0.997$ & {\tiny N/A} \\ \hline
% 2017
\multirow{2}{*}{2017} & \multirow{2}{*}{Arora et al. \cite{Arora2017}} & \multirow{2}{*}{\parbox{0.21\textwidth}{Benign and malicious (from eleven families) Android apps for training, benign and malicious (from six families) Android apps for testing (different from those used for training)}} & \multirow{2}{*}{\parbox{0.12\textwidth}{\centering 17.Arora\_{}0 (training) and 17.Arora\_{}1 (testing)}} & \multirow{2}{*}{\parbox{0.16\textwidth}{Devices (one Android device)}} & \multirow{2}{*}{\parbox{0.10\textwidth}{IP packets}} & \multirow{2}{*}{\parbox{0.16\textwidth}{Machine learning (supervised learning)}} & Naive Bayes, $22$ network-layer features (no feature selection) & {\tiny N/A} & {\tiny N/A} & {\tiny N/A} & $0.873$ \\ \cline{8-12}
& & & & & & & Naive Bayes, $9$ network-layer features chosen by feature selection & {\tiny N/A} & {\tiny N/A} & {\tiny N/A} & $0.833$ \\ \hline
\end{tabular}
}
\end{sidewaystable*}

\revOne{
\subsection{User Action Identification}
\label{sec:validation_user_action_identification}

% [Notes for the paragraph]
In Table~\ref{tab:validation_user_action_identification}, we compare the validation methods and results of works that aim at identifying user actions.
Given the nature of this goal and the difficulty to build a ground truth, all the works leverage their own dataset to validate their analysis. 
Among such works, only one relies on dictionaries (i.e., Coull et al. in~\cite{Coull2014}), while the others use machine learning techniques. 
In what follows, we provide additional observations about datasets, considered actions, and experiments.

Watkins et al. in~\cite{Watkins2013} do not consider actual actions for their analysis.
More precisely, they focus on the resource consumption of actions, dividing these into three categories: (i) CPU intensive; (ii) I/O intensive; and (iii) non-CPU intensive.
%For this reason, only three actions are considered for this work, while other works  
The reported accuracy is the average accuracy across the two employed devices (i.e., 93\% and 95\%, respectively).

% [Watkins et al. \cite{Watkins2013}]
% Rather than on three actions, the focus is on three *types* of action: (i) CPU intensive; (ii) I/O intensive; and (iii) non-CPU intensive.
% The accuracy reported in the maxi-table is the average accuracy across the two employed devices (93% and 95%, respectively).

Coull et al. in~\cite{Coull2014} consider only the iMessage app and choose the following actions: ``start typing'', ``stop typing'', ``send text'', ``send attachment'', and ``read receipt''. 
The reported result (i.e., accuracy  $> 0.99$) is related to the identification of such actions, except ``read receipt" which is often confused with ``start typing". 
Moreover, that results assume that traffic from iMessage app has been correctly detected.
The authors carry out other experiments related to message attributes which are not reported in Table~\ref{tab:validation_user_action_identification}:
\begin{itemize}
\item The first experiment aims to infer the language (among six languages: Chinese, English, French, German, Russian, and Spanish) of the exchanged messages. It uses a multinomial naive Bayes classifier with $10$-fold cross-validation. 
Assuming to have correctly identified an iMessage action on a mobile device running iOS, the language classification achieves more than 80\% accuracy by considering the first 50 packets.
\item The second experiment aims to infer the length of the exchanged messages. 
It uses linear regression and $10$-fold cross-validation. 
This method is able to achieve an average error of 6.27 characters for text messages, and an absolute error of at most 10 bytes for attachment transfers.
\end{itemize}

% [Coull et al. \cite{Coull2014}]

% Considered user actions: "start typing", "stop typing", "send text", "send attachment", and "read receipt".

% Regarding the results reported in the maxi-table, they refer to a scenario in which the attacker has correctly inferred that the target mobile device is running iOS. Moreover, the reported accuracy refers to all the considered actions except "read receipt", which is often confused with "start typing".

% First experiment related to message attributes (not reported in the maxi-table): to infer the language (among six languages: Chinese, English, French, German, Russian, and Spanish) of the exchanged messages. Machine learning: multinomial naive Bayes, $10$-fold cross-validation. Results: assuming to have correctly identified an iMessage action on a mobile device running iOS, the language classification achieves more than 80% accuracy by considering the first 50 packets.

% Second experiment related to message attributes (not reported in the maxi-table): to infer the length of the exchanged messages. Machine learning: linear regression with least squares estimation, $10$-fold cross-validation. Results: an average error of 6.27 characters for text messages, and an absolute error of at most 10 bytes for attachment transfers.

% [Park et al. \cite{Park2015}]

Park et al. in~\cite{Park2015}  consider eleven user actions performed on the  KakaoTalk app: ``join a chat room'', ``leave the chat room'', ``receive a message'', ``send a message'', ``add a friend'', ``hide a friend'', ``block a user'', ``unblock a blocked user'', ``re-add a blocked friend'', ``view a user's profile'', and ``synchronize friend list''.
The precision, recall, and F-measure reported in Table~\ref{tab:validation_user_action_identification} are the average across all the considered user actions.
%
% [Park et al. \cite{Park2015}]
%
% Considered user actions: "join a chat room", "leave the chat room", "receive a message", "send a message", "add a friend", "hide a friend", "block a user", "unblock a blocked user", "re-add a blocked friend", "view a user's profile", and "synchronize friend list".
%
% The precision, recall, and F-measure reported in the maxi-table are the average across all the considered user actions.
%
% [Fu et al. \cite{Fu2016}]
%
The results achieved by Fu et al. in~\cite{Fu2016} refer to the best-performing classifier (i.e., the random forest).
}

\begin{sidewaystable*}
%\color{blue}
\caption{The achieved results in the surveyed works that deal with user action identification.}
\label{tab:validation_user_action_identification}
\centering
\scalebox{0.76}{
\begin{tabular}{|c|l|c|m{0.16\textwidth}|m{0.10\textwidth}|m{0.16\textwidth}|c|m{0.10\textwidth}|c|c|c|c|c|}
\hline
\textbf{Year} & \multicolumn{1}{c|}{\textbf{Paper}} & \textbf{Dataset} & \multicolumn{1}{c|}{\textbf{Point of Capturing}} & \multicolumn{1}{c|}{\textbf{Content}} & \multicolumn{1}{c|}{\textbf{Analysis Technique}} & \textbf{Platform} & \multicolumn{1}{c|}{\textbf{App(s)}} & \textbf{\# Actions} & \rot{\textbf{Accuracy }} & \rot{\textbf{Precision }} & \rot{\textbf{Recall }} & \rot{\textbf{F-measure }} \\ \hline
\multirow{6}{*}{2013} & \multirow{6}{*}{Watkins et al. \cite{Watkins2013}} & \multirow{6}{*}{13.Watkins\_{}0} & \multirow{6}{*}{\parbox{0.16\textwidth}{APs (one access point, two Android devices)}} & \multirow{6}{*}{\parbox{0.10\textwidth}{Inter-packet time of ICMP responses}} & \multirow{6}{*}{\parbox{0.16\textwidth}{Machine learning ($0.5/0.5$, neural-fuzzy classifier)}} & \multirow{6}{*}{Android} & Adobe Reader & \multirow{6}{*}{$3$} & \multirow{6}{*}{$0.940$} & \multirow{6}{*}{{\tiny N/A}} & \multirow{6}{*}{{\tiny N/A}} & \multirow{6}{*}{{\tiny N/A}} \\ \cline{8-8}
& & & & & & & Angry Birds & & & & & \\ \cline{8-8}
& & & & & & & AppMgr III & & & & & \\ \cline{8-8}
& & & & & & & Music Player & & & & & \\ \cline{8-8}
& & & & & & & Temple Run & & & & & \\ \cline{8-8}
& & & & & & & ZArchiver & & & & & \\ \hline
2014 & Coull et al. \cite{Coull2014} & 14.Coull\_{}0 & Devices & Packet sizes within APNS connection & Dictionary, kCross($10$) & iOS & iMessage & $5$ & $0.990+$ & {\tiny N/A} & {\tiny N/A} & {\tiny N/A} \\ \hline
2015 & Park et al. \cite{Park2015} & 15.Park\_{}0 & Wired (forwarding server, one Android device) & IP and TCP headers & Machine learning (agglomerative hierarchical clustering, random forest, kCross($10$)) & Android & KakaoTalk & $11$ & $0.997$ & $0.976$ & $0.977$ & $0.977$ \\ \hline
\multirow{10}{*}{2016} & \multirow{7}{*}{Conti et al. \cite{Conti2016}} & \multirow{7}{*}{16.Conti\_{}0} & \multirow{7}{*}{\parbox{0.16\textwidth}{Wired (forwarding server, one Android device)}} & \multirow{7}{*}{\parbox{0.10\textwidth}{IP and TCP headers}} & \multirow{7}{*}{\parbox{0.16\textwidth}{Machine learning (agglomerative hierarchical clustering, random forest)}} & \multirow{7}{*}{Android} & Dropbox & $9$ & {\tiny N/A} & $0.950$ & $0.920$ & $0.920$ \\ \cline{8-13}
& & & & & & & Evernote & $7$ & {\tiny N/A} & $1.000$ & $1.000$ & $1.000$ \\ \cline{8-13}
& & & & & & & Facebook & $8$ & {\tiny N/A} & $0.990$ & $0.980$ & $0.990$ \\ \cline{8-13}
& & & & & & & Gmail & $5$ & {\tiny N/A} & $0.830$ & $0.850$ & $0.860$ \\ \cline{8-13}
& & & & & & & Google+ & $11$ & {\tiny N/A} & $0.900$ & $0.940$ & $0.920$ \\ \cline{8-13}
& & & & & & & Tumblr & $11$ & {\tiny N/A} & $0.990$ & $0.990$ & $0.990$ \\ \cline{8-13}
& & & & & & & Twitter & $7$ & {\tiny N/A} & $0.980$ & $0.970$ & $0.970$ \\ \cline{2-13}
& \multirow{2}{*}{Fu et al. \cite{Fu2016}} & 16.Fu\_{}0 & APs (one virtual access point per device, fifteen Android devices) & \multirow{2}{*}{\parbox{0.10\textwidth}{Size and timing of IP packets}} & \multirow{2}{*}{\parbox{0.16\textwidth}{Machine learning (hierarchical clustering, random forest, hidden Markov model)}} & \multirow{2}{*}{Android} & WeChat & $8$ & $0.960+$ & {\tiny N/A} & {\tiny N/A} & {\tiny N/A} \\ \cline{3-4} \cline{8-13}
& & 16.Fu\_{}1 & APs (one virtual access point per device, five Android devices) & & & & WhatsApp & $6$ & $0.976$ & {\tiny N/A} & {\tiny N/A} & {\tiny N/A} \\ \cline{2-13}
& Saltaformaggio et al. \cite{Saltaformaggio2016} & 16.Saltaformaggio\_{}0 & APs (one access point, two iOS and five Android devices) & IP headers & Machine learning ($k$-means clustering, multi-class support vector machine) & Android, iOS & $22$ popular apps available on both platforms & $35$ & {\tiny N/A} & $0.780$ & $0.760$ & {\tiny N/A} \\ \hline
\end{tabular}
}
\end{sidewaystable*}

\subsection{Operating System Identification}
\label{sec:validation_OS_identification}

In Table~\ref{tab:validation_OS_identification}, we report datasets and results for the works that deal with operating system identification.
As we can notice, all four works employ supervised learning techniques.
% [Notes for the paragraph]

Coull et al. in~\cite{Coull2014} aim at determining whether the user is using iMessage on iOS (i.e., mobile) or OS X (i.e., desktop) and they are able to distinguish such operating systems with a perfect accuracy (i.e., 100\%) after observing only five packets.

In their experiments, Ruffing et al. in~\cite{Ruffing2016} observe that, in case of heavy multitasking, the OS detection accuracy can reach 100\% with only 30 seconds of network traffic.
Considering Android and iOS, the authors also evaluate whether their approach is suitable to discriminate different versions of the same OS.
Moreover, the OS detection accuracy can reach 98\% and 50\% on fifteen-minutes-long traces from the Skype and Youtube apps, respectively.

\subsection{Position Estimation}
\label{sec:validation_position_estimation}

In Table~\ref{tab:validation_position_estimation} we summarize the information related to the datasets used by the works that deal with position estimation.

Husted et al. in~\cite{Husted2010} show that, in a metropolitan population of users equipped with 802.11g mobile devices, having only 10\% of tracking population is sufficient to track the position of the remaining 90\% of users. 
Besides, the tracking benefits from extending the broadcasting range of the mobile devices (better with newer 802.11 standards).

Musa et al. in~\cite{Musa2012} evaluate the accuracy of their trajectory estimation method by considering three deployments of monitors and leveraging GPS as ground truth. 
Using monitors spaced over $400$ meters apart, the proposed system achieves a mean error of less than $70$ meters.

\begin{sidewaystable*}
%\color{blue}

\caption{The achieved results in the surveyed works that deal with operating system identification.}
\label{tab:validation_OS_identification}
\centering
\scalebox{0.75}{
\begin{tabular}{|c|l|c|m{0.16\textwidth}|m{0.16\textwidth}|m{0.16\textwidth}|m{0.13\textwidth}|c|c|c|c|c|}
\hline
\textbf{Year} & \multicolumn{1}{c|}{\textbf{Paper}} & \textbf{Dataset} & \multicolumn{1}{c|}{\textbf{Point of Capturing}} & \multicolumn{1}{c|}{\textbf{Content}} & \multicolumn{1}{c|}{\textbf{Analysis Technique}} & \multicolumn{1}{c|}{\textbf{Analysis Details}} & \textbf{Platform} & \rot{\textbf{Accuracy }} & \rot{\textbf{Precision }} & \rot{\textbf{Recall }} & \rot{\textbf{F-measure }} \\ \hline
% 2014
\multirow{3}{*}{2014} & \multirow{2}{*}{Chen et al. \cite{Chen2014}} & \multirow{2}{*}{14.Chen\_{}0} & \multirow{2}{*}{\parbox{0.16\textwidth}{Wired (switch), monitors (nine monitors for two days at OSDI 2006, eight monitors for five days at SIGCOMM 2008), APs (one access point)}} & \multirow{2}{*}{\parbox{0.16\textwidth}{Layer-2+ headers plus DHCP and DNS payloads (switch), size and header of IP packets (monitors), IP and TCP headers (APs)}} & \multirow{2}{*}{\parbox{0.16\textwidth}{Machine learning (supervised learning)}} & \multirow{2}{*}{\parbox{0.12\textwidth}{Naive Bayes}} & \parbox{0.06\textwidth}{\rule{0.0cm}{0.6cm}\\ \centering Android\\ \rule{0.0cm}{0.5cm}} & {\tiny N/A} & $1.000$ & $0.800$ & {\tiny N/A} \\ \cline{8-12}
& & & & & & & \parbox{0.08\textwidth}{\rule{0.0cm}{0.6cm}\\ \centering iOS\\ \rule{0.0cm}{0.5cm}} & {\tiny N/A} & $1.000$ & $1.000$ & {\tiny N/A} \\ \cline{2-12}
& Coull et al. \cite{Coull2014} & 14.Coull\_{}0 & Devices & Packet sizes within APNS connection & Machine learning (supervised learning) & Binomial naive Bayes, kCross($10$) & iOS & $1.000$ & {\tiny N/A} & {\tiny N/A} & {\tiny N/A} \\ \hline
% 2016
\multirow{2}{*}{2016} & \multirow{2}{*}{Ruffing et al. \cite{Ruffing2016}} & \multirow{2}{*}{16.Ruffing\_{}0} & \multirow{2}{*}{\parbox{0.16\textwidth}{Monitors (one monitor for three months, two Android devices, two iOS devices, a Windows Phone device, and a Symbian device)}} & \multirow{2}{*}{\parbox{0.16\textwidth}{Timing of 802.11 frames}} & \multirow{2}{*}{\parbox{0.16\textwidth}{Machine learning (supervised learning, frequency spectrum analysis)}} & Consider only the traces lasting up to $30$ seconds & \multirow{2}{*}{\parbox{0.08\textwidth}{\centering Android, iOS, Windows Phone, Symbian}} & $0.700$ & {\tiny N/A} & {\tiny N/A} & {\tiny N/A} \\ \cline{7-7} \cline{9-12}
& & & & & & Consider only the traces lasting five minutes or more & & $0.900+$ & {\tiny N/A} & {\tiny N/A} & {\tiny N/A} \\ \hline
% 2017
\multirow{2}{*}{2017} & \multirow{2}{*}{Malik et al. \cite{Malik2017}} & 17.Malik\_{}0 & \multirow{2}{*}{\parbox{0.16\textwidth}{APs (one access point, one Android device, one iOS device, one Windows Phone device)}} & Inter-packet time of ICMP responses & \multirow{2}{*}{\parbox{0.16\textwidth}{Machine learning (supervised learning)}} & \multirow{2}{*}{\parbox{0.12\textwidth}{Random forest, kCross($10$)}} & \multirow{2}{*}{\parbox{0.08\textwidth}{\centering Android, iOS, Windows Phone}} & $0.752$ & {\tiny N/A} & {\tiny N/A} & {\tiny N/A} \\ \cline{3-3} \cline{5-5} \cline{9-12}
& & 17.Malik\_{}1 & & Inter-packet time of IP packets related to video streaming & & & & $0.736$ & {\tiny N/A} & {\tiny N/A} & {\tiny N/A} \\ \hline
\end{tabular}
}
\bigskip
\caption{The datasets and analysis techniques of the surveyed works that deal with position estimation.}
\label{tab:validation_position_estimation}
\centering
\scalebox{0.80}{
\begin{tabular}{|c|l|c|m{0.40\textwidth}|m{0.160\textwidth}|m{0.16\textwidth}|}
\hline
\textbf{Year} & \multicolumn{1}{c|}{\textbf{Paper}} & \textbf{Dataset} & \multicolumn{1}{c|}{\textbf{Point of Capturing}} & \multicolumn{1}{c|}{\textbf{Content}} & \multicolumn{1}{c|}{\textbf{Analysis Technique}} \\ \hline
% 2010
2010 & Husted et al. \cite{Husted2010} & 10.Husted\_{}0 & Simulator (3D simulation, mobile devices in dense metropolis) & 802.11 probe requests & Trilateration \\ \hline
% 2012
\multirow{3}{*}{2012} & \multirow{3}{*}{Musa et al. \cite{Musa2012}} & 12.Musa\_{}0 & Monitors (five monitors, nine months, streets near a campus) & \multirow{3}{*}{\parbox{0.16\textwidth}{802.11 probe requests}} & \multirow{3}{*}{\parbox{0.16\textwidth}{Machine learning (hidden Markov model, Viterbi's map-matching algorithm)}} \\ \cline{3-4}
& & 12.Musa\_{}1 & Monitors (six monitors, twelve hours, fairly busy city roads) & & \\ \cline{3-4}
& & 12.Musa\_{}2 & Monitors (seven monitors, twelve hours, arterial city road) & & \\ \hline
\end{tabular}
}
\bigskip
\caption{The achieve results in the surveyed works that deal with user fingerprinting.}
\label{tab:validation_user_fingerprinting}
\centering
\scalebox{0.80}{
\begin{tabular}{|c|l|c|m{0.25\textwidth}|m{0.10\textwidth}|m{0.16\textwidth}|c|c|c|c|c|}
\hline
\textbf{Year} & \multicolumn{1}{c|}{\textbf{Paper}} & \textbf{Dataset} & \multicolumn{1}{c|}{\textbf{Point of Capturing}} & \multicolumn{1}{c|}{\textbf{Content}} & \multicolumn{1}{c|}{\textbf{Analysis Technique}} & \rot{\textbf{Accuracy }} & \rot{\textbf{Precision }} & \rot{\textbf{Recall }} & \rot{\textbf{F-measure }} & \rot{\textbf{DAR}} \\ \hline
% 2014
\multirow{2}{*}{2014} & \multirow{2}{*}{Verde et al. \cite{Verde2014}} & 14.Verde\_{}0 & Wired (gateway router, $26$ mobile users on same Wi-Fi AP, one month) & \multirow{2}{*}{\parbox{0.10\textwidth}{NetFlow records}} & \multirow{2}{*}{\parbox{0.16\textwidth}{Machine learning (hidden Markov model, supervised learning)}} & {\tiny N/A} & $0.950$ & $0.930$ & $0.940$ & {\tiny N/A} \\ \cline{3-4} \cline{7-11}
& & 14.Verde\_{}1 & Wired (tier-2 router of metropolitan Wi-Fi network, $200{,}000$ users, one day) & & & {\tiny N/A} & $0.958$ & $0.956$ & $0.954$ & {\tiny N/A} \\ \hline
% 2016
2016 & Vanrykel et al. \cite{Vanrykel2016} & 16.Vanrykel\_{}0 & Wired (two VPN servers) & HTTP messages & Graph building & {\tiny N/A} & {\tiny N/A} & {\tiny N/A} & {\tiny N/A} & $0.570$ \\ \hline
\end{tabular}
}
%\bigskip
%\caption{The performance of the machine learning classifier leveraged by Crussell et al. in~\cite{Crussell2014} to carry out ad fraud detection.}
%\label{tab:validation_ad_fraud_detection}
%\centering
%\scalebox{0.88}{
%\begin{tabular}{|c|l|c|m{0.16\textwidth}|m{0.12\textwidth}|m{0.10\textwidth}|m{0.16\textwidth}|c|c|c|c|}
%\hline
%\textbf{Year} & \multicolumn{1}{c|}{\textbf{Paper}} & \textbf{Dataset} & \multicolumn{1}{c|}{\textbf{Analyzed Apps}} & \multicolumn{1}{c|}{\textbf{Point of Capturing}} & \multicolumn{1}{c|}{\textbf{Content}} & \multicolumn{1}{c|}{\textbf{Analysis Technique}} & \rot{\textbf{Accuracy }} & \rot{\textbf{Precision }} & \rot{\textbf{Recall }} & \rot{\textbf{F-measure }} \\ \hline
% 2014
%2014 & Crussell et al. \cite{Crussell2014} & 14.Crussell\_{}0 & $130{,}339$ Android apps from nineteen marketplaces, and $35{,}087$ Android apps that probably contain malware & Emulators & Layer-2+ data & Machine learning (random forest, kCross($3$)) & $0.859$ & {\tiny N/A} & $0.718$ & {\tiny N/A} \\ \hline
%\end{tabular}
%}
\end{sidewaystable*}

\revOne{
\subsection{User Fingerprinting}
\label{sec:validation_user_fingerprinting}

% [Notes for the paragraph]

In Table~\ref{tab:validation_user_fingerprinting}, we report validation methods and results of the two works about user fingerprinting. 
Despite the common goal, these two analyses are quite different.
Indeed, Verde et al. in~\cite{Verde2014} collect datasets of NetFlow records, while Vanrykel et al. in~\cite{Vanrykel2016} rely on a dataset of HTTP messages.
Besides, the former work employ machine learning techniques while the latter performs a graph-based analysis.
% [Verde et al. \cite{Verde2014}]

Regarding Verde et al. in~\cite{Verde2014}, the results achieved on the ``14.Verde\_0'' dataset refers to the best performing classifier (i.e., random forest).
On the other hand, the results achieved on the ``14.Verde\_1'' dataset are the average across the five targeted mobile users. 
To build a reliable profile for those users, their network traffic is captured from a Wi-Fi access point (under the control of the authors).
Vanrykel et al. in~\cite{Vanrykel2016} are able to link 57\% of the unencrypted mobile traffic collected to a specific user/device using graph-based analysis on HTTP messages.

\subsection{Ad Fraud Detection}
\label{sec:validation_ad_fraud_detection}

The only work that investigates ad fraud is the one by Crussell et al. in~\cite{Crussell2014}. 
The authors validate their proposal on a dataset of network traffic (i.e., layer-2+) collected from mobile device emulators on which they run $130{,}339$ Android apps from nineteen marketplaces, and $35{,}087$ Android apps that probably contain malware.  
To build the ground truth, the authors manually labeled the page requests of the domains related to ad providers.
The authors used a random forest classifier applying a three-fold cross-validation on such labeled dataset. 
The achieved results are an accuracy of $0.859$ and a recall of $0.718$.
As an additional finding, the authors discover that around $30\%$ of apps with ads request to display an ad while running in the background, and $27$ apps generate clicks without user interaction.
}

%Layer-2+ data 
%Machine learning (random forest, kCross($3$)) 

% [Notes for the paragraph]

% [Crussell et al. \cite{Crussell2014}]

% In the maxi-table, there is the performance of the classifier that the authors use to carry out their investigation of the considered apps.

% To build the ground truth for the classifier, the authors manually labeled the page requests of the domains related to ad providers.

% Using the trained classifier, the authors discover that about 30% of apps with ads request to display an ad while running in the background, and 27 apps generate clicks without user interaction.

\revOne{
\subsection{Sociological Inference}
\label{sec:validation_sociological_inference}

In this section, we report the datasets used for validation by Barbera et al. in~\cite{Barbera2013}, the only work that carries out sociological inference. 
% [Barbera et al. \cite{Barbera2013}]
The authors collect datasets containing 802.11 probe requests. 
The datasets are related to an event or place in which the monitor(s) are deployed:
\begin{itemize}
\item Datasets \textit{P1} and \textit{P2} at a political meeting (five monitors);
% Datasets "P1" and "P2":
% - Point of capturing (5 monitors, political meeting);
% - Content: 802.11 probe requests.
\item Datasets \textit{V1} and \textit{V2} at a Pope's mass (five monitors);
% Datasets "V1" and "V2":
% - Point of capturing (5 monitors, Pope's mass);
% - Content: 802.11 probe requests.
\item Dataset \textit{M} at a big shopping mall (five monitors);
% Dataset \textit{M}:
% - Point of capturing (5 monitors, big mall);
% - Content: 802.11 probe requests.
\item Dataset \textit{TS} at a train station (five monitors);
% Datasets \textit{TS}:
% - Point of capturing (5 monitors, train station);
% - Content: 802.11 probe requests.
\item Dataset \textit{U} at a university's campus (one monitor);
% Datasets \textit{U}:
% - Point of capturing (1 monitor, campus);
% - Content: 802.11 probe requests.
\item Dataset \textit{others} at city streets and squares (one monitor).
% Datasets \textit{Others}:
% - Point of capturing (1 monitors, city streets and aggregation places);
% - Content: 802.11 probe requests.
\end{itemize}
Relying on these datasets, the authors provide the findings that are discussed in Section~\ref{sec:sociological_inference}.
% The authors' findings are reported in the "goals" section.

\subsection{Tethering Detection}
\label{sec:validation_tethering_detection}

Only one work aims to detect tethering and it is proposed by Chen et al. in~\cite{Chen2014}.
% [Chen et al. \cite{Chen2014}]
The authors build the first dataset using several points of capturing: a wired network equipment (i.e., a network switch), Wi-Fi monitors (i.e., nine monitors for two days at OSDI 2006, eight monitors for five days at SIGCOMM 2008), and an access point.
In particular, they capture DHCP and DNS payloads, and layer-2+ headers from the network switch; size and header of IP packets from monitors; and IP and TCP headers from the access point.  
The second dataset contains one week of network traffic (i.e., IP packets) collected at the Internet gateway of a campus Wi-Fi network serving $12,600$ users.

The authors develop an ad hoc probabilistic classifier to carry out their analysis.
The achieved results on the first dataset are $0.68$-$0.85$ recall with precision fixed at $0.95$, and $0.78$-$0.89$ recall with precision stable at $0.8$.
Besides, the results on the second dataset are  $0.86$ precision, $0.74$ recall, and $0.8$ F-measure.
}

% Dataset: "14.Chen_0".
% Point of capturing: wired (switch), monitors (nine monitors for two days at OSDI 2006, eight monitors for five days at SIGCOMM 2008), APs (one access point).
% Content: layer-2+ headers plus DHCP and DNS payloads (switch), size and header of IP packets (monitors), IP and TCP headers (APs).

% Dataset: "14.Chen_1".
% Point of capturing: wired (Internet gateways of campus Wi-Fi network, 12,600 users, 1 week).
% Content: IP packets.

% Analysis technique: machine learning (supervised learning via probabilistic classifier).

% Results on "14.Chen_0" dataset: 0.68-0.85 recall with target precision fixed at 0.95, and 0.78-0.89 recall with target precision fixed at 0.8.

% Results on "14.Chen_1" dataset: 0.86 precision, 0.74 recall, and 0.8 F-measure.

\revOne{
\subsection{Website Fingerprinting}
\label{sec:validation_website_fingerprinting}

The work by Spreitzer et al. in~\cite{Spreitzer2016} deal with website fingerprinting.
This work relies on a dataset containing statistics provided by a mobile browser in terms of transmitted and received bytes of TCP connections.
The authors develop a fingerprinting system that employs a machine learning classifier based on Jaccard's index.
Under a normal Internet connection, such fingerprinting system can correctly infer 97\% of $2{,}500$ page visits out of a set of $500$ monitored pages.
Instead, with the traffic routed through Tor by using the Orbot proxy combined with the Orweb browser, the proposal identifies $95\%$ of $500$ page visits out of a set of $100$ monitored pages. 
}
% Dataset: "16.Spreitzer_0".
% Point of capturing: devices (five Android devices).
% Content: Tx/Rx TCP bytes of the browser app.

% Analysis techniques: machine learning (based on Jaccard index).

% The fingerprinting app can correctly infer 97% of 2,500 page visits out of a set of 500 monitored pages, and 95% of 500 page visits out of a set of 100 monitored pages when the traffic is routed through Tor by using the Orbot proxy in combination with the Orweb browser.

\revOne{
\section{Countermeasures against Mobile Traffic Analysis}
\label{sec:countermeasures}
In this section, we present possible countermeasures proposed in the literature to thwart mobile traffic analysis. %tackle
In the first instance, we discuss how encryption on different layers can affect the surveyed work in Section~\ref{sub:countermeasures_encryption}. 
We will show that part of the surveyed work is able to cope with encryption. 
%(Original, edited by Alberto) Nonetheless, part of the surveyed work is able to cope with encryption. 
Hence, we survey the state-of-the-art countermeasures and their effectiveness and limitations in Section~\ref{sub:countermeasures_other}.
}

\revOne{
\subsection{Encryption}
\label{sub:countermeasures_encryption}
The first countermeasures in place are network traffic encryption methods. 
Such methods aim to guarantee users' privacy against information leaks and DPI. 
Encryption can be applied at different levels of network protocol stack, such as network (e.g., IPsec) and transport (e.g., SSL/TLS) layers. 
From the network analysis perspective, the primary effect of the encryption at a given layer is to make unavailable the information of the above layers. 
This means that SSL/TLS encryption will hide the transport-layer payloads, but TCP/UDP headers will be still available; IPsec encryption, instead, will also hide the TCP/UDP headers, leaving only IP headers for analysis.
% (Original, edited by Alberto) This means that while TCP header is still available with SSL/TLS encryption applied but not the payload, but if IPsec encryption is applied at the network layer we will not be able to access neither TCP header fields nor the payload.
It is worth noticing that for the sake of simplicity we use the term ``IPsec encryption'' to refer all the methods that make available IP headers only, such as IPsec (in both transport and tunnel mode), Virtual Private Networks (VPNs) and Tor (The Onion Routing).
% (Original, edited by Alberto) It is worth noticing that for the sake of simplicity we refer as IPsec encryption the methods that make available IP headers only, such as an IPsec (in both transport and tunnel mode), Virtual Private Networks (VPN) and Tor (The Onion Routing).

As a preliminary overview, for each surveyed work we listed in Table~\ref{tab:all_papers}, we pointed out whether its analyses were applicable in presence of the aforementioned encryption methods. 
% (Original, edited by Alberto) As a preliminary overview, we list the surveyed work in Table~\ref{tab:all_papers} pointing out whether it is able to deal with the aforementioned encryption methods.
In what follow, we discuss in detail whether encryption affects or not the network analysis techniques adopted by the surveyed work.
%TODO Riccardo: summary of papers that cope with which encryptions
On one hand, $40$ works in this survey are still able to carry out their analysis if SSL/TLS encryption is in place.
On the other hand, only $21$ works do not rely on information that is hidden by IPsec encryption.
The effect of encryption on the analysis strictly depends on the point of capturing.
For this reason, for each point of capturing (ordered by the number of related works, as in Section~\ref{sec:classification_by_PoC}) we report which analysis and why is affected by which type of encryption.
In particular, given a point of capturing we discuss the works by year of publication.

%%%%%%%%%%%%%%%%%%%%TODO I ARRIVED HERE TODAY!!!!!!
\subsubsection{Wired networks}
In this section, we discuss the impact of encryption when the point of capturing is wired networks. We divide the presentation of the works into small and large scale networks.

%\noindent 
\paragraph{Small Scale}
%%%%SMALL SCALE WIRED NETWORKS
%In the remaining of this section, we clarify how the works reported deal with encryption.

\begin{itemize}
\item Rao et al. in~\cite{Rao2011} study the network traffic of the Android and iOS apps for Netflix and YouTube.
They successfully inspect the HTTP messages to get the encoding rate of the videos, therefore both services stream videos in clear (at least, they did so at the time the authors collected their datasets).

\item The analysis carried out by Baghel et al. in~\cite{Baghel2012} needs to inspect the transport-layer headers, therefore it does not work if IPsec is employed to hide the payload of IP packets.
The Android malware detector by Wei et al. \cite{Wei2012} requires to access DNS data, which is not possible if the traffic is encrypted.

\item To carry out PII leakage detection, Rao et al. in~\cite{Rao2013} and Ren et al. in~\cite{Ren2016} inspect HTTP traffic, which is sent in clear, and also HTTPS traffic, which is decrypted using SSLsplit. This approach cannot work, however, if the traffic of a given app is protected by IPsec.

%The mobile user fingerprinting framework presented by Verde et al. in \cite{Verde2014} takes NetFlow records as input. Since NetFlow records can be extracted even from encrypted traffic, the proposed solution is encryption-agnostic.
\item Chen et al. in~\cite{Chen2015} focus on the properties of the network traffic of malicious Android apps, and their findings are mainly related to the application layer.
For this reason, such findings are limited to the data that the analyzed apps sent in clear during the capturing process.

\item The user action identification frameworks developed by Conti et al. in~\cite{Conti2016}, and Park and Kim in~\cite{Park2015}, the app identification solution proposed by Taylor et al. in~\cite{Taylor2017} and the PII leakage detection method by Cheng et al. in~\cite{Cheng2017} leverage the information available in IP and TCP headers.
As a consequence, such approaches are by design resilient against SSL/TLS, but cannot cope with encryption via IPsec.

\item To study the network behavior of several Android and iOS free apps, Nayam et al. in~\cite{Nayam2016} inspect the HTTP messages, and employ a proxy server to deal with HTTPS traffic.
Although such approach does not work with apps that employ IPsec to hide their network transmissions, it seems that all the analyzed apps do not leverage such type of encryption.

\item The PII leakage detection and user fingerprinting framework proposed by Vanrykel et al. in~\cite{Vanrykel2016} is focused on unencrypted mobile traffic only, since it requires to inspect HTTP messages.

\item Wang et al. in~\cite{Wang2016} present two Android malware detection models which leverage TCP- and HTTP-related information, respectively. The latter cannot work for apps that encrypt their network traffic using SSL/TLS, and both cannot cope with apps that employ IPsec for their data transmissions.

\item The PII leakage detection solution by Continella et al.~\cite{Continella2017} requires to access the HTTP messages. Although a man-in-the-middle approach is adopted to deal with HTTPS traffic, the framework cannot cope with network traffic protected by IPsec.
\end{itemize}

%\begin{itemize}
%%%%LARGE SCALE WIRED NETWORKS
\paragraph{Large Scale}
\begin{itemize}
\item Afanasyev et al. in~\cite{Afanasyev2010} focus part of their analysis on the applications that generate mobile and non-mobile traffic. In particular, they need to inspect transport- and application-layer headers. For this reason, the reported findings do not cover the encrypted traffic present in the collected network traces.
The same holds for the mobile traffic characterization by Chen et al.~\cite{Chen2012}.

\item The analysis carried out by Maier et al. in~\cite{Maier2010} requires to access transport- and application-layer information, therefore it cannot deal with encrypted traffic.

\item The study by Finamore et al. in~\cite{Finamore2011} focuses on YouTube traffic carried over HTTP and does not consider the users that watch videos via a secure connection (i.e., HTTPS).

\item The app identification via payload signatures proposed by Lee et al. in~\cite{Lee2011} cannot work with apps that encrypt their network traffic.
Moreover, in case of encryption the authors' studies of mobile traffic characteristics and mobile users' habits are severely limited.

\item The tethering detection technique proposed by Chen et al. in~\cite{Chen2014} requires to inspect the information available in TCP headers, therefore it does not work if IPsec is employed to hide the payload of IP packets.

\item As we previously mentioned in Section~\ref{sec:wired_network_equipments_small-scale}, the mobile user fingerprinting framework by Verde et al. in~\cite{Verde2014} is encryption-agnostic even in their experiment carried out on a large-scale network, since it takes NetFlow records as input.

\item A few of the findings about mobile traffic reported by Wei et al. in~\cite{Wei2017} are based on application-layer information, which is unavailable in case of traffic encryption.
\end{itemize}

\subsubsection{Mobile Devices}
%%%MOBILE DEVICES
Regarding the works that capture network traffic directly within mobile devices, we provide the following observations about the effect of encryption.
\begin{itemize}
\item Network traffic statistics (e.g., the amount of received bytes through the cellular network) are not affected by encryption, therefore the methods that leverage them are encryption-agnostic. These works are the ones proposed by Ham and Choi in~\cite{Ham2012}, {Shabtai et al. in~\cite{Shabtai2012},} Shabtai et al. in~\cite{Shabtai2014}, Fukuda et al. in~\cite{Fukuda2015}, Soikkeli and Riikonen~\cite{Soikkeli2015}, Spreitzer et al. in~\cite{Spreitzer2016}, and Arora and Peddoju in~\cite{Arora2017}.
It is worth to notice that this assertion is not trivial for the work in~\cite{Spreitzer2016} since its authors leverage the TCP bytes sent/received by the browser app, which it is still available even if the traffic is encrypted via IPsec. 
\item Falaki et al. in~\cite{Falaki2010} carry out both traffic characterization (Section~\ref{sec:traffic_characterization}) and usage study (Section~\ref{sec:usage_study}).
The former is focused on the TCP protocol, therefore it does not cover the traffic protected by IPsec, whenever present in the collected network traces.
The latter leverages the per-app transmitted/received bytes, which are network traffic statistics (i.e., they are not affected by encryption).
\item Shepard et al. in~\cite{Shepard2010} provide a few findings about the network traffic of iOS devices.
Their analysis is focused on the TCP protocol, hence it does not cover the network traffic protected by IPsec.
\item Su et al. in~\cite{Su2012} propose a classifier for Android malware detection that cannot process the network traffic encrypted via IPsec since one of the leveraged features is the average TCP session duration, which is not computable without accessing TCP headers.
\item To identify Android apps, Wei et al. in~\cite{Wei2012ProfileDroid} perform the following operations: (i) inspect the IP addresses of the captured packets; (ii) compute the amount of transmitted/received data; and (iii) discriminate between HTTP and HTTPS traffic.
The first two operations are encryption-agnostic, while the third one is not possible in case an app communicates through IPsec. % to hide its network transmissions.
%However, it seems that the analyzed apps do not use such type of encryption.
\item To apply clustering for PII leakage detection, Kuzuno and Tonami in~\cite{Kuzuno2013} use two metrics that are based on information within HTTP messages which are not accessible when any form of encryption is in place.
\item To evaluate their solution for Android app identification, Qazi et al. in~\cite{Qazi2013} set up a monitored access point serving a few mobile devices.
The network traffic flowing through the AP is captured, and netstat logs from the devices are gathered. Such logs are then used to match the network flows observed at the AP with the TCP transmissions from the mobile devices, hence this methodology does not work if IPsec is employed.
\item The user action and OS identification methods devised by Coull and Dyer in~\cite{Coull2014} are designed to work with the network traffic of iMessage (Apple's instant messaging service), which uses encryption by default.
\item Le et al. in~\cite{Le2015} carry out both app identification (see Section~\ref{sec:app_identification}) and PII leakage detection (see Section~\ref{sec:PII_leakage_detection}). The former requires to access the flags of TCP segments, which are hidden if IPsec is employed. The latter needs to inspect application-layer data, which is infeasible if the traffic is encrypted.
\item The app for PII leakage detection by Song and Hengartner~\cite{Song2015} employs a man-in-the-middle approach to inspect TLS traffic, but it cannot deal with IP packets whose payloads are encrypted by IPsec.
\item The Android malware detection solution by Zaman et al.~\cite{Zaman2015} needs to access the URLs within HTTP messages, which are not available if the traffic is encrypted.
\item Mongkolluksamee et al. in~\cite{Mongkolluksamee2016} extract the TCP and UDP data from the collected network traffic of the apps to be profiled. After that, they inspect the headers to reconstruct the captured network flows and compute their statistics (e.g., per-flow total amount of transferred bytes).
This approach cannot be applied to apps that encrypt their network traffic using IPsec.
\item The malware detection framework by Narudin et al.~\cite{Narudin2016} requires to inspect HTTP messages, therefore it cannot work with encrypted traffic.
\item By employing the mobile traffic characterization framework proposed by Espada et al.~\cite{Espada2017}, it is possible to check whether the network traffic of an Android app satisfies a given property. The effect of encryption on the analysis depends on the properties to be verified. Regarding the presented case study (Spotify), the authors successfully access HTTP headers and compute traffic statistics (e.g., number of sent/received TCP segments).
\end{itemize}

\subsubsection{Wi-Fi Access Points}

In this section we consider works in which network traffic is captured from access points in a controlled or not controlled environment. We provide the following observations about the effect of encryption.

\noindent \textit{APs in a controlled environment} ---
Stevens et al. in~\cite{Stevens2012} study thirteen popular ad libraries for Android.
For each library, the authors build a simple app that makes ad requests, then they execute it on a mobile device while capturing the network traffic at the access point to which that device is associated. Since only one of the considered ad libraries leverages encryption to protect its network traffic, the authors apply deep packet inspection to investigate the leakage of the user's PII.
Qazi et al. in~\cite{Qazi2013} set up a wireless access point running OpenFlow and instruct it to extract features from the network traffic of the associated mobile devices.
Since it requires to inspect transport-layer information, the proposed framework cannot process network traffic protected by IPsec.

The framework for user action identification presented by Watkins et al. in~\cite{Watkins2013} exploits the inter-packet time of the responses to ICMP packets sent to the target mobile device by a laptop connected via cable to the same network, therefore it is not affected by traffic encryption.
The OS identification method by Chen et al. \cite{Chen2014} needs to access the headers of TCP segments, so it cannot work if IPsec is employed to hide the payload of IP packets.
The app identification solution by Yao et al. \cite{Yao2015} requires to access HTTP messages, which is not possible if the traffic is encrypted.
The app identification framework proposed by Alan and Kaur \cite{Alan2016} provides three different classifiers. Two of them only leverage the size of IP packets, thus they can take encrypted traffic as input. Instead, the other classifier requires to inspect the content of TCP headers, therefore it works on network traffic encrypted via SSL/TLS, but it does not via IPsec.

The solutions proposed by Saltaformaggio et al. in~\cite{Saltaformaggio2016} (user action identification), as well as Tadrous and Sabharwal in~\cite{Tadrous2016} (traffic characterization), are encryption-agnostic: the former requires only to inspect IP headers; the latter needs only the size and header information of 802.11 frames.
The framework for OS identification presented by Malik et al. in~\cite{Malik2017} exploits the inter-packet time of the packets (either ICMP responses or IP packets related to video streaming) coming from the target mobile device, therefore it is not affected by traffic encryption.
Finally, the proposal by Fu et al. in~\cite{Fu2016} rely on IP headers only, thus it is resilient to traffic encryption.

%%%%%%%%%%%%%%%%%%%%MULTIPLE AP
\noindent \textit{APs in a uncontrolled environment} ---
Gember et al. in~\cite{Gember2011} carry out a comparison between mobile and non-mobile devices with regard to network traffic properties and habits of users.
Since the analysis is mainly focused on transport and application layers, most of the authors' findings are related to non-encrypted traffic.
Whenever the encryption at network (IPsec) or transport layers (SSL/TLS) is employed, the HTTP information becomes inaccessible, thus making the discrimination process (if not the entire analysis) infeasible.

\subsubsection{Wi-Fi Monitors}
As we discuss in Section~\ref{sec:Wi-Fi_monitors}, Wi-Fi monitors scan radio bands to capture IEEE 802.11 frames which can be encrypted at data-link layer using Wired Equivalent Privacy (WEP) or Wi-Fi Protected Access (WPA).

The attacks by Musa and Eriksson in~\cite{Musa2012} and Barbera et al. in~\cite{Barbera2013} cannot be affected by encryption because they rely on probe requests. Due to their nature, probe requests are transmitted in clear. 
The analyses carried out by Wang et al. in~\cite{Wang2015} and Ruffing et al. in~\cite{Ruffing2016} are encryption-agnostic since they leverage only size and/or timing of the captured 802.11 frames. 
However, this statement does not hold for Chen et al. in~\cite{Chen2014}, since their analysis requires to access IP payloads.

\subsubsection{Mobile Device Emulators}
\label{sub:encryption_emulator}
Mobile device emulators are employed as sandboxes for ad fraud detection and malware analysis. 
All the three works that rely on emulators aim to inspect HTTP messages, thus the proposed solutions would not work anymore if encryption is applied. 

Crussell et al. in~\cite{Crussell2014} carry out ad fraud detection relying on emulators. 
The proposed framework is not resilient to encryption since it needs to inspect the HTTP and DNS data generated by apps. 
However, the authors' analysis covers most of the available ad libraries. 
This means that such libraries do not usually employ any form of encryption for their data transfers, and simply rely on plain HTTP.

The ANDRUBIS framework proposed by Lindorfer et al. in~\cite{Lindorfer2014} relies on Android emulators to carry out dynamic malware analysis. 
Such framework focuses its analysis on high-level protocols (e.g., DNS, HTTP, IRC), which is not feasible if the analyzed app encrypts its network traffic.
Yao et al. in~\cite{Yao2015} propose an app identification method on three mobile platforms (i.e., Android, iOS, and Symbian). 
Unfortunately, since the system requires to inspect HTTP {messages}, it does not work if an app leverages HTTPS or lower-layers encryption.

Narudin et al. in~\cite{Narudin2016} and Chen et al. in \cite{Chen2017} propose machine learning to detect Android malware and a method for app identification, respectively. 
Both these works rely on HTTP messages inspection, hence they cannot cope with encrypted network traffic.

\subsubsection{Network Simulators}
\label{sub:encryption_simulator}

The work on position estimation by Husted and Myers in~\cite{Husted2010} is the only work that relies on software network simulators to generate mobile traffic.
This work is not affected by encryption since it focuses on propagation of probe requests, which are not encrypted.
}

\revOne{
\subsection{Other Countermeasures}
\label{sub:countermeasures_other}
As a countermeasure to privacy invasive mobile traffic analyses (i.e., app identification, user action identification, website fingerprinting, and user fingerprinting), encryption alone is not enough to neutralize them. 
Indeed, such analyses often focus on network flows behavior or packets exchange patterns.
A research field that provides solutions to thwart these kinds of analysis is the one that investigates countermeasures against websites fingerprinting.
In this research field, researchers consider an adversary that is able to observe timing, direction, and size of packets within an encrypted connection when a browser loads a webpage~\cite{Liberatore2006,Herrmann2009, Panchenko2011, panchenko2016website}. 
Similarly to our IPsec scenario, the considered attacks rely on IP headers only since they assume to carry out traffic analysis in presence of an anonymity network such as a VPN or Tor.
Under these settings, the countermeasures proposed to tackle traffic analysis can be divided in two categories: padding- and distribution-based countermeasures.
In what follows, we describe such kind of countermeasures. Moreover, we discuss what impact their application could bring on mobile network traffic.

\paragraph{Padding-based Countermeasures}
This category of countermeasures considers an active modification of the network traffic at packet level. 
Packet padding is a technique that consists in appending extra information to a packet payload in order to obfuscate its original size. 
Implementations of SSL $3.0$ and TLS $1.0+$ already apply a random padding between $0$ and $255$ bytes to encrypted packets. 
Unfortunately, SSL/TLS padding provides a limited protection since a padding of at most $255$ bytes, compared with the Maximum Transmission Unit (MTU) does not introduce a significant level of noise.
For this reason, other padding techniques propose to add dummy bytes to reach a packet size that is multiple of $128$ bytes (i.e., linear padding), the nearest power of two (i.e., exponential padding), or MTU.

Researchers propose more sophisticated techniques that not only apply padding on packets' sizes, but also to timing.
Dyer et al. in~\cite{dyer2012peek} present BuFLO (Buffered Fixed-Length Obfuscation), which apply padding in such a way that packets are sent with a fixed size. Moreover, BuFLO also fixes the packet transmission rate to cope with timing attacks. 
Cai et al. in~\cite{cai2014systematic} propose Tamaraw which is based on BuFLO, but apply a different padding according to the direction of the packet. 
More recently, Wang et al. in~\cite{wang2017walkie} present Walkie Talkie which modify a web browser to buffer packets, add padding, and transmit them in bursts (i.e., half-duplex mode).

\paragraph{Distribution-based Countermeasures}
A different approach to counter traffic analysis aims to intervene on the distribution of packets in a network flow.
Wright et al. in~\cite{wright2009traffic} proposed \textit{traffic morphing}, a distribution-based countermeasure which transforms the original distribution of network packets to a pre-defined target distribution.
In practice, traffic morphing reshapes network flows by truncating and padding packets, not only modifying packets sizes but also changing the very number of packets. 

%luo2011httpos
Another example of traffic morphing is called Glove, which is proposed by Nithyanand et al. in~\cite{Nithyanand2014}. 
In first instance, Glove regroups websites which network flows are similar into clusters.
Hence, Glove applies the minimum dummy traffic in such a way that all websites in a cluster will have the same shape, thus they cannot be distinguished from each other. 
This means that an attacker cannot identify a specific website, but only the cluster to which it belongs.

\paragraph{Countermeasures Applied to Mobile Traffic}
In the mobile scenario, we have to take into account that we are considering devices that are powered by batteries. 
Another aspect that has to be considered is that mobile users have a limited volume of Internet traffic given the subscription with a telephonic company (usually a fixed amount of gigabytes).
On one hand, the aforementioned padding-based countermeasures generate a bandwidth overhead that ranges between 31\% and 145\%, with a time overhead that ranges between 34\% and 180\% for~\cite{wang2017walkie} and~\cite{dyer2012peek}, respectively.
On the other hand, distribution-based countermeasures change the shape of a network flow. 
This procedure has computational costs due to packets aggregation and segmentation while still relying in part on padding. 

Countermeasures that introduce a high overhead in terms of bandwidth, time, and computational cost are not feasible on mobile device since such overhead directly result in additional monetary cost, worse user experience, and increased battery consumption.
Among the considered countermeasures, the most mobile friendly seems to be Walkie Talkie~\cite{wang2017walkie}, since it offers a reasonable trade off between bandwidth and time overhead.
Another possible solution is to deploy a countermeasure (even with high overhead) on the access point rather than on the mobile device, offloading to the access point the padding and computational burdens.
Unfortunately, this solution does not protect from attacks that eavesdrop the traffic between mobile devices and access points. 
}

\revOne{
\section{Challenges and Future of Traffic Analysis Targeting Mobile Devices}
\label{sec:present_future}
In this section, we first provide an overall discussion about today's challenges and pitfalls that emerged from the surveyed works in Section~\ref{sec:challenges}.
Then, we outline possible future research directions of mobile traffic analysis in Section~\ref{sub:future}.

\subsection{Challenges and Pitfalls}
\label{sec:challenges}

In state-of-the-art works on mobile traffic analysis, researchers encounter several challenges while devising their analyses.  
A first challenge involves the discrimination between mobile and non-mobile traffic in large-scale networks.
As a common practice to overcome this challenge, researchers leverage the information available from network traffic.
In case of unencrypted traffic, HTTP messages contain the \textit{user-agent} field from which it is possible to obtain information such as mobile device and operating system~\cite{Maier2010,Finamore2011,Chen2012}. 
In presence of traffic encryption, it is still possible to rely on the default Time-To-Live (TTL) value of IP packets.
Another method uses DHCP logs to map an assigned IP address with a MAC address, then to identify mobile devices by extracting the Organizationally Unique Identifier (OUI) from those MAC addresses. 
In the aforementioned methods, researchers assume that user-agent fields and MAC addresses are not spoofed, and TTL values are the default ones given by the operating systems.

Another challenge is collecting a dataset that includes a solid ground truth, a fundamental task to obtain truthful results.
%Dataset collection of mobile network traffic is a quite challenging task, especially when it considers real mobile devices.
While dataset collection process and data format strictly depend on the point of listening, network traffic labeling requires additional information that has to be retrieved from a different source.
For some specific goals of the analysis, traffic labeling can be simple but, especially when it considers real mobile devices, it can become challenging.

% solid ground truth for traffic classification tasks.
In general, it is possible to build a solid ground truth whether one of these two conditions occur:
(i) the goal of the analysis is related to the property of the device itself;
(ii) the researchers can gain full control of the experiment.
On one hand, the former condition can be applied to operating system identification and user fingerprinting. 
In fact, knowing the pre-assigned IP address or the MAC address of a mobile device is enough to label the traffic as belonging to a specific user or operating system.
On the other hand, two examples of the latter condition are works that perform app identification and user action identification, since the goal of these analyses is related to OS internal processes and human-device interaction.
For this reason, researchers need to acquire full control of the experiment in order to build a reliable ground truth. 
While for user action identification tasks researchers have to provide a solid time correlation between a user action and the resulting network traffic, building the ground truth for app identification tasks is more challenging, since they need to know exactly which app generated which network packet or flow.
To cope with this problem, a good practice on Android devices is to rely on a logging app (e.g., NetworkLog~\cite{NetworkLog}, DELTA~\cite{spolaor2017delta}) that associates the Process ID (PID) of an app to each network packet or flow it generates. 
%As far as concerns malware detection tasks

Common pitfalls of mobile traffic analysis are mostly related to the experimental design and data collection.
In app identification tasks, researchers have to take into account that apps often rely on common third-party libraries (e.g., ad libraries in free apps), thus some traffic patterns generated by an app are not useful to discriminate it among the other considered apps.
Hence, the analysis has to cope with such ambiguous traffic and filter it out.

%This problem during  and coverage
A possible error is not considering enough data sources in order to build a representative dataset for the goal of the analysis. 
The work by Malik et al. in~\cite{Malik2017} is a clear example of this kind of pitfall.
In this work, the authors aim at distinguishing between three operating systems but they collect network traffic of only three devices (i.e., one for each targeted operating system).  
The collected dataset is not representative enough for operating system identification because the traffic may be influenced by both device model and OS version (i.e., bias on the data).
Hence, the accuracy is related to the recognition of a specific device (i.e., device model, OS, and its version altogether) among the three considered ones.
%the sample have to be representative enough.

Another similar pitfall is to perform an analysis under a close-world assumption. 
This assumption does not consider aspects or possible events in the real world.
This means that in a classification task it must be taken into account not only the classes on which the analysis is focused, but also other possible classes.
As an example, in app identification it has to consider, in addition to the targeted apps, the other applications on the market (e.g., as a stand-alone class).
Another pitfall is related to mobile malware detection.
Proposed works on this topic use emulators instead of real devices.
Unfortunately, emulators can be detected using sandbox detection techniques.
Hence a malware that detects a sandbox could not execute its malicious payload, thus evading malware analysis~\cite{bordoni2017mirage}. 

\subsection{Possible Future Directions}
\label{sub:future}
In this survey, we can observe an imbalance in terms of the number of works per goal of the analysis. 
Indeed, some goals are not covered enough by the state of the art of mobile traffic analysis. 
This makes room for interesting research directions that could have severe economic and privacy implications.

Research on user fingerprinting, sociological inference, and position estimation could be vastly improved since those fields are seriously related to user privacy.
Indeed, information obtained from such types of analysis can be used to track the user's movements and infer user sensible information (e.g., nationality, relationships).  
Surprisingly, only one work investigates ad fraud detection.
Ad providers are economically affected by such frauds since they result in a financial loss.
Another interesting research direction could be investigating whether there exist differences between mobile and desktop websites in terms of network traffic.  

%\subsection{Application to IoT}
Solutions proposed for mobile traffic analysis can also be applied to the emerging Internet of Things (IoT) paradigm.
%In fact, IoT networks  heterogeneous devices (including mobile devices). 
IoT devices are interconnected and they can both sense the real-world status (i.e., sensors) and intervene to change it (i.e., actuators). 
Similarly to mobile devices, the communications between IoT devices generate network traffic and they are carried through the same protocol stack. 
For this reason, we believe that the surveyed analyses on network traffic of mobile devices can become valuable in the IoT domain.
As an example, user action identification techniques could be applied to infer information about the real world from the actions performed by an actuator.
Moreover, OS or app identification techniques could be used to infer an IoT device's model or firmware version.
}

\newpage
\section{Conclusions}
\label{sec:conclusions}

In this paper, we surveyed the state of the art of the methods for analyzing {the} network traffic generated by mobile devices. 
In particular, we are the first that surveyed the works in which the mobile traffic is captured from alternative sources to cellular networks: Wi-Fi monitors and access points{;} wired networks{;} logging apps installed on mobile devices{; and} networks of mobile devices simulated via software. 
For each point of capturing, %capturing points
we described its characteristics, the {number} %population size
of mobile users that it monitors, as well as the issues related to the capturing process. 
Moreover, we observed that the most {frequently} used {approach} %approaches
to capture mobile traffic {is logging at either} %are
wired networks {or mobile devices themselves.}%and directly on a mobile device (e.g., tcpdump).

We provide a systematic classification of the state of the art according to the goal of the analysis that targets the network traffic of mobile devices. In particular, we realized that most of the {works focus} %work focuses
on studying the features of {the} network traffic generated by mobile devices. Other popular goals are app and user action %actions
identification, usage study, and Personal Identifiable Information (PII) leakage and malware detection. 
While a lot of work has been done on such goals, promising topics, such as user fingerprinting and sociological inference, still offer much room for further investigation.
 
We also categorized the {works on mobile traffic analysis} %work on the traffic analysis methods
according to the targeted mobile platforms. 
We observed that Android is not only the most popular mobile platform{,} but also the {most targeted by the analysis methods} %most targeted platform by the analysis methods
(i.e., $42$ out of $45$ works which are not platform-independent). %not platform-independent works).
In fact, we demonstrated that the openness of the Android platform is a double-edged sword: on the one hand, it provides mobile users with a large number of apps that enable the most disparate functionalities; on the other hand, it helps malicious developers distribute malware, and more generally, apps that behave ambiguously with regard to the security of mobile devices and the privacy of mobile users.
 
We observed that most of the surveyed works rely on machine learning techniques, thus we outline the procedure to carry out a machine-learning-based analysis as a tutorial for new researchers. 
In particular, we also  observed 
%We report that most of the surveyed works rely on machine learning techniques, with 
a prevalence of frameworks based on supervised learning, clustering, or a combination of both. %the combination of both of them.
For each framework, we reported and discussed actual application and performance.
We also report possible countermeasures to tackle against mobile traffic analysis and preserve user privacy.
Finally, we discuss challenges, pitfalls, and possible future research directions in the field of mobile traffic analysis.
%performance achieved.

\section*{Acknowledgment}
Mauro Conti is supported by a Marie Curie Fellowship funded by the European Commission (agreement PCIG11-GA-2012-321980). 
This work is partially supported by the EU TagItSmart! Project (agreement H2020-ICT30-2015-688061), the EU-India REACH Project (agreement ICI+/2014/342-896), the grant n. 2017-166478 (3696) from Cisco University Research Program Fund and Silicon Valley Community Foundation, and by the grant "Scalable IoT Management and Key security aspects in 5G systems" from Intel. QianQian Li is supported by Fondazione Cassa di Risparmio di Padova e Rovigo.

\bibliographystyle{myIEEEtran}
\bibliography{papers}

\newpage

\begin{IEEEbiography}[{\includegraphics[width=1in,height=1.25in,clip,keepaspectratio]{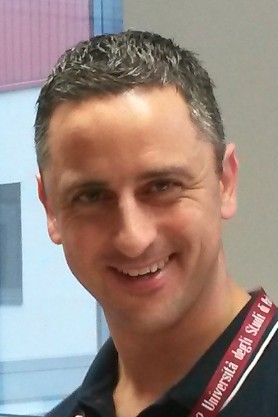}}]{Mauro Conti}
Mauro Conti is an Associate Professor at the University of Padua, Italy. 
He obtained his Ph.D. from Sapienza University of Rome, Italy, in 2009. 
After his Ph.D., he was a Post-Doc Researcher at Vrije Universiteit Amsterdam, The Netherlands. 
In 2011 he joined as Assistant Professor the University of Padua, where he became Associate Professor in 2015. 
In 2018, he became Full Professor for Computer Science and Computer Engineering. 
He has been Visiting Researcher at GMU (2008, 2016), UCLA (2010), UCI (2012, 2013, 2014, 2017), TU Darmstadt (2013), UF (2015), and FIU (2015, 2016). 
He has been awarded with a Marie Curie Fellowship (2012) by the European Commission, and with a Fellowship by the German DAAD (2013). 
His main research interest is in the area of security and privacy. 
In this area, he published more than 200 papers in topmost international peer-reviewed journals and conference. 
He is Associate Editor for several journals, including IEEE Communications Surveys \& Tutorials and IEEE Transactions on Information Forensics and Security. 
He was Program Chair for TRUST 2015, ICISS 2016, WiSec 2017, and General Chair for SecureComm 2012 and ACM SACMAT 2013.

\end{IEEEbiography}

%\vspace{-10mm}

\begin{IEEEbiography}[{\includegraphics[width=1in,height=1.25in,clip,keepaspectratio]{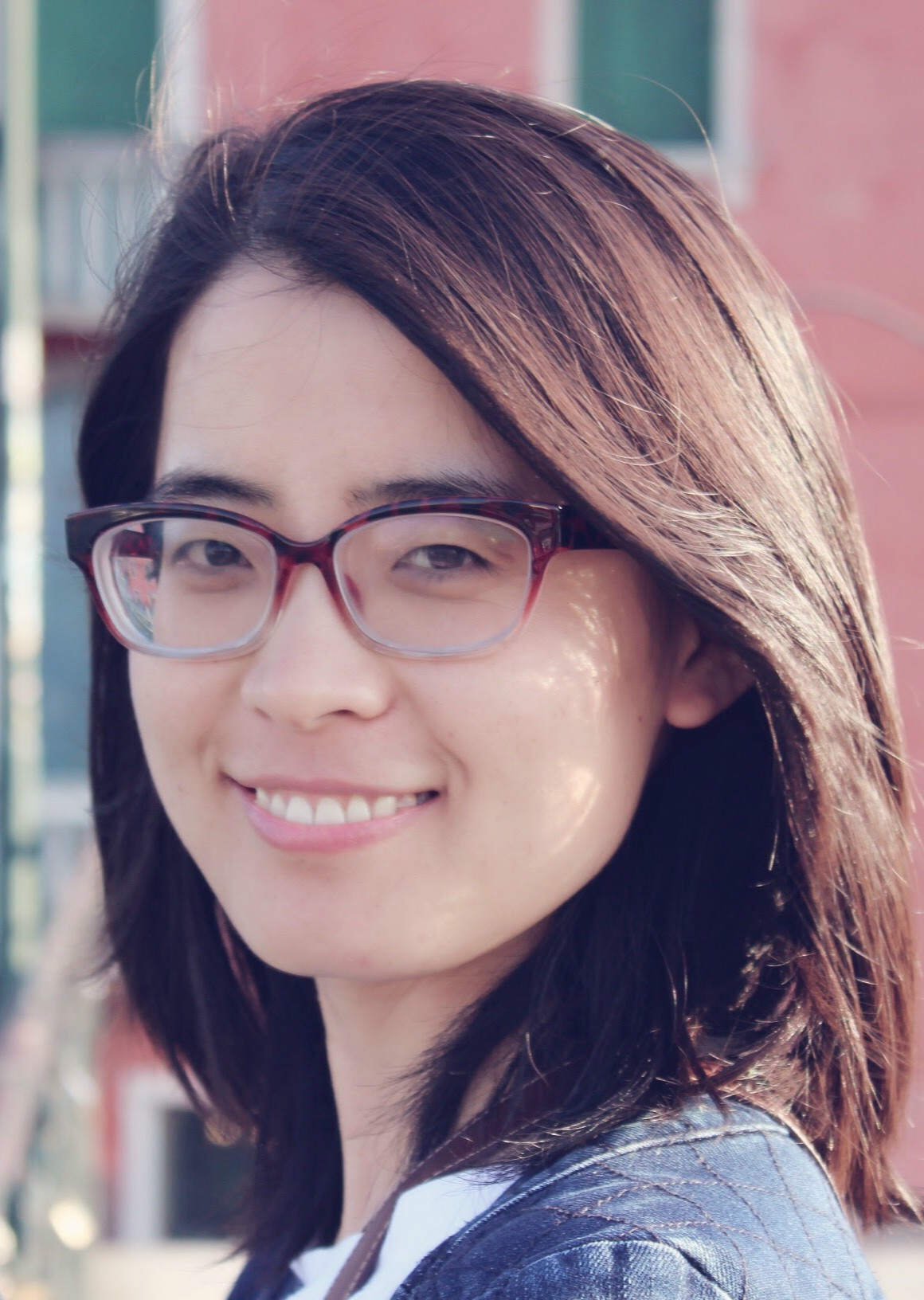}}]{QianQian Li}
obtained her Ph.D. in Brain, Mind, and Computer Science at the University of Padua, Italy, in 2018. 
She finished her Master's Degree in Computer Application and Technology in 2013 at ShanDong Normal University, China. 
In November 2014, she started her Ph.D. under the supervision of Prof. Mauro Conti. 
Her main research is focused on security challenges in human--computer interfaces and software-defined networking.
\end{IEEEbiography}

%\vspace{-10mm}

\begin{IEEEbiography}[{\includegraphics[width=1in,height=1.25in,clip,keepaspectratio]{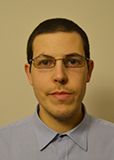}}]{Alberto Maragno}
obtained his Master's Degree in Computer Science from the University of Padua, Italy, in 2016. 
His research activity focuses on security and privacy in the mobile domain. 
In particular, he is interested in traffic analysis techniques targeting mobile devices, as well as in mobile operating system security and mobile app privacy assessment.
\end{IEEEbiography}

%\vspace{-10mm}

\begin{IEEEbiography}[{\includegraphics[width=1in,height=1.25in,clip,keepaspectratio]{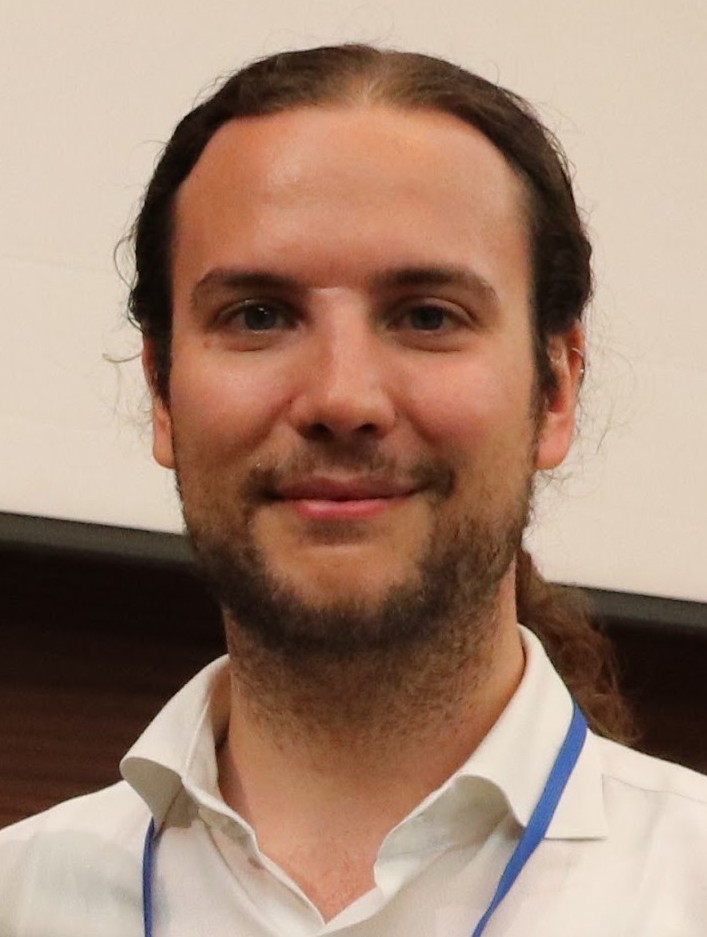}}]{Riccardo Spolaor}
obtained his Ph.D. in Brain, Mind, and Computer Science at the University of Padua, Italy, in 2018. 
He obtained his Master's Degree in Computer Science in 2014 from the same university, with a thesis about a smartphone privacy attack inferring user actions via traffic analysis. 
In November 2014, he started his Ph.D. under the supervision of Prof. Mauro Conti. 
He has been Visiting Ph.D. Student at Radboud University (2015), Ruhr-Universit{\"a}t Bochum (2016), and University of Oxford (2016, 2017, and 2018). 
His main research interests are privacy and security issues on mobile devices. 
In particular, he applies machine learning techniques to infer user information relying on side-channel analysis. 
Most of the research that he carried out up to now is about the application of machine learning classifiers to network traffic and energy consumption traces.
\end{IEEEbiography}
%\balance

\end{document}